\documentclass[cernpreprint,coverpage=false,atlasdraft=false,texlive=2017,UKenglish,subfigure,texmf]{atlasdoc}
 
\usepackage{atlaspackage}
\usepackage{atlasbiblatex}
 
\usepackage{atlasphysics}

\graphicspath{{logos/}{figures/}}

\usepackage{atlasunit}
\usepackage{atlasprocess}
 
\usepackage{atlasjetetmiss}
 
\usepackage{multirow}

\AtlasTitle{Optimisation of large-radius jet reconstruction for the ATLAS detector in 13~TeV proton--proton collisions}

\AtlasJournalRef{Eur. Phys. J. C 81 (2021) 4, 334}
\AtlasDOI{10.1140/epjc/s10052-021-09054-3}
 
\AtlasAbstract{
 
Jet substructure has provided new opportunities for searches and measurements at the LHC, and has seen continuous development since the optimization of the large-radius jet definition used by ATLAS was performed during Run 1. A range of new inputs to jet reconstruction, pile-up mitigation techniques and jet grooming algorithms motivate an optimisation of large-radius jet reconstruction for ATLAS. In this paper, this optimisation procedure is presented, and the performance of a wide range of large-radius jet definitions is compared. The relative performance of these jet definitions is assessed using metrics such as their pileup stability, ability to identify hadronically decaying $W$ bosons and top quarks with large transverse momenta. A new type of jet input object, called a `unified flow object' is introduced which combines calorimeter- and inner-detector-based signals in order to achieve optimal performance across a wide kinematic range. Large-radius jet definitions are identified which significantly improve on the current ATLAS baseline definition, and their modelling is studied using $pp$ collisions recorded by the ATLAS detector at $\sqrt{s}=13~\TeV$ during 2017.
}
 
\author{The ATLAS Collaboration}
 
\AtlasRefCode{JETM-2018-06}

\PreprintIdNumber{CERN-EP-2020-134}
\hypersetup{pdftitle={ATLAS document},pdfauthor={The ATLAS Collaboration}}

\begin{document}
 
\maketitle
 
\tableofcontents

\section{Introduction}\label{sec:intro}
High-energy particle collisions such as those produced in the Large Hadron Collider (LHC) at CERN can result in the production of massive particles (e.g.\ $W$/$Z$/$H$ bosons and top quarks) with large Lorentz boosts. When such particles decay, their decay products become collimated, or `boosted', in the direction of the progenitor particle. For massive particles that are sufficiently boosted, it is advantageous to reconstruct their hadronic decay products as a single large-radius (\largeR) jet. Such \largeR jets capture a characteristic, multi-pronged jet substructure from the two-body or three-body decays of hadronically decaying $W$, $Z$ and $H$ bosons and top quarks, which is distinct from the radiation pattern of a light-quark- or gluon-initiated jet.
 
The substructure of boosted particle decays~\cite{Asquith:2018igt,Larkoski:2017jix} allows powerful new approaches to be utilised in searches for physics beyond the Standard Model (BSM)~\cite{EXOT-2015-04,SUSY-2016-10,EXOT-2016-25,SUSY-2017-01,HDBS-2018-31,CMS-B2G-18-002,CMS-B2G-17-019,CMS-B2G-17-018,CMS-EXO-18-012,CMS-HIG-19-003} at high energy scales, and has enabled novel measurements of Standard Model processes~\cite{STDM-2017-04, STDM-2017-33, STDM-2017-14, Aad:2020zcn,CMS-TOP-19-005,CMS-SMP-16-010,CMS-TOP-17-013,Adam:2020kug,Aaij:2017fak,Aaij:2019ctd,Acharya:2019djg,Zardoshti:2020cwl}.
 
The reconstruction of boosted hadronic systems is complicated by the presence of soft radiation from several sources, which degrades performance when reconstructing jet substructure observables. In particular, soft radiation from the underlying event and uncorrelated radiation from additional $pp$ interactions concurrent with the hard-scattering event of interest (pile-up interactions) can degrade the jet mass resolution and other jet substructure quantities, which are critical to boosted object identification. These effects are amplified by the use of a large radius for jet reconstruction~\cite{Dasgupta:2007wa,Cacciari:2008gn,Cacciari:2010te,Soyez:2012hv}, which incorporates more uncorrelated energy. During Run~1, the average number of pile-up interactions per LHC bunch crossing was roughly 20. This number increased to $\sim34$ in the Run~2 dataset, although some events during this period were recorded with up to 70 pile-up interactions. The average number of pile-up collisions is expected to increase further during Run~3 and will reach $\sim200$ pile-up interactions during high-luminosity LHC operations~\cite{ApollinariG.:2017ojx}. As experimental conditions become more challenging, the choices made when reconstructing \largeR jets will need to evolve to maintain optimal performance.
 
There is no single way to reconstruct a jet, and several choices must be made at the level of a physics analysis to define the jets which will be used. Jets at the LHC are typically reconstructed from some set of input objects (`jet inputs', or simply `inputs' throughout) using a sequential recombination algorithm with a user-specified radius parameter ($R$). Once a jet input type is chosen, it may be preprocessed before jet reconstruction, for example, to mitigate the effects of pile-up.  After jet reconstruction, a grooming algorithm may be applied to the jets which preferentially removes soft and/or wide-angled radiation from the reconstructed jet, to further suppress contributions from pile-up and the underlying event and to enhance the resolution of the jet mass and other substructure observables.
 
\LargeR jets are typically reconstructed by ATLAS using the \akt algorithm~\cite{Cacciari:2008gp} and a radius parameter $R=1.0$. The choice of recombination scheme and radius parameter has been studied previously~\cite{PERF-2012-02}, and is not revisited in these studies. ATLAS \largeR jet reconstruction has so-far been based on topological cluster inputs reconstructed only using calorimeter-based energy measurements. These clusters provide excellent energy resolution, but do not accurately represent the positions of individual particles within jets with large transverse momentum (\pt), particularly in areas where the energy density is large or the calorimeter granularity is coarse. This can result in degraded performance when the resolution of individual particles becomes relevant, for instance, when reconstructing the mass of showers which are so collimated that they are not spatially resolved by the ATLAS calorimeter's granularity. In order to better reconstruct the angular distributions of charged particles within jets, several particle-flow (PFlow) algorithms which were developed and commissioned by ATLAS during Run~2 are considered. These include a PFlow implementation designed to improve $R=0.4$ jet performance at low \pt~\cite{PERF-2015-09}, and a variant designed to reconstruct jet substructure at the highest transverse momenta, called Track-CaloClusters (TCCs)~\cite{ATL-PHYS-PUB-2017-015,HDBS-2018-31}. In this work, a union of PFlow and TCCs called `Unified Flow Objects' (UFOs) is established to provide optimal performance across a wider kinematic range than is possible with either particle-flow objects (PFOs) or TCCs alone, which are each found to perform well in distinct kinematic regions. Jet inputs may also be preprocessed using one or several of the many input-object-level pile-up mitigation techniques which have been developed, such as constituent subtraction~\cite{Berta:2014eza, Berta:2019hnj}, Voronoi subtraction~\cite{Soyez:2018opl}, SoftKiller~\cite{Cacciari:2014gra}, and pile-up per particle identification (PUPPI)~\cite{Bertolini:2014bba}. Various input types and pile-up mitigation algorithms can be combined to create pile-up-robust inputs to jet reconstruction, adding additional complexity to the search for optimal performance.
 
Grooming algorithms are another tool which may be used to remove undesirable radiation from jets after they have been reconstructed. The performance of several grooming algorithms was studied by ATLAS in detail using Run~1 data~\cite{PERF-2015-03} and during preparations for Run~2~\cite{ATL-PHYS-PUB-2014-004}, including the jet trimming~\cite{Krohn:2009th}, pruning~\cite{Ellis:2009me}, and mass drop filtering~\cite{Butterworth:2008iy} algorithms. Based on these studies, \largeR jets groomed with the trimming algorithm using parameter choices of $\Rsub=0.2$ and $\fcut=5$\% were found to be optimal for ATLAS with Run~2 conditions. Since the completion of these studies, several additional jet grooming algorithms have been proposed, including the modified mass drop (mMDT)~\cite{Dasgupta:2013ihk} and soft-drop (SD)~\cite{Larkoski:2014wba} algorithms, and their recent extensions: bottom-up soft-drop (BUSD) and recursive soft-drop (RSD)~\cite{Dreyer:2018tjj}.
 
The development of new input objects, pile-up mitigation techniques and jet grooming algorithms by the experimental and phenomenological communities motivates a thorough reoptimisation of the \largeR jet definition used by ATLAS. In this paper, the jet tagging and substructure performance of 171 distinct combinations of the different jet inputs, pile-up mitigation techniques and grooming algorithms is evaluated using Run~2 conditions. The performance of different jet definitions is compared in the context of several metrics, which quantify their tagging performance, their pile-up stability, and the sensitivity of their mass response to different jet substructure topologies. The performance in data is also studied to ensure the validity of the conclusions from the Monte Carlo studies.
 
The remaining sections of this document are structured as follows. The ATLAS detector is described in Section~\ref{sec:detector}, along with aspects of the 2017 $pp$ dataset and details of the simulated events used to perform these studies. An overview of the jet reconstruction techniques surveyed by these studies is provided in Section~\ref{sec:objects}. Several metrics are used to determine the optimal jet definition, as well as to understand the behaviour of individual algorithms. Due to the large number of possible \largeR jet definitions, a two-stage optimisation is performed to determine which of these exhibit the best performance. In the first stage, presented in Section~\ref{sec:nocalib}, the metrics which will be used to evaluate the relative performance of all jet definitions are established by studying the performance of a limited set of jet definitions. The observations made from these comparisons motivate a union of the existing particle-flow and TCC input objects; this new input object type is presented in Section~\ref{sec:ufos}. The results of the complete survey of jet definitions are presented in Section~\ref{sec:results}. UFO-based definitions which perform consistently well are selected for further study. This smaller list of jet definitions, each of which improves on the current ATLAS baseline \largeR jet definition, is calibrated using simulated events, and a more detailed comparison of their performance in terms of their tagging performance and jet \pt and mass resolutions as well as their performance in data is made in Section~\ref{sec:calib}. In an appendix, more details of the interaction between pile-up interactions and topological cluster formation are provided.
 
\section{The ATLAS detector, data and simulated events}\label{sec:detector}
 
The ATLAS detector~\cite{PERF-2007-01,ATLAS-TDR-2010-19,PIX-2018-001} consists of three principal subsystems\footnote{ATLAS uses a right-handed coordinate system with its origin at the nominal interaction point (IP) in the centre of the detector and the $z$-axis along the beam pipe. The $x$-axis points from the IP to the centre of the LHC ring, and the $y$-axis points upward. Cylindrical coordinates $(r,\phi)$ are used in the transverse plane, $\phi$ being the azimuthal angle around the $z$-axis. The pseudorapidity is defined in terms of the polar angle $\theta$ as $\eta=-\ln\tan(\theta/2)$.}. The inner detector (ID) provides tracking of charged particles within $|\eta|<2.5$ using silicon pixel and microstrip detectors, as well as a transition radiation tracker which provides a large number of hits in the ID's outermost layers in addition to particle identification information. This subsystem is immersed in an axial magnetic field generated by a 2~T solenoid. A sampling calorimeter surrounds the ID and barrel solenoid, providing energy measurements of electromagnetically and hadronically interacting particles within $|\eta| < 4.9$, and is followed by a muon spectrometer.
 
The electromagnetic showers of electrons and photons are measured with a high-granularity liquid argon (LAr) calorimeter, consisting of a barrel module within $|\eta|<1.475$ and two endcaps from $1.365 < |\eta| < 3.2$. Hadronic showers are measured using a steel/scintilator tile calorimeter within $|\eta|<1.7$ and with a pair of LAr/copper endcaps within $1.5 < |\eta| < 3.2$. In the forward region, a LAr/copper and LAr/tungsten forward calorimeter measures showers of both kinds within $3.2 < |\eta| < 4.9$.
 
The muon spectrometer is based one barrel and two endcap superconducting toroidal magnets. Precision chambers provide measurements for all muons within $|\eta|<2.7$, and separate trigger chambers allow the online selection of events with muons within $|\eta| < 2.4$.
 
As writing events to disk at the nominal LHC collision rate of 40~MHz is currently unfeasible, a two-level trigger system is used to select events for analysis. The hardware-based Level-1 trigger accepts events at a rate of $\sim$100~kHz using a subset of available detector information. The software-based High-Level Trigger then reduces the event rate to $\sim$1~kHz, which is retained for further analysis.
 
Studies presented in this paper utilise a dataset of proton--proton collisions delivered by the LHC in 2017 with centre-of-mass-energy $\sqrt{s}=13$~\TeV\ and collected with the ATLAS detector. Data containing high-\pt dijet events were selected using a single-jet trigger, and the leading \akt $R=1.0$ jet is required to have \pt above 600~\GeV. All data are required to meet standard ATLAS quality criteria~\cite{DAPR-2018-01}; data taken during periods when detector subsystems were not functional, which contain significant contamination from detector noise, or where there were detector read-out problems are discarded. The resulting dataset has an integrated luminosity of 44.3~\ifb\ and an associated luminosity uncertainty of 2.4\%~\cite{ATLAS-CONF-2019-021}, obtained using the LUCID-2 detector~\cite{LUCID2} for the primary luminosity measurements.
 
The simulated event samples used to perform these studies were generated using \PYTHIA~8.186~\cite{Sjostrand:2006za,Sjostrand:2007gs} with the NNPDF2.3 LO~\cite{Ball:2012cx} set of parton distribution functions (PDF), a \pt-ordered parton shower, Lund string hadronisation~\cite{Andersson:1983ia,Sjostrand:1984ic}, and the A14 set of tuned parameters (tune)~\cite{ATL-PHYS-PUB-2014-021}. These samples provide `background' jets which originate from high-energy quark and gluon scattering (using a \TwoToTwo matrix element), and `signal' jets originating from high-\pt $W$ boson and top quark decays across a wide kinematic range. The signal $W$ jets were produced using a BSM spin-1 $W' \rightarrow WZ \rightarrow qqqq$ model including only hadronic $W$ and $Z$ decays. The signal top quark jets are taken from a BSM $Z' \rightarrow tt$ model, where the top quarks may decay either hadronically or semileptonically. In order to remove dependence on the specific BSM physics models used to generate these jets, the \pt spectrum of signal jets is always reweighted to match that of background jets~\cite{PERF-2015-04}. Straightforward particle-level containment definitions are used to ensure that the signal jets provide samples of two- and three-pronged jet topologies: the decay partons of the $W$ boson or top quark are required to be within $\Delta R = 0.75$ of the particle-level jet axis. Top jets containing leptonic $W$ boson decays are rejected using particle-level information.
 
All simulated events were passed through the complete ATLAS detector simulation~\cite{SOFT-2010-01} based on $\GEANT4$~\cite{Agostinelli:2002hh} using the FTFP\_BERT\_ATL model~\cite{SOFT-2010-01}. The effect of pile-up was modelled by overlaying the hard-scatter event with minimum-bias $pp$ collisions generated by \PYTHIA~8.210 with the A3 tune~\cite{ATL-PHYS-PUB-2016-017} and the NNPDF2.3 LO PDF set. The number of pile-up vertices was reweighted to match the data events, which have an average of $38$ simultaneous interactions per bunch crossing in the 2017 dataset. Pile-up events are overlaid such that each subdetector reconstructs the effect of signals from adjacent bunch crossings (`out-of-time' pile-up) as well as those from the same bunch crossing as the hard-scatter event (`in-time' pile-up)~\cite{PERF-2014-03}.
 

\section{Objects and algorithms}\label{sec:objects}
This section provides a brief overview of different jet input object, pile-up mitigation and grooming options. All jets discussed in these studies are reconstructed using the \akt algorithm as implemented in \Fastjet~\cite{Cacciari:2011ma} with radius parameter $R=1.0$. All jets used in these results are required to have a minimum \pt of 300~\GeV, and to be within $\eta<1.2$.
 
The complete set of jet input object types, pile-up mitigation and grooming algorithms surveyed is summarised in Table~\ref{tab:summary:algos}. In some cases, additional algorithms or settings were studied but were not found to produce results which differed significantly from those presented here. Notes have been made in Section~\ref{sec:nocalib} when appropriate regarding these omitted jet definitions, and they are indicated in Table~\ref{tab:summary:algos} by an asterisk (*).
 
\begin{table}[ht]
\centering
\caption{Summary of pile-up mitigation algorithms, jet inputs, and grooming algorithms, the abbreviated names used throughout this work, and the relevant parameters tested for each algorithm. UFOs are introduced in Section~\ref{sec:ufos}. Algorithms marked with an asterisk (*) were studied, but were not found to produce results significantly different from other configurations. Such results are not presented in these studies.}
\begin{tabular}{|c||c|c|c|}
\hline
& \textbf{Algorithm} & \textbf{Abbreviation} & \textbf{Settings} \\ \hline \hline
\multirow{3}{*}{\textbf{Jet input objects}} & \multirow{2}{*}{Topological Clusters} & \multirow{2}{*}{Topoclusters} & \multirow{2}{*}{N/A} \\
& & &  \\ \cline{2-4}
& \multirow{2}{*}{Particle-Flow}          & \multirow{2}{*}{PFlow} & \multirow{2}{*}{N/A} \\
& & &        \\ \cline{2-4}
& \multirow{2}{*}{Track-CaloClusters}    & \multirow{2}{*}{TCCs} & \multirow{2}{*}{N/A}  \\
& & &         \\ \cline{2-4}
& \multirow{2}{*}{Unified Flow Objects}    & \multirow{2}{*}{UFOs} & \multirow{2}{*}{N/A} \\
& & &        \\
\hline \hline
\multirow{3}{*}{\textbf{Pile-up mitigation algorithms}} & \multirow{3}{*}{Constituent Subtraction} & \multirow{3}{*}{CS}
& $A_g = 0.01$ \\
&                        & & $\DeltaR_\text{max} = 0.25$ \\
&                        & & $\alpha = 0$ \\ \cline{2-4}
& Voronoi Subtraction (*)& VS
& N/A \\ \cline{2-4}
& SoftKiller             & SK
& $\ell = 0.6$ \\ \cline{2-4}
& \multirow{4}{*}{Pile-up Per Particle Identification} & \multirow{4}{*}{PUPPI}
& $R_\text{min} = 0.001$ \\
&                      & & $R_0 = 0.3$ \\
&                      & & $a = 200~\MeV$ \\
&                      & & $b = 14~\MeV$ \\
\hline \hline
\multirow{5}{*}{\textbf{Jet grooming algorithms}}
& \multirow{2}{*}{Soft-Drop} & \multirow{2}{*}{SD}
& $\zcut$ = 0.1     \\
&                      & & $\beta$ = 0, 1, 2(*)    \\ \cline{2-4}
& \multirow{2}{*}{Bottom-up Soft-Drop} & \multirow{2}{*}{BUSD}
& $\zcut$ = 0.05, 0.1   \\
&                      & & $\beta$ = 0, 1, 2(*)  \\ \cline{2-4}
& \multirow{3}{*}{Recursive Soft-Drop} & \multirow{3}{*}{RSD}
& $\zcut$ = 0.05, 0.1   \\
&                      & & $\beta$ = 0, 1, 2(*)  \\
&                      & & $N$ = 3, 5(*), $\infty$  \\ \cline{2-4}
& \multirow{2}{*}{Pruning}  & \multirow{2}{*}{N/A}
& \zcut = 0.15    \\
&                      & & \Rcut = 0.25     \\    \cline{2-4}
& \multirow{2}{*}{Trimming} & \multirow{2}{*}{N/A}
& \fcut = 5\%, 9\%   \\
&                      & & \Rsub = 0.1, 0.2 \\
\hline 
\end{tabular}
\label{tab:summary:algos}
\end{table}

\subsection{Jet input objects}
 
\subsubsection{Stable generator-level particles}\label{sec:objects:truth}
 
Particle-level jets, or `truth jets', are reconstructed in simulated events at generator level. All detector-stable particles from the hard-scattering process with a lifetime $\tau$ in the laboratory frame such that $c\tau > 10$ mm are used. Particles that are expected to leave only negligible energy depositions in the calorimeter, i.e.\ muons and neutrinos, are excluded.
 
Ungroomed particle-level jets are used as the reference objects for selections throughout these studies in order to ensure that the same set of reconstructed jets are selected for comparison, regardless of the jet input objects used in reconstruction or grooming algorithm applied. In studies of simulated jets, unless otherwise specified, ungroomed particle-level jets are geometrically matched ($\DeltaR$ < 0.75) to ungroomed reconstructed jets, and kinematic selections are applied to the ungroomed particle-level jet four-vector.
 
Particle-level jets are also taken as the reference for simulation-based ATLAS jet calibrations, and for studies of the jet energy and mass resolution. In this circumstance, they are groomed using the same algorithm and parameters as the reconstructed jets to which they are being compared (Section~\ref{sec:calib}).
 
\subsubsection{Inner detector tracks}
 
Tracks are reconstructed from charged-particle hits in the inner detector. In order to ensure that only well-reconstructed tracks from the hard scattering are used, track quality criteria are applied. The `loose' quality working point is used, which places requirements on the number of silicon hits in each subdetector~\cite{ATL-PHYS-PUB-2015-051}. Tracks are associated to the primary vertex (PV) of the hard interaction by placing a requirement on the track distance of closest approach to the PV along the $z$ axis, $|z_{0}\sin{\theta}| < 2.0~\text{mm}$. The PV is selected as the vertex with the highest scalar $\pt^2$ sum of tracks associated with it using transverse and longitudinal impact parameter requirements. In addition, tracks are required to have \pt > 500~\MeV\ and to be within the tracking volume (|$\eta$| < 2.5).
 
\subsubsection{Topological clusters}
 
Jets reconstructed from ATLAS calorimeter information are built from `topoclusters'~\cite{PERF-2014-07}, which are three-dimensional groupings of topologically connected calorimeter cells. Topoclusters are formed using iterated `seed' and `collect' steps based on the absolute value of the signal significance in a cell relative to the expected noise, $\sigma_\text{noise}$, which considers both electronic noise and stochastic noise from pile-up interactions. Cells with signal significance over $4\sigma_\text{noise}$ in an event are allowed to seed topocluster formation, and their neighbouring cells with significance over $2\sigma_\text{noise}$ are subsequently included. This step is repeated until all adjacent cells have a significance below $2\sigma_\text{noise}$, at which point all neighbouring cells are added to the cluster ($0\sigma_\text{noise}$). If this process results in a cluster with two or more local energy maxima, a splitting algorithm is used to separate the showers. The energies of the resulting set of clusters are calibrated at the electromagnetic (EM) scale, and all clusters are taken to be massless.
 
An additional calibration using the local cell weighting (LCW) scheme is applied to form clusters whose energy is calibrated at the correct particle-level scale~\cite{PERF-2014-07}. This weighting scheme classifies energy depositions as either electromagnetic- or hadronic-like using a variety of cluster moments, and accounts for the non-compensating response of the calorimeter, out-of-cluster energy, and for energy deposited in the dead material within the detector.
 
Finally, the angular coordinates ($\eta$ and $\phi$) of topoclusters are recalculated relative to the primary vertex of the event, instead of the geometric centre of the ATLAS detector. A detailed description of topocluster reconstruction and calibration is provided in Ref.~\cite{PERF-2014-07}.
 
\subsubsection{Particle-flow objects (PFOs)}\label{sec:objects:pfo}
 
Particle-flow (PFlow) reconstruction combines track- and calorimeter-based measurements and results in improved jet energy and mass resolution, and improved pile-up stability relative to jets reconstructed from topoclusters alone~\cite{PERF-2015-09,Aad:2020flx}. Double-counting of contributions from the momentum measurement of charged particles in the inner detector and their energy measurement from the calorimeters is removed using a cell-based energy subtraction.
 
The PFlow algorithm first attempts to match each selected track to a single topocluster in the calorimeter, using topoclusters calibrated to the EM scale, and tracks selected using the ``tight'' quality working point~\cite{ATL-PHYS-PUB-2015-051}. The track momentum and the topocluster position are used to compute the expected energy deposition in the calorimeter by the particle that created the track. It is not uncommon for a single particle to deposit energy in multiple topoclusters. For each track/topocluster system, the PFlow algorithm evaluates the probability that the particle's energy was deposited in more than one topocluster, and may include additional topoclusters in the track/topocluster system if they are necessary to reconstruct the full shower energy. The expected energy deposited in the calorimeter by the particle that produced the track is subtracted, cell-by-cell, from the associated topoclusters. If the associated calorimeter energy following this subtraction is consistent with the expected shower fluctuations of a single particle, the remaining calorimeter energy is removed.
 
Topoclusters which are not matched to any tracks are assumed to contain energy deposited by neutral particles and are left unmodified. In the cores of jets, particles are often produced at higher energies and in dense environments, decreasing the advantages of using inner-detector-based measurements of charged particles. To account for this degradation of inner tracker performance, the shower subtraction is gradually disabled for tracks with momenta below $100~\GeV$ if the energy $E_\text{clus}$ deposited in the calorimeter in a cone of size $\DeltaR=0.15$ around the extrapolated track trajectory satisfies
\[
\frac{E^{\textrm{clus}}-\langle E_{\textrm{dep}} \rangle}{\sigma(E_{\textrm{dep}})}>33.2\times\log_{10}(40~\GeV / ~\pT^{\textrm{trk}})\,,
\]
where $E_{\textrm{dep}}$ is the expected energy deposition from a charged pion. The subtraction is completely disabled for tracks with $\pt>100$~\GeV\ when this condition is satisfied.
 
After the PFlow algorithm has run to completion, the collection of particle-flow objects (PFOs) consists of tracks, and both modified and unmodified topoclusters. Charged PFOs which are not matched to the PV are removed in order to reduce the contribution from pile-up; this procedure is referred to as `Charged Hadron Subtraction' (CHS)~\cite{CMS-JME-18-001,CMS-PRF-14-001}.
 
\subsubsection{Track-CaloClusters (TCCs)}\label{sec:objects:tccs}
 
Track-CaloClusters (TCCs)~\cite{ATL-PHYS-PUB-2017-015} were developed in the context of searches for massive BSM diboson resonances~\cite{HDBS-2018-31}. These constituents combine calorimeter- and inner-detector-based measurements in a manner which is optimised for jet substructure reconstruction performance in the highest-\pt jets. Unlike PFlow, which uses the expected energy depositions of single particles to determine the contributions of individual tracks to clusters, the TCCs use the energy information from topoclusters and angular information from tracks.
 
The TCC algorithm starts by attempting to match each `loose' track in the event (from both the hard-scatter and pile-up vertices) to topoclusters calibrated to the local hadronic scale in the calorimeter. In the case where one track matches one topocluster, the $\pt$ of the TCC object is taken from the topocluster, while its $\eta$ and $\phi$ coordinates are taken from the track. In more complex situations where multiple tracks are matched to multiple topoclusters, several TCC objects are created (where the TCC multiplicity is equal to the track multiplicity): each TCC object is given some fraction of the momentum of the topocluster, where that fraction is determined from the ratios of momenta of the matched tracks. TCC angular properties ($\eta$, $\phi$) are taken directly from the unmodified inner detector tracks, and their mass is set to zero.
 
As in PFlow reconstruction, unmatched topoclusters are included in the TCC objects as unmodified neutral objects.

\subsection{Jet-input-level pile-up mitigation algorithms}
 
Prior to jet reconstruction, the set of input objects may be preprocessed by one or by a combination of several input-level pile-up mitigation algorithms. When reconstructing jets from topoclusters, these algorithms are applied to the entire set of inputs. When incorporating tracking information, the PV provides an additional, powerful method to reject charged particles from pile-up interactions. In this case, these additional pile-up mitigation algorithms are applied only to the neutral PFOs or TCCs in an event before jet finding.
 
\subsubsection{Constituent Subtraction (CS)}
 
Constituent Subtraction~\cite{Berta:2014eza} is a per-particle method of performing area subtraction~\cite{Soyez:2009cw} on jet input objects. The catchment area~\cite{Cacciari:2008gn} of each input object is defined using ghost association: massless particles called `ghosts' are overlaid on the event uniformly, with \pt satisfying
\begin{equation*}
\pt^{\text{g}}= A_{\text{g}} \times \rho,
\end{equation*}
where $A_g$, the area of the ghosts, is set to 0.01 and $\pT^g$ corresponds to the expected contribution from pile-up radiation in a small $\Delta \eta$--$\Delta\phi$ area of $0.1\times0.1$. For each event, the pile-up energy density $\rho$ is estimated as the median of the $\pt/A$ distribution of the $R=0.4$ \kt~\cite{Ellis:1993tq} jets in the event. These jets are reconstructed without a \pt requirement, but are required to be within $|\eta| < 2.0$. The total \pt of all of the ghosts is equal to the expected average pile-up contribution, based on the estimated value of $\rho$.
 
After the ghosts have been added, the distance $\Delta R_{i,k}$ between each cluster $i$ and ghost $k$ is given by\footnote{In the original formulation, there is also the option to make a $\pt^\alpha$-dependent distance metric. Only values of $\alpha=0$ were considered in Ref.~\cite{Berta:2014eza}, and so only this configuration is considered in these studies.}
\begin{equation*}
\Delta R_{i,k} = \sqrt{(\eta_i - \eta_k)^2+(\phi_i - \phi_k)^2}.
\end{equation*}
The cluster--ghost pairs are then sorted in order of ascending $\Delta R_{i,k}$, and the algorithm proceeds iteratively through each $(i,k)$ pair,
modifying the \pt of each cluster and ghost by
\begin{center}
\begin{tabular}{c l}
If $p_{\text{T,}i} \geq p_{\text{T},k}$: & $p_{\text{T,}i}\longrightarrow p_{\text{T,}i} - p_{\text{T,}k}$, \\
& $p_{\text{T,}k}\longrightarrow 0$; \\
otherwise: & $p_{\text{T,}k} \longrightarrow p_{\text{T,}k} - p_{\text{T,}i}$, \\
& $p_{\text{T,}i}\longrightarrow 0$.
\end{tabular}
\end{center}
until $\Delta R_{i,k} > \Delta R_{\text{max}}$, where $\Delta R_{\text{max}}$ is a free parameter of the algorithm taken to be 0.25 in this study, based on studies of $R=0.4$ jet performance~\cite{ATLAS-CONF-2017-065}. Any ghosts remaining after the subtraction are eliminated.
 
In the authors' description of this algorithm, a correction is also applied for the mass of the input object. Since all neutral ATLAS jet inputs are defined to be massless, this correction is unnecessary in the ATLAS implementation.
 
\subsubsection{SoftKiller (SK)}
 
The SoftKiller (SK)~\cite{Cacciari:2014gra} algorithm applies a \pt cut to input objects. This cut is chosen on an event-by-event basis such that the value of $\rho$ after the selection is approximately zero. To achieve this, the event is divided into an $\eta$--$\phi$ grid of user-specified length scale, chosen to be $\ell=0.6$, based on studies of $R=0.4$ jet performance~\cite{ATLAS-CONF-2017-065}. The \pt cut is determined in order to make half of the grid spaces empty after it is applied (input objects are removed from all grid cells, not just the half which are empty following SK).
 
To account for detector-level effects, where input objects may not consist purely of hard-scatter or pile-up contributions (see appendix), the best performance is achieved by applying some form of area subtraction to input objects before applying SK. In these studies, SK is always applied to inputs after the CS algorithm; this combination is indicated as `CS+SK'.
 
An alternative approach to assigning areas to jet input objects is based on Voronoi tesselation~\cite{Soyez:2018opl} and was studied both in isolation and in conjunction with the SoftKiller algorithm. Both variants of this alternative were found to perform similarly to the CS+SK results presented here.
 
\subsubsection{Pile-up Per Particle Identification (PUPPI)}
 
`Pile-up per particle identification', or PUPPI~\cite{Bertolini:2014bba}, is a pile-up-mitigation algorithm which assigns each input object $i$ a likelihood to have originated from a pile-up interaction based on its kinematic properties and proximity to charged hard-scatter particles matched to the event's PV. This likelihood is given by
\[
\alpha_{i} = \log\left(\sum_{j} \frac{\pt^{j}}{\Delta R^{ij}} \times \Theta\left(R_\mathrm{min} \le \Delta R_{ij} \le R_{0}\right)\right),
\]
where the index $j$ tracks the charged inputs matched to the PV, $R_{0}$ is the maximum radial distance at which inputs may be matched to each other, $R_\mathrm{min}$ is the minimum radial distance of matching, $\Delta R^{ij}$ is the angular distance between an input object and a charged hard-scatter particle, and $\Theta$ is the Heaviside step function. The value of $R_\mathrm{min}$ is generally taken to be very small, and is chosen to be $0.001$ in these studies.  The value of $R_{0}$ is chosen to be $0.3$.
 
Once $\alpha$ has been calculated for all input objects, then the following quantity is determined:
\[
\chi_{i}^2 = \Theta\left(\alpha_i - \bar{\alpha}_\mathrm{PU}\right) \times \frac{\left(\alpha_i - \bar{\alpha}_\mathrm{PU}\right)^2 }{\sigma_\mathrm{PU}^2},
\]
where $\bar{\alpha}_\mathrm{PU}$ is the mean value of $\alpha$ for all charged pile-up input objects in the event, and $\sigma_\mathrm{PU}$ is the RMS of that same distribution. The four-momentum of each neutral input $i$ is then weighted by
\[
w_{i} = F_{\chi^{2},\mathrm{NDF=1}}\left(\chi_{i}^{2}\right),
\]
where $F_{\chi^{2}}$ is the cumulative distribution of the $\chi^2$ distribution, eliminating all neutral inputs $i$ whose
calculated value of $\alpha_{i}$ is less than $\bar{\alpha}_\mathrm{PU}$.
 
In order to suppress additional noise, a $\pt$ cut is applied to the remaining input objects after they have been reweighted. This cut is dependent on the number of reconstructed primary vertices (\Npv), and is determined by
\[
p_\mathrm{T,cut} = a + b \times \Npv
\]
where the parameters $a$ and $b$ are user-specified. For these studies, the parameters are chosen to be $a = 200$~\MeV\ and $b = 14$~\MeV, based on studies of the $R=0.4$ PFlow jet energy resolution.
 
While PUPPI could technically be applied to topoclusters, the principles of the algorithm depend strongly on the matching of neutral input objects
to nearby charged particles from the hard-scatter event. It is therefore more effective for particle-flow-type algorithms. Due to the large number of free parameters,
and since it has only been optimised for ATLAS PFlow jets with $R=0.4$, PUPPI is only applied to PFlow jets.

\subsection{Grooming algorithms}
 
\subsubsection{Trimming}
 
Trimming~\cite{Krohn:2009th} was designed to remove contamination from soft radiation in the jet by excluding regions of the jet where the energy flow originates mainly from the underlying event, pile-up, or initial-state radiation (ISR), in order to improve the resolution of the jet energy and mass measurements. In Run~1~\cite{PERF-2012-02}, it was also found to be effective in mitigating the effects of pile-up on \largeR jets. To trim a \largeR jet, the jet constituents are reclustered into subjets of a user-specified radius \Rsub using the \kt algorithm. Subjets with \pt less than some user-specified fraction \fcut of the \pt of the original ungroomed jet are discarded: their constituents are removed from the final groomed jet.
 
\subsubsection{Pruning}
 
Pruning~\cite{Ellis:2009me} proposes a modification of the jet clustering sequence, which removes splittings that are assessed as likely to pull in soft radiation from pile-up interactions and the underlying event. This is achieved by determining a `pruning radius' such that hard prongs fall into separate subjets, while discarding softer radiation outside of these prongs. The constituents of the \largeR jet are reclustered using the Cambridge--Aachen (C/A) algorithm~\cite{Dokshitzer:1997in,Wobisch:1998wt} to form an angle-ordered cluster sequence. At each step of the clustering sequence, the softer subjet is discarded if it is either too soft or wide-angled, enforced by requiring
\[\Delta R_{12} \ge \Rcut \times 2\frac{M_{12}}{p_{\text{T,}12}},\]
\[z \le \zcut,\]
where $\Delta R_{12}$, $M_{12}$, and $p_{\text{T,}12}$ are respectively the angular distance, the mass, and the transverse momentum of the subjet pair at a given step in the clustering sequence, and $z = \text{min} \left( p_{\text{T,1}}, p_{\text{T,2}} \right) / \left(p_{\text{T,1}}+p_{\text{T,2}  }\right)$. The parameters $\Rcut$ and $\zcut$ are user-defined, and respectively control the amount of wide-angled and soft radiation which is removed by the pruning algorithm.
 
\subsubsection{Soft-Drop (SD)}
 
Soft-drop~\cite{Larkoski:2014wba} is a technique for removing soft and wide-angle radiation from a jet. In this algorithm, the constituents of the \largeR jet are reclustered using the C/A algorithm, creating an angle-ordered jet clustering history. Then, the clustering sequence is traversed in reverse (starting from the widest-angled radiation and iterating towards the jet core). At each step in the clustering sequence, the kinematics of the splitting are tested with the condition
\[\frac{\mathrm{min}(p_{\text{T,}1}, p_{\text{T,}2})}{p_{\text{T,}1} + p_{\text{T,}2}} \textless \zcut \Big( \frac{\Delta R_{12}}{R} \Big)^{\beta}\,,\]
where the subscripts $1$ and $2$ respectively denote the harder and softer branches of the splitting, and the parameters $\zcut$ and $\beta$ dictate the amount of soft and wide-angled radiation which is removed. If the splitting fails this condition, the lower-\pt branch of the clustering history is removed, and the declustering process is repeated on the higher-\pt branch. If the condition is satisfied, the process terminates and the remaining constituents form the groomed jet.
 
If $\beta$ = 0, SD suppresses radiation purely based on the \pt, while larger values of $\beta$ allow more soft radiation to remain within the groomed jet when it is sufficiently collinear. SD with $\beta=0$ is equivalent to the modified Mass Drop Tagger (MDT) algorithm~\cite{Marzani:2017mva,PERF-2012-02}. SD grooming has an intrinsic quality which is not shared by the trimming or pruning algorithms: certain jet substructure observables are calculable beyond leading-logarithm accuracy following the application of SD~\cite{Marzani:2017mva,Marzani:2017kqd,Frye:2016aiz,Frye:2016okc,Kang:2018vgn,Kang:2018jwa,Hoang:2017kmk}.
 
\subsubsection{Recursive Soft-Drop (RSD) and Bottom-Up Soft-Drop (BUSD)}
 
The standard soft-drop algorithm aims to find the first hard splitting in the jet clustering history in order to define a groomed jet. In the case of a multi-pronged decay, this treatment may not be sufficient to remove enough soft radiation from the jet, since the SD condition may be satisfied before removing all of this energy. A recursive extension of the SD algorithm (`recursive soft-drop,' or RSD) has been proposed~\cite{Dreyer:2018tjj}, in which the algorithm continues recursively along the harder branch of the C/A clustering sequence until $N$ hard splittings have been found. The case of $N$=1 is equivalent to the standard SD algorithm, while for larger values of $N$, a larger fraction of the jet may be traversed by the grooming algorithm. When $N=\infty$, the entire C/A sequence is traversed by the grooming algorithm regardless of the number of hard splittings found.
 
Bottom-up soft-drop (BUSD)~\cite{Dreyer:2018tjj} instead incorporates the SD criteria within the jet clustering algorithm, similar to pruning. In these studies, the `local' version of BUSD is implemented, which is applied after initial jet reconstruction. Using this approach, jets are reconstructed with the \antikt algorithm, and then reclustered using a modified version of the C/A algorithm, where particles $i$ and $j$ with the smallest distance $d_{ij} = \Delta R_{ij} / R_{0}$ are combined to create a new pseudojet given by
\[
p_{ij} = \left\{
\begin{array}{ll}
\mathrm{max}(p_{i}, p_{j})~\mathrm{,~if~the~\textrm{soft-drop}~condition~fails,} \\
p_{i} + p_{j}~\mathrm{,~otherwise}.  \\
\end{array}
\right.
\]
The results of applying local BUSD are expected to be similar to those of RSD with $N=\infty$, since both algorithms begin with the same set of constituents per jet and groom the entire C/A clustering sequence.
 
Other configurations for the SD family of algorithms were studied, including $\beta=2$ grooming, but were not found to give results significantly different from those reported in detail.

\section{Performance metrics}\label{sec:nocalib}
In order to survey the relative performance of all considered \largeR jet definitions, several metrics must be established which probe relevant aspects of their behaviour in the context of \largeR jet reconstruction and calibration by ATLAS. It is not feasible to calibrate each of the definitions studied (even with a simulation-based approach, as in Section~\ref{sec:calib}), and so these metrics have been chosen in order to be robust against differences caused by calibration. The metrics selected include the tagging performance of high-\pt $W$ bosons and top quarks, the stability of the jets in the presence of pile-up interactions, and the degree to which a jet definition's mass scale depends on the signal- or background-like substructure of the jet.
 
In this section, the behaviour of each metric is illustrated using a reduced list of jet definitions that have been selected to highlight the interplay between different aspects of jet reconstruction. For each metric, jets reconstructed from topological clusters, particle-flow and track-calocluster input objects are compared, with and without pile-up mitigation. Two grooming algorithms are also compared for each jet input: trimming with $\Rsub=0.2$ and $\fcut=0.05$, and soft-drop with $\beta=1.0$ and $\zcut = 0.1$. The trimming algorithm is chosen because it is the current baseline definition used by ATLAS. The soft-drop algorithm is chosen as an alternative which has demonstrated good performance, as is shown in Section~\ref{sec:results}.
 
Results of the complete survey of all jet definitions summarised in Table~\ref{tab:summary:algos} are provided in Section~\ref{sec:results}.
 
\subsection{Tagging performance}\label{sec:nocalib:tagging}
 
Many analyses using \largeR jets rely on a tagger to distinguish between different types of jets, such as distinguishing between the decay of a high-\pt, hadronically decaying top quark and a jet originating from a high-energy quark or gluon. Such boosted-particle taggers range in complexity from simple mass cuts to complex machine-learning algorithms~\cite{JETM-2018-03,CMS-JME-13-006,CMS-JME-18-002}. While the complete optimisation of a jet tagger is outside the scope of this work, it is important to compare the tagging performance of different jet definitions in terms of their background rejection (defined as the reciprocal of the background-jet tagging efficiency) at fixed signal-jet tagging efficiency. This may be done using a simple tagger based on the jet mass and a jet substructure (JSS) observable. In order to study the tagging performance for different jet topologies, taggers are created for high-\pt $W$ bosons and top quarks by combining the jet mass with another jet substructure observable which is sensitive to either two- or three-pronged signal jet topologies.
 
The jet mass, as defined by
\[
\massjet = \sqrt{\left(\sum_{i \in \text{jet}} E_i\right)^2-\left(\sum_{i \in \text{jet}} \vec{p}_i\right)^2},
\]
where $i$ are the constituents of the jet, is typically one of the most powerful variables that can be used to discriminate between different types of jets.
 
To tag boosted $W$ decays, which have a two-pronged structure, the $\DTwo$ observable~\cite{Larkoski:2013eya,Larkoski:2014gra,Larkoski:2015kga} is used with a choice of angular exponent $\beta=1.0$.
This observable is a ratio of three-point to two-point energy--energy correlation functions which has been used by ATLAS in $W$ taggers since Run~1~\cite{PERF-2015-03,JETM-2018-03}.
 
For boosted top quark decays, which have a three-pronged structure, \tauthrtwo with the winner-take-all axis configuration~\cite{Thaler_2011,Thaler_2012} is used.  This observable is a ratio of two $N$-subjettiness variables, which tests the compatibility of a jet's substructure with a particular $N$-pronged hypothesis.
ATLAS has incorporated \tauthrtwo into its top taggers, whether simple or complex, since Run~1~\cite{PERF-2015-04,JETM-2018-03}.
 
Unlike a mass-only tagger, where more aggressive grooming can improve the jet mass resolution at the cost of grooming away additional information contained within a jet's soft radiation, a mass+JSS tagger relies on such soft radiation to achieve better background rejection. Such taggers are a more realistic approximation to the expected future tagging performance of any given jet definition (which will use more sophisticated techniques), and are amenable to this survey of many jet definitions.
 
For both the $W$ and top taggers, the tagging algorithm proceeds similarly: first, a fixed signal-efficiency ($\epsilon_\text{sig}$) mass window is selected, where the window is defined to be the minimum mass range which contains 68\% of the signal mass distribution. This window should select the signal jet mass peak. A one-sided cut is then applied to \DTwo or \tauthrtwo, and background rejection ($1/\epsilon_\text{bkg}$) is compared at a fixed signal efficiency taken to be $\epsilon_\text{sig}=50\%$. This signal efficiency working point is representative of taggers used by ATLAS in physics analysis, and the results were not found to depend strongly on the working point which was selected. The relative performance of various jet definitions in terms of their background rejection at a fixed signal efficiency point was noted to typically provide a consistent ordering of jet definitions before and after applying a simulation-based calibration, and so this metric was selected instead of possible alternatives such as the Receiver Operating Characteristic (ROC) curve integral.
 
The background rejection for the boosted $W$ boson tagger is shown as a function of signal tagging efficiency in Figure~\ref{nocalib:tag:w} for two \pt bins: a low-\pt bin (300~\GeV~< $\pt^\text{true, ungroomed}$ < 500~\GeV), and a high-\pt bin (1000~\GeV~< $\pt^\text{true, ungroomed}$ < 1500~\GeV), where kinematic requirements are placed on the \pt of the ungroomed particle-level jet which is associated with the detector-level jet under study (Section~\ref{sec:objects:truth}). The low-\pt bin represents the regime where the $W$ decay products are boosted just enough to be contained within a single \largeR jet, while the high-\pt bin represents the regime where the decay products are more collimated and may begin to merge. The performance in these two regions is expected to be different due to detector effects and algorithmic differences. Similarly, the background rejection of the top tagger is shown in Figure~\ref{nocalib:tag:top}, except the lower \pt bin is chosen to be 500~\GeV~< $\pt^\text{true, ungroomed}$ < 1000~\GeV, since the larger mass of the top quark results in less collimation of its decay products.
 
Better alternatives to the baseline topocluster jet definition are clearly visible. At low \pt, PFlow reconstruction results in the best performance for $W$ boson and top tagging, while TCCs have a lower background rejection than topocluster jets. At high \pt, TCCs provide a significantly better background rejection than the other options, although PFlow still provides an improvement over topocluster reconstruction.
 
The application of CS+SK pile-up mitigation has very little effect for the high-\pt jets, but for the low-\pt $W$ tagger, it significantly improves the background rejection for soft-drop jets, which are more susceptible to pile-up than trimmed jets. This effect is seen for all three jet input types, but it is pronounced for topocluster inputs, which do not use tracking information to remove pile-up. Top tagging performance benefits more from adopting soft-drop grooming than $W$ tagging: background rejection increases when tagging top quarks regardless of the input object type or \pt bin when soft-drop is chosen.
 
\begin{figure}[htbp]
\begin{center}
\subfigure[]{\includegraphics[width=0.49\textwidth]{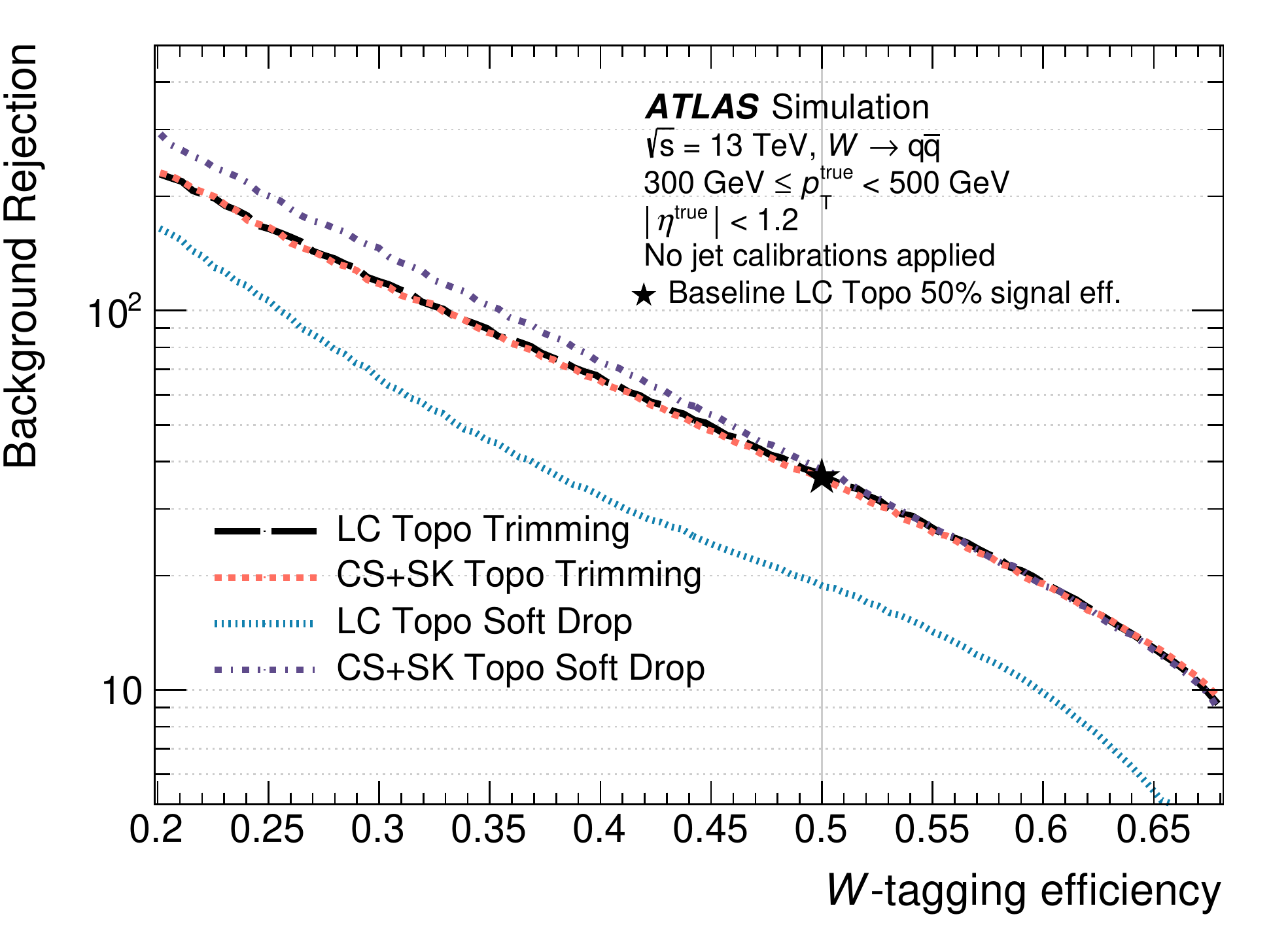}}
\subfigure[]{\includegraphics[width=0.49\textwidth]{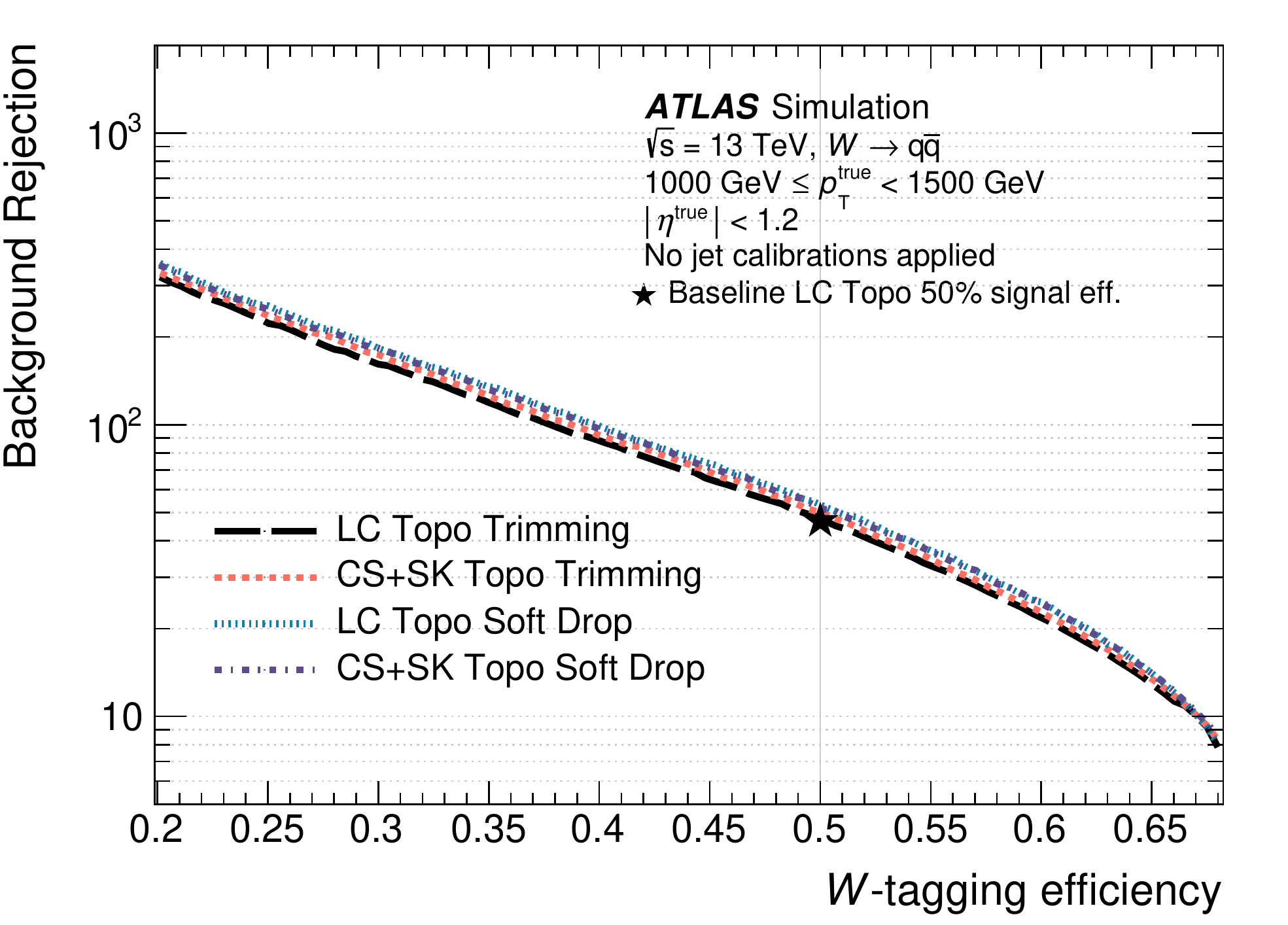}}\\
\subfigure[]{\includegraphics[width=0.49\textwidth]{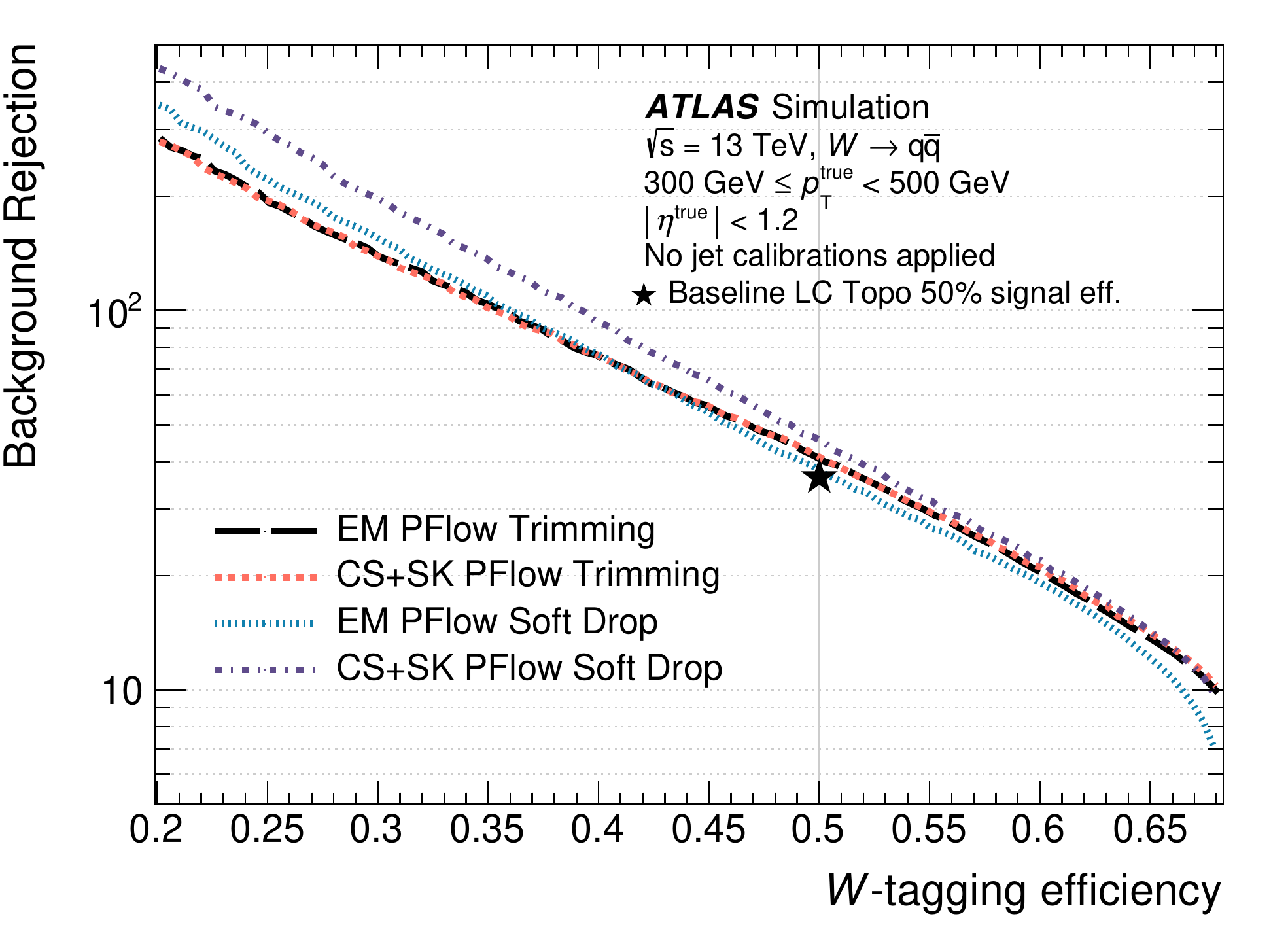}}
\subfigure[]{\includegraphics[width=0.49\textwidth]{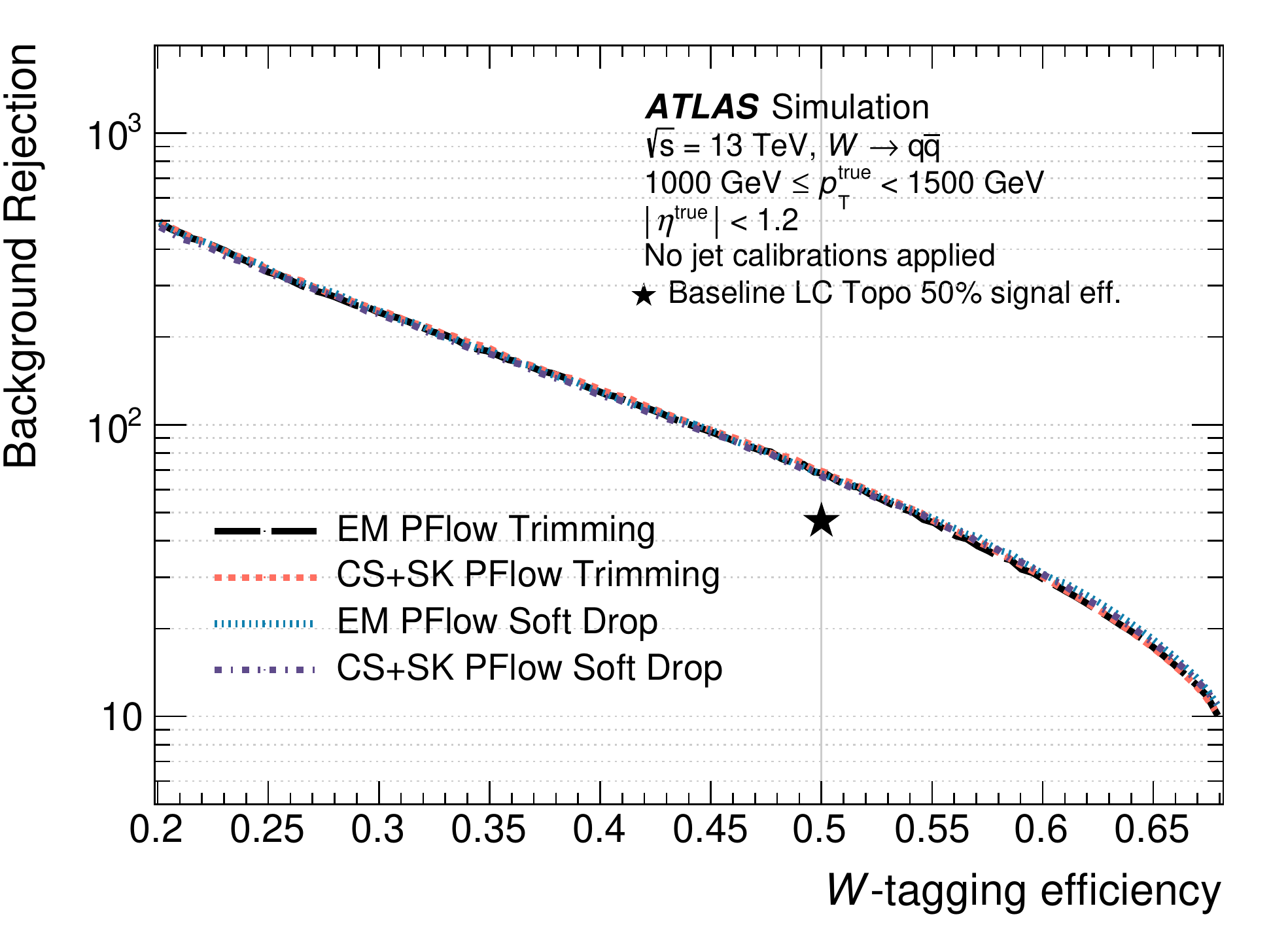}}\\
\subfigure[]{\includegraphics[width=0.49\textwidth]{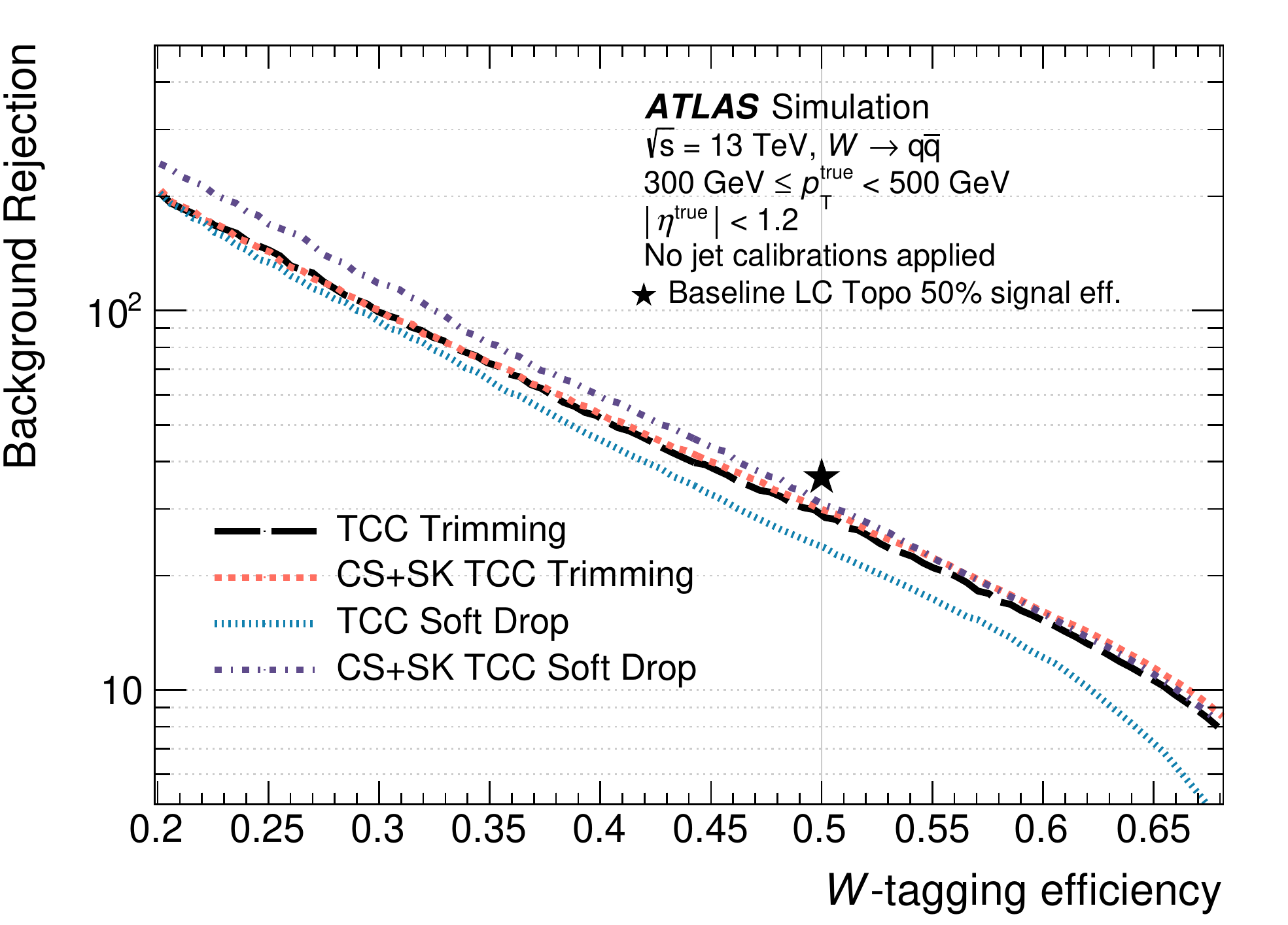}}
\subfigure[]{\includegraphics[width=0.49\textwidth]{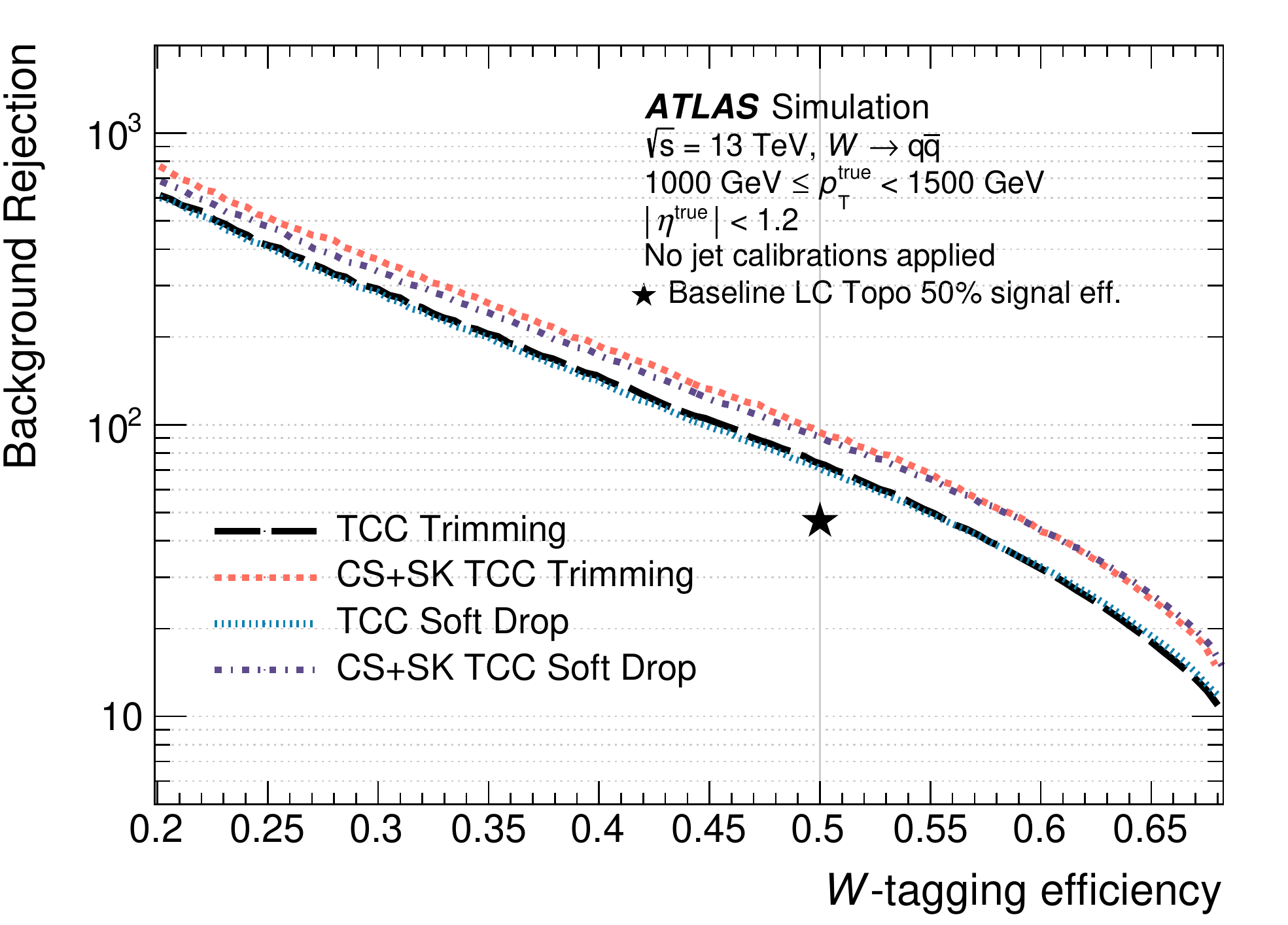}}
\end{center}
\caption{Background rejection as a function of signal efficiency for a tagger using the jet mass and \DTwo for $W$ boson jets at (a,c,e) low \pt, and (b,d,f) high \pt. Several different jet input object types are shown: (a,b) topoclusters, (c,d) particle-flow objects and (e,f) track-caloclusters. Jet \pt and $\eta$ cuts before tagging are made using the ungroomed particle-level \largeR jet matched to each of the groomed reconstructed \largeR jets. Jets groomed with the trimming ($\Rsub=0.2$, $\fcut=0.05$) and soft-drop ($\beta=1.0$, $\zcut = 0.1$) algorithms are shown. The background rejection factor of the baseline topocluster-based trimmed collection at a fixed signal tagging efficiency of 50\% is indicated with a $\star$.}
\label{nocalib:tag:w}
\end{figure}
 
\begin{figure}[htbp]
\begin{center}
\subfigure[]{\includegraphics[width=0.49\textwidth]{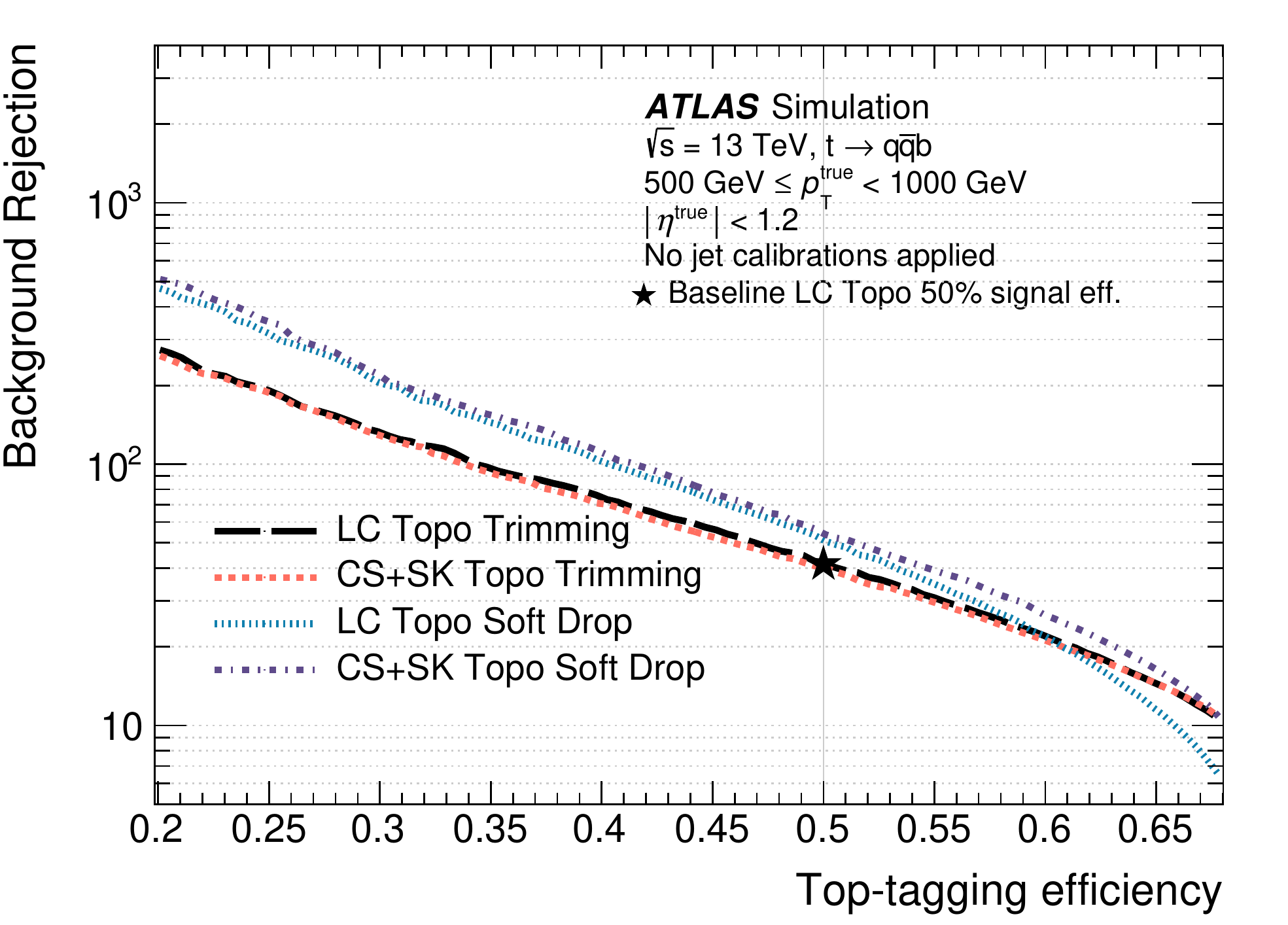}}
\subfigure[]{\includegraphics[width=0.49\textwidth]{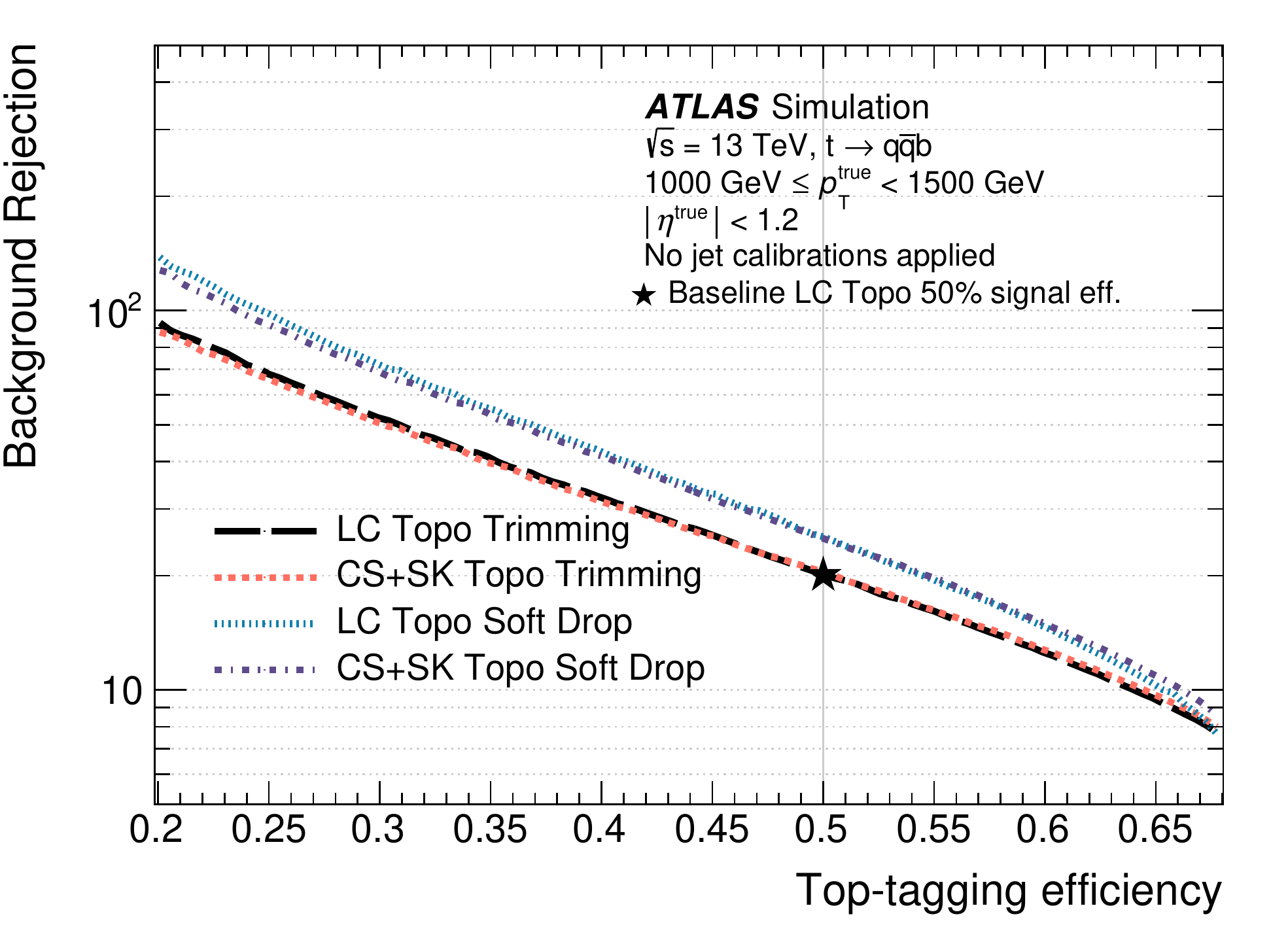}}\\
\subfigure[]{\includegraphics[width=0.49\textwidth]{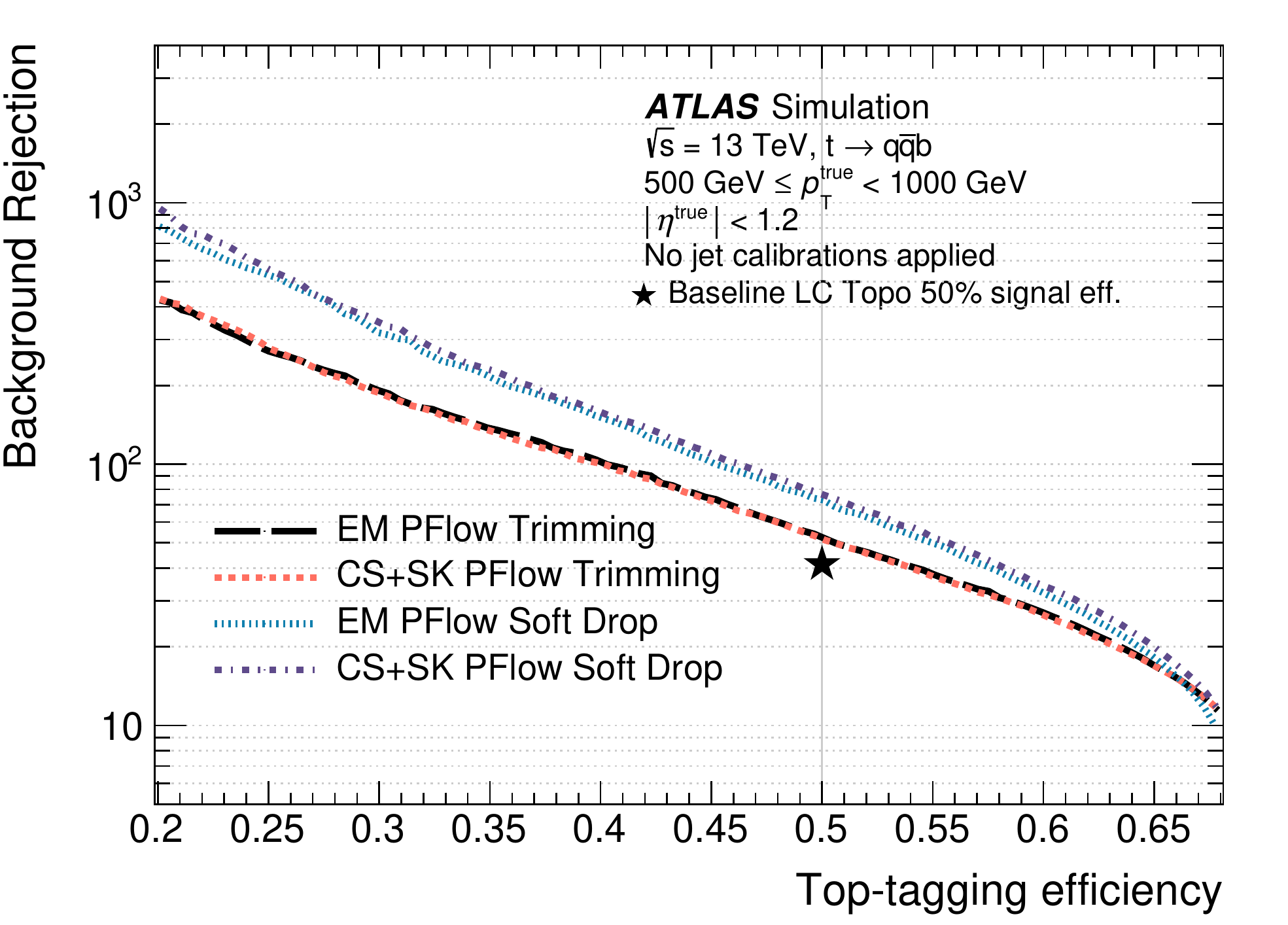}}
\subfigure[]{\includegraphics[width=0.49\textwidth]{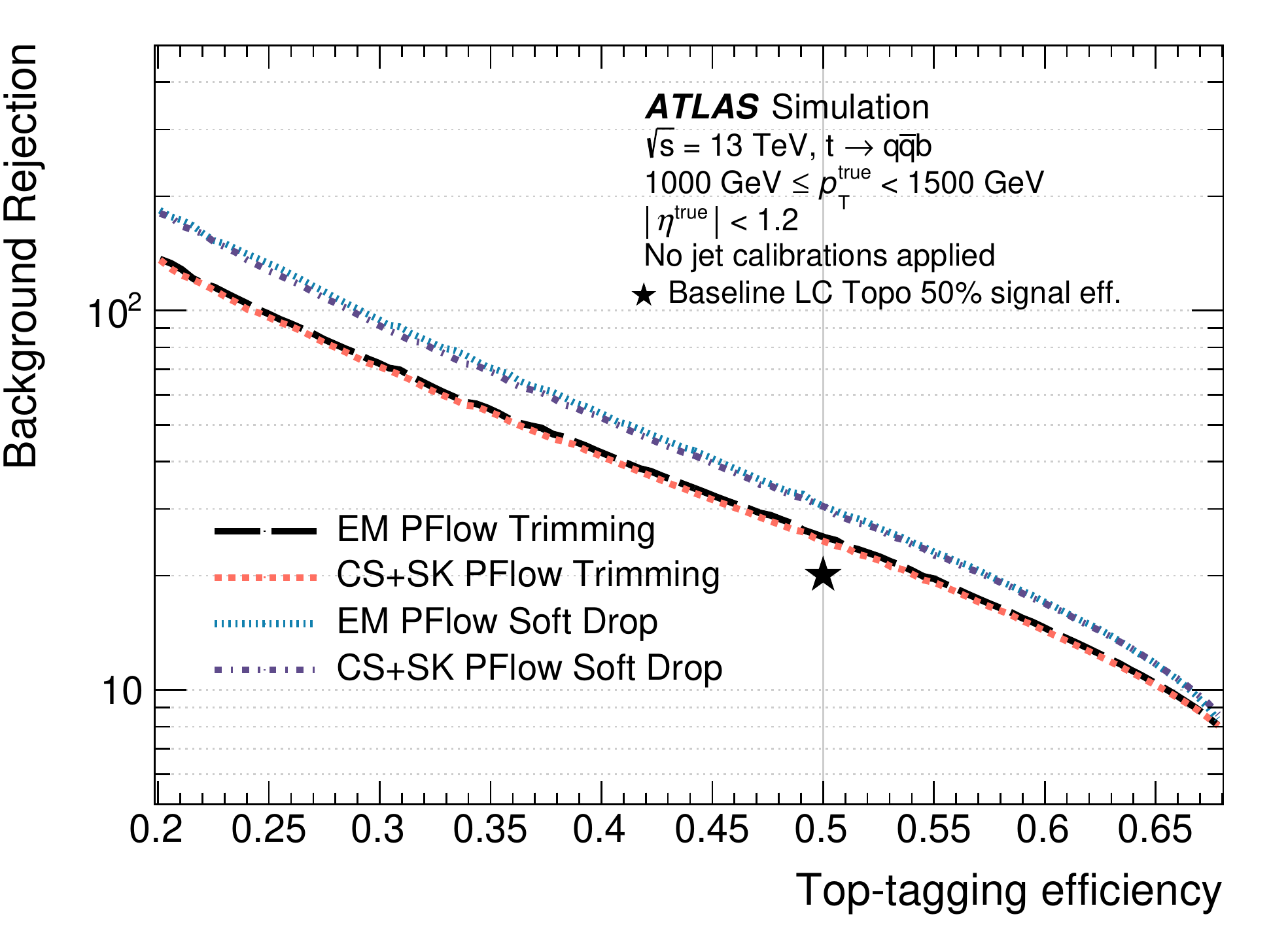}}\\
\subfigure[]{\includegraphics[width=0.49\textwidth]{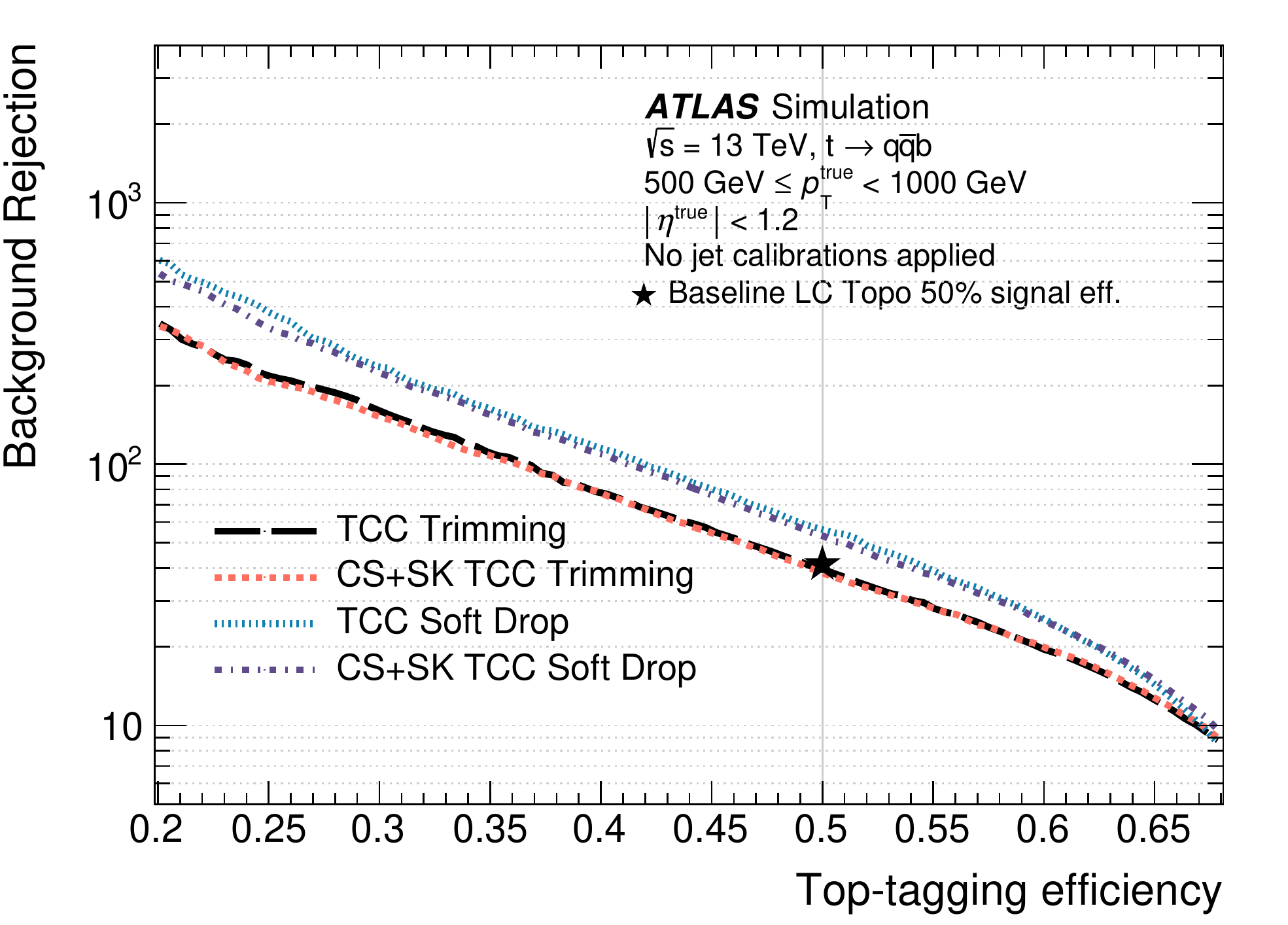}}
\subfigure[]{\includegraphics[width=0.49\textwidth]{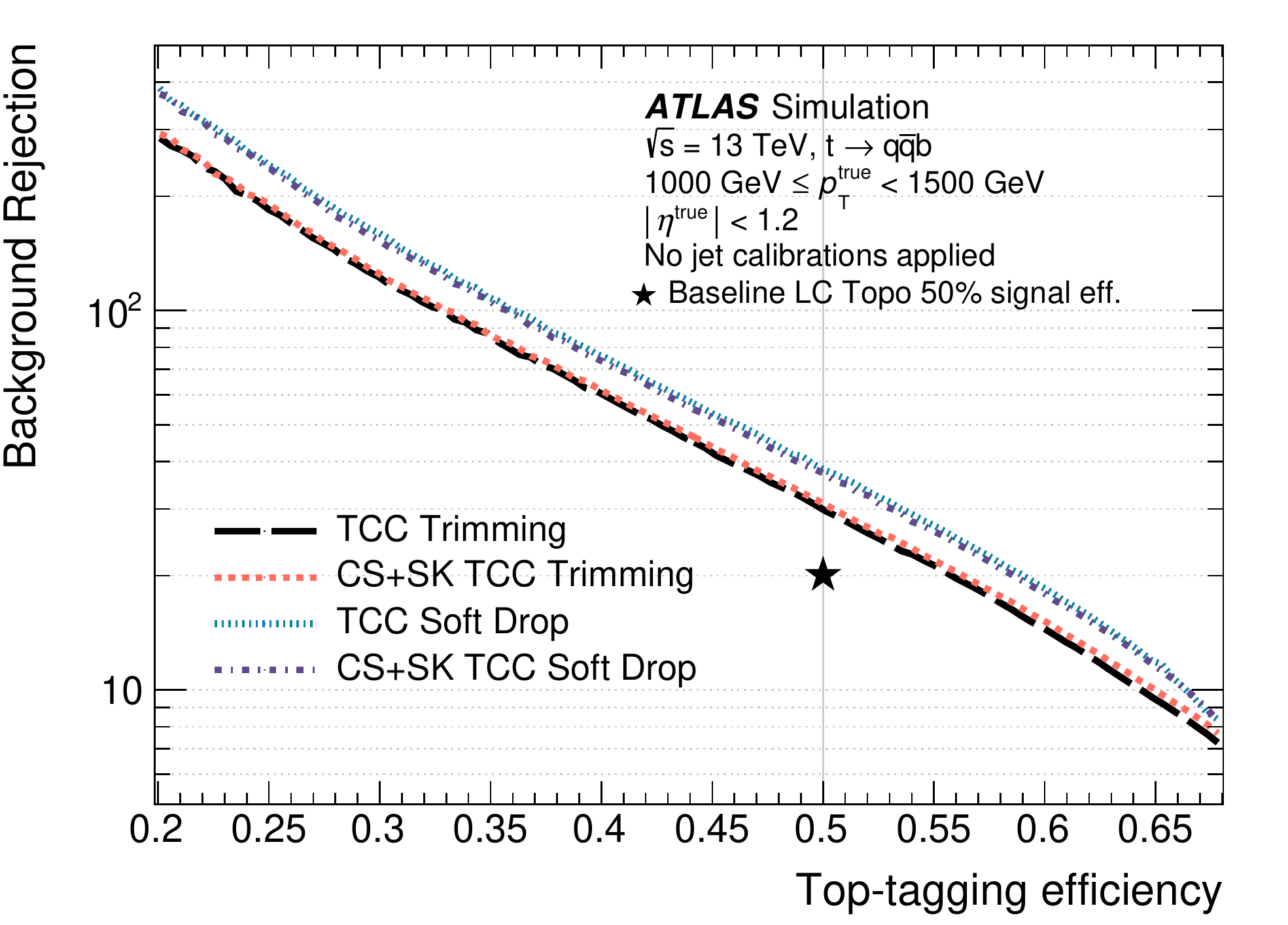}}
\end{center}
\caption{Background rejection as a function of signal efficiency for a tagger using the jet mass and \tauthrtwo for top quark jets at (a,c,d) low \pt, and (b,d,f) high \pt. Several different jet input object types are shown: (a,b) topoclusters, (c,d) particle-flow objects and (e,f) track-caloclusters.  Jet \pt and $\eta$ cuts before tagging are made using the ungroomed particle-level \largeR jet matched to each of the groomed reconstructed \largeR jets. Jets groomed with the trimming ($\Rsub=0.2$, $\fcut=0.05$) and soft-drop ($\beta=1.0$, $\zcut = 0.1$) algorithms are shown. The background rejection factor of the baseline topocluster-based trimmed collection at a fixed signal tagging efficiency of 50\% is indicated with a $\star$.}
\label{nocalib:tag:top}
\end{figure}
 
\FloatBarrier
\subsection{Pile-up stability}
 
Two metrics are used to study the pile-up stability of jet definitions in order to determine which definitions are sufficiently insensitive to pile-up. The first quantifies the effect on the jet mass scale by studying how the $W$ boson mass peak position changes as a function of pile-up, and provides a handle with which to assess the impact of pile-up on a jet's hard structure. The second quantifies the impact on substructure observables by studying the pile-up dependence of $W$ boson tagging efficiency, in order to quantify how pile-up contributions alter the soft radiation patterns within jets.
 
A related study of the effects of pile-up on topocluster reconstruction is presented in an appendix of this publication, utilising a new technique which propagates particle-level information about hard-scatter and pile-up energy depositions through the ATLAS reconstruction procedure.
 
\FloatBarrier
\subsubsection{Pile-up stability of the $W$ boson jet mass peak position}\label{sec:nocalib:pumass}
 
Jet substructure observables such as the jet mass are particularly sensitive to pile-up; the contribution of pile-up to the jet mass scales approximately with the jet radius cubed~\cite{STDM-2011-19}. Figure~\ref{nocalib:pileup:wmass} shows a subset of the trimmed mass distribution of $W$ jets in bins of \Npv for various jet input object types, demonstrating that pile-up can visibly alter the average value and width of the jet mass distribution. This effect is quantified using a simple metric. In bins of \Npv, the core of the $W$ mass peak is iteratively fit with a Gaussian distribution. The trend of the fitted peak position versus \Npv is then fit with a line. The slope of this line is a measure of the sensitivity of the jet mass to PU: a larger magnitude indicates larger pile-up sensitivity. The position of the $W$ jet mass peak was found to be a more resilient metric when studying the performance of uncalibrated jet definitions than other possible choices, such as properties of the jet mass response.
 
The results of this fitting procedure are provided in Figure~\ref{nocalib:pileup:wmass:fits} for the reduced set of jet definitions.  The application of CS+SK pile-up mitigation is shown to stabilise trends in topocluster and PFlow jets, even for jet grooming algorithms which are most sensitive to the effects of pile-up such as soft-drop with topocluster jets.  The fitted value of the $W$ boson mass peak position decreases as a function of \Npv for TCCs. This is related to TCC cluster splitting: as the number of pile-up interactions increases, the number of pile-up tracks also increases. Since these tracks are included in the energy-sharing step of the TCC algorithm, topoclusters are divided into more parts, and more energy is removed. Unlike PFlow and topocluster jet reconstruction, the pile-up stability of TCCs deteriorates after the application of CS+SK. Uncorrected PFlow and TCC jet reconstruction are less sensitive to pile-up than topocluster inputs, since they are able to remove the charged pile-up component via CHS.

\begin{figure}[htbp]
\begin{center}
\subfigure[]{
\includegraphics[width=0.45\textwidth]{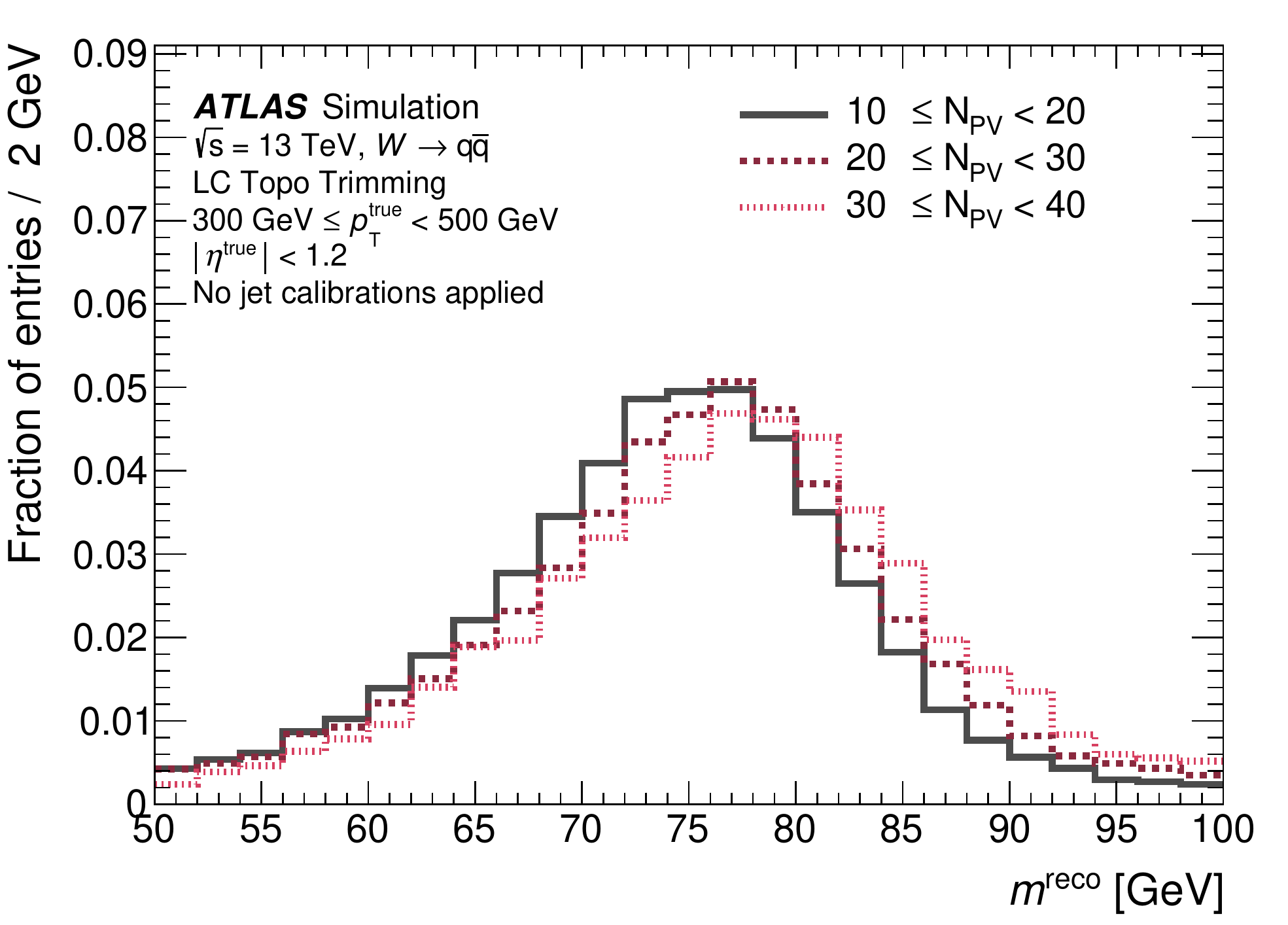}}
\subfigure[]{\includegraphics[width=0.45\textwidth]{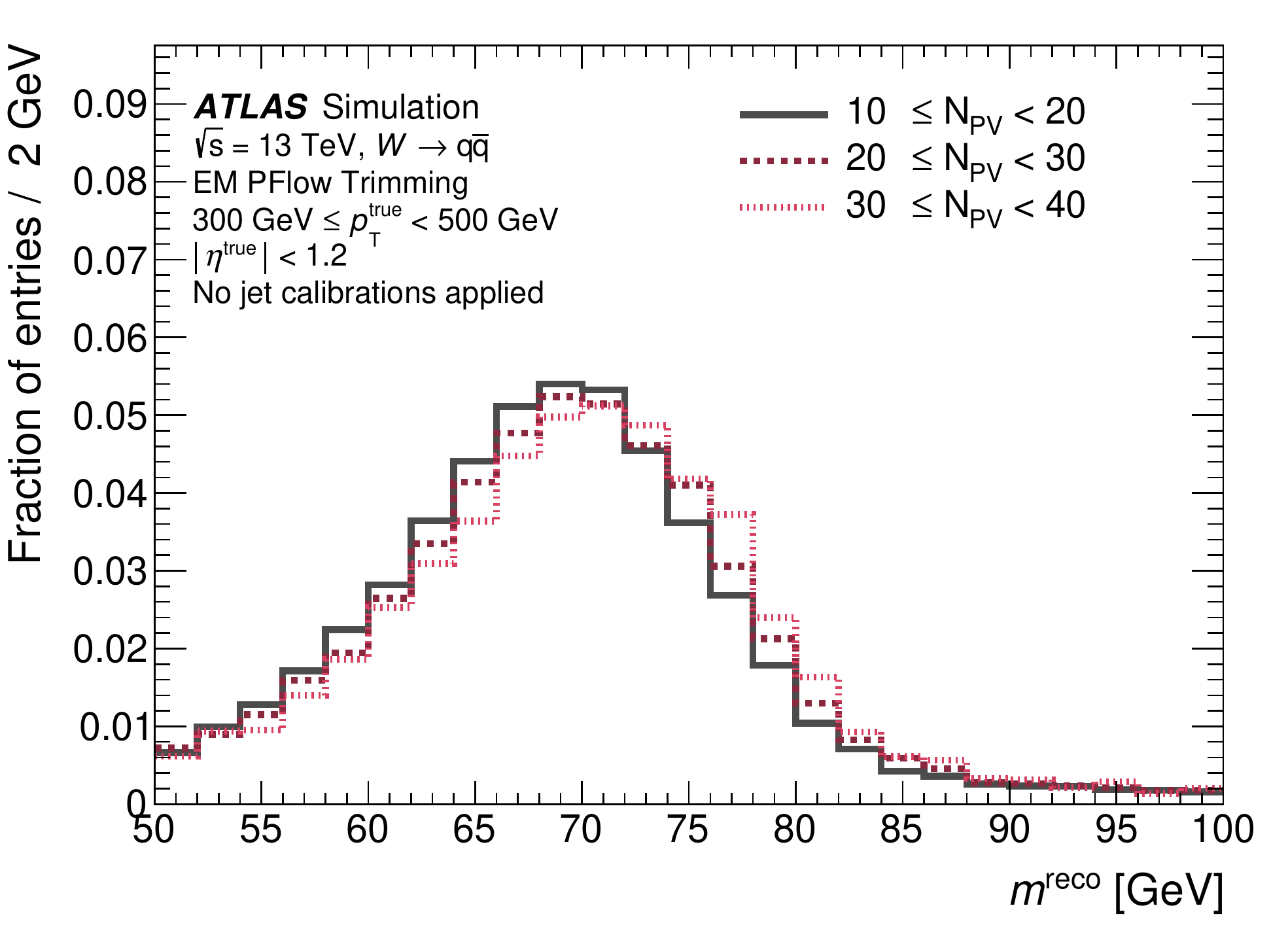}}\\
\subfigure[]{\includegraphics[width=0.45\textwidth]{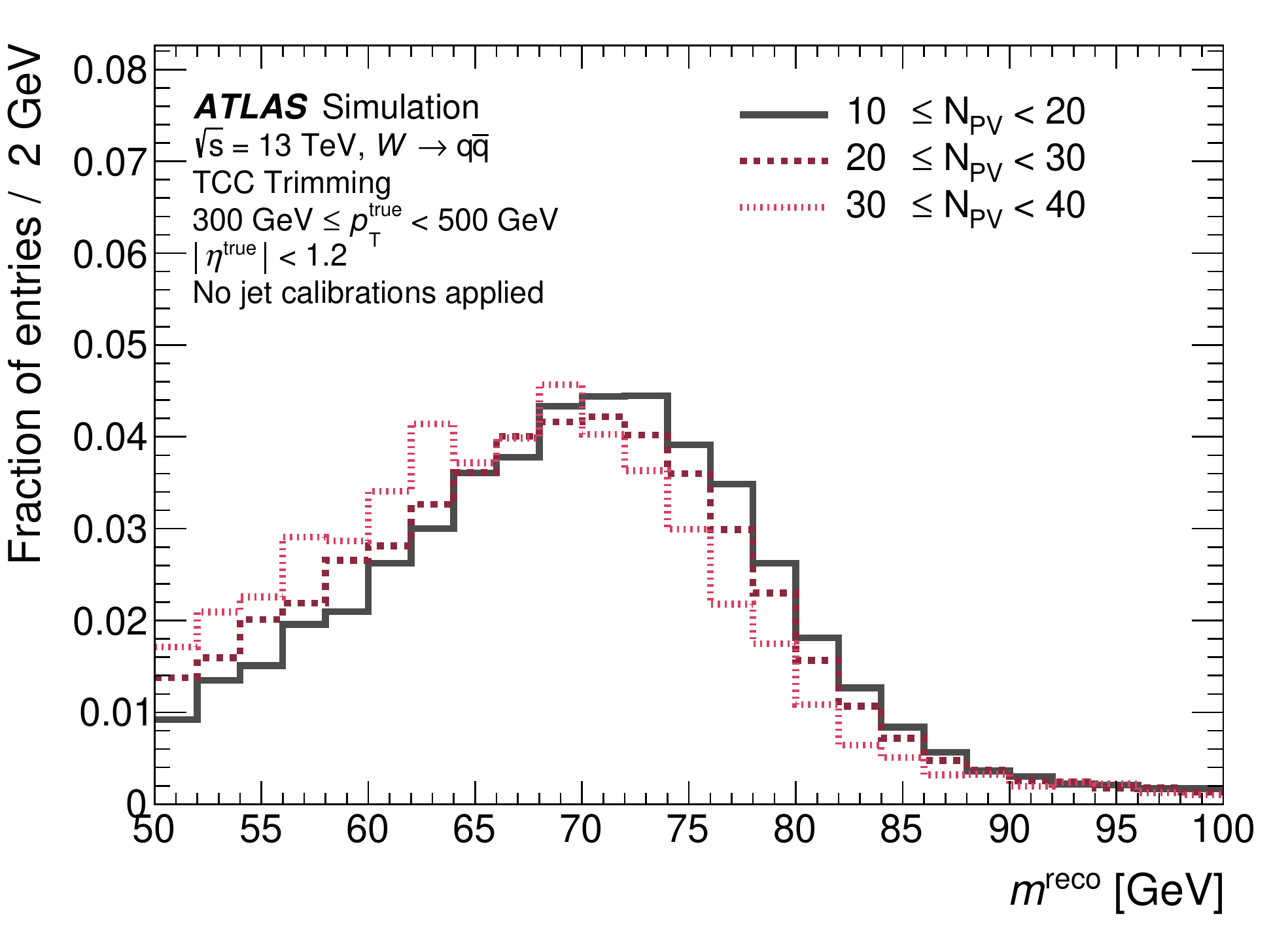}}\\
\end{center}
\caption{Pile-up dependence of the $W$ boson jet mass reconstructed using (a) topoclusters, (b) particle-flow objects and (c) track-caloclusters. Distributions are shown for the trimming grooming algorithm ($\Rsub=0.2$, $\fcut=0.05$), with unmodified  jet input objects. Jet \pt and $\eta$ cuts before tagging are made using the ungroomed particle-level \largeR jet matched to each of the groomed reconstructed \largeR jets.}
\label{nocalib:pileup:wmass}
\end{figure}
 
\begin{figure}[htbp]
\begin{center}
\subfigure[]{\includegraphics[width=0.49\textwidth]{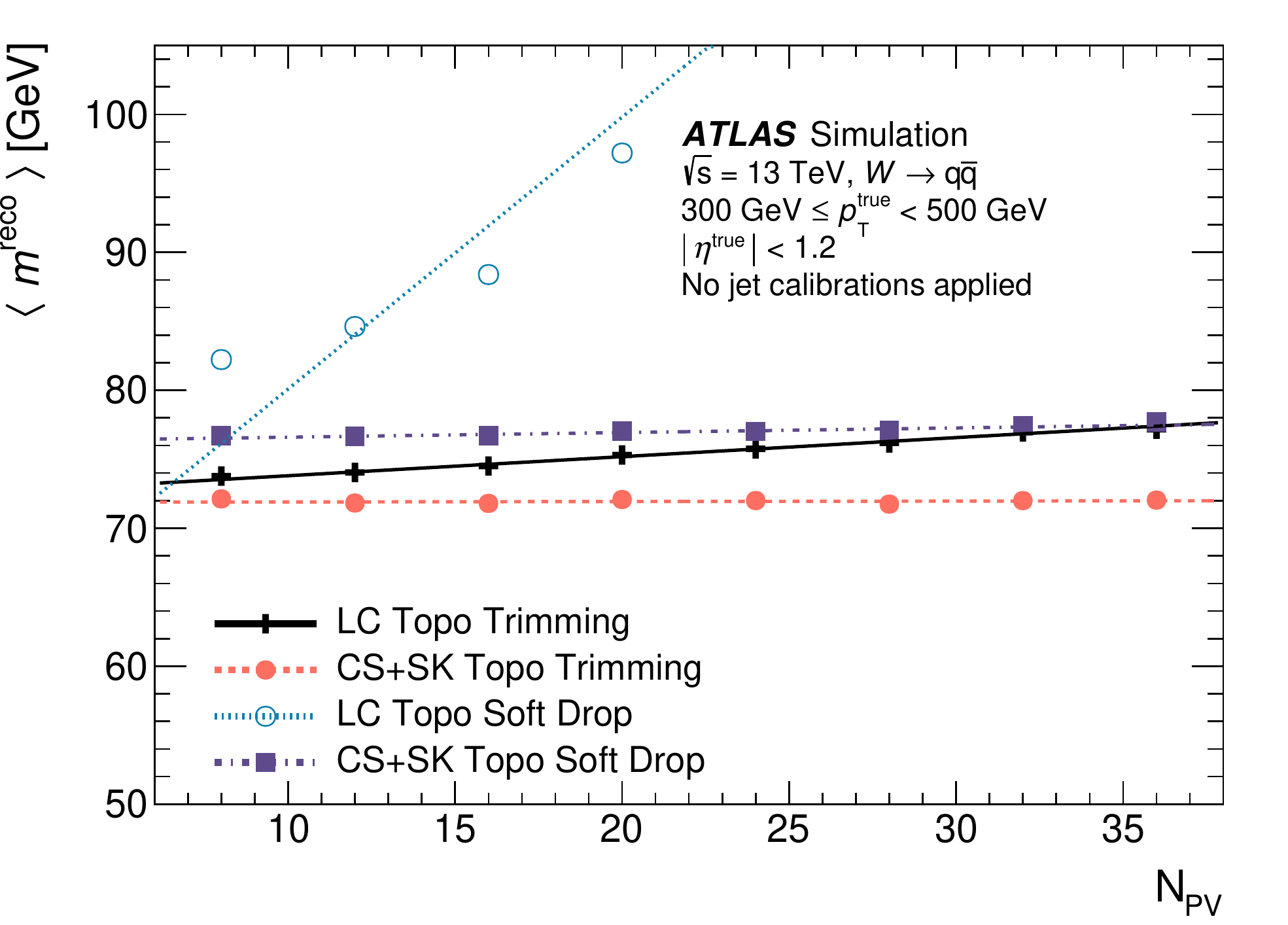}}
\subfigure[]{\includegraphics[width=0.49\textwidth]{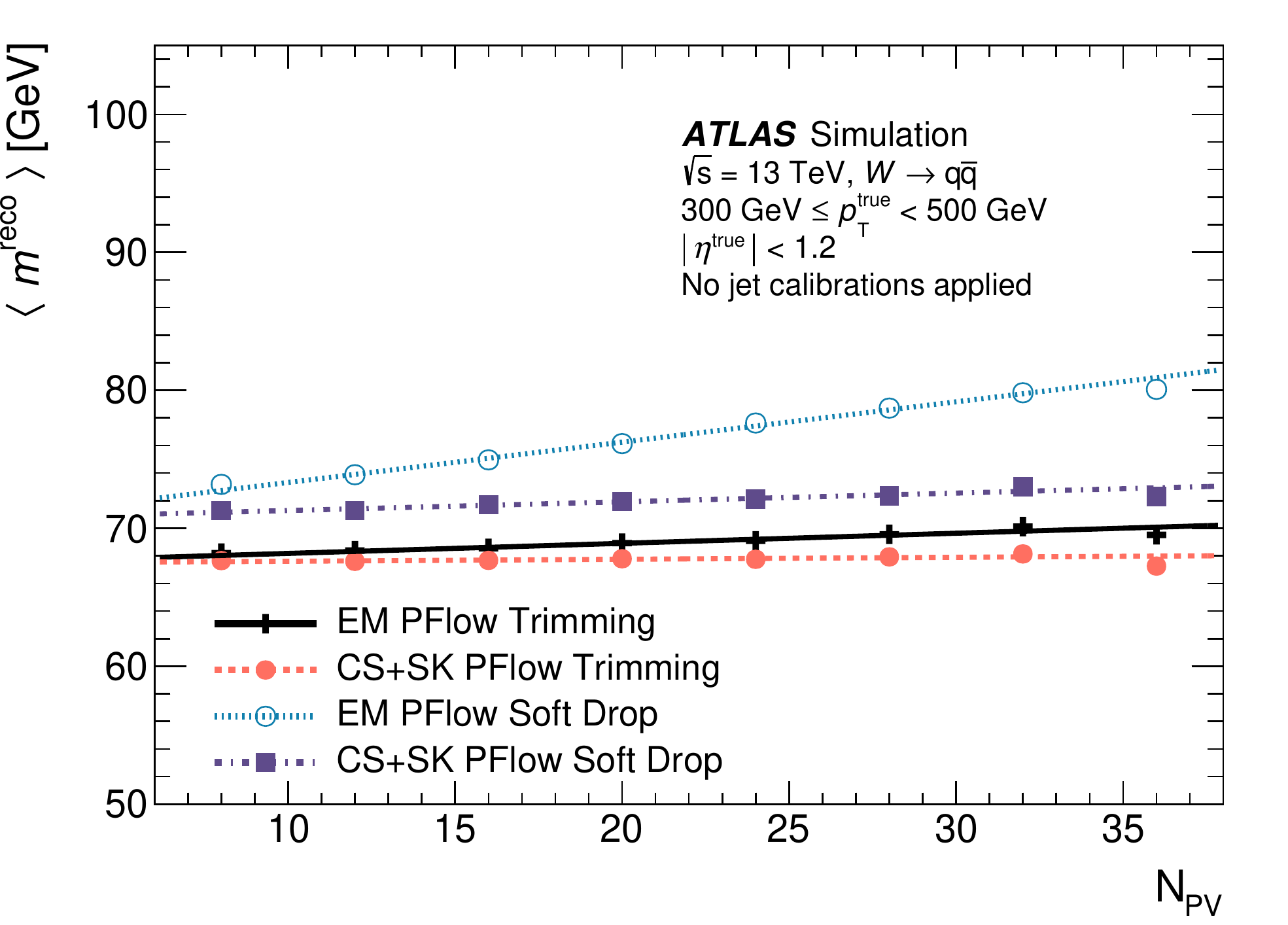}}
\subfigure[]{\includegraphics[width=0.49\textwidth]{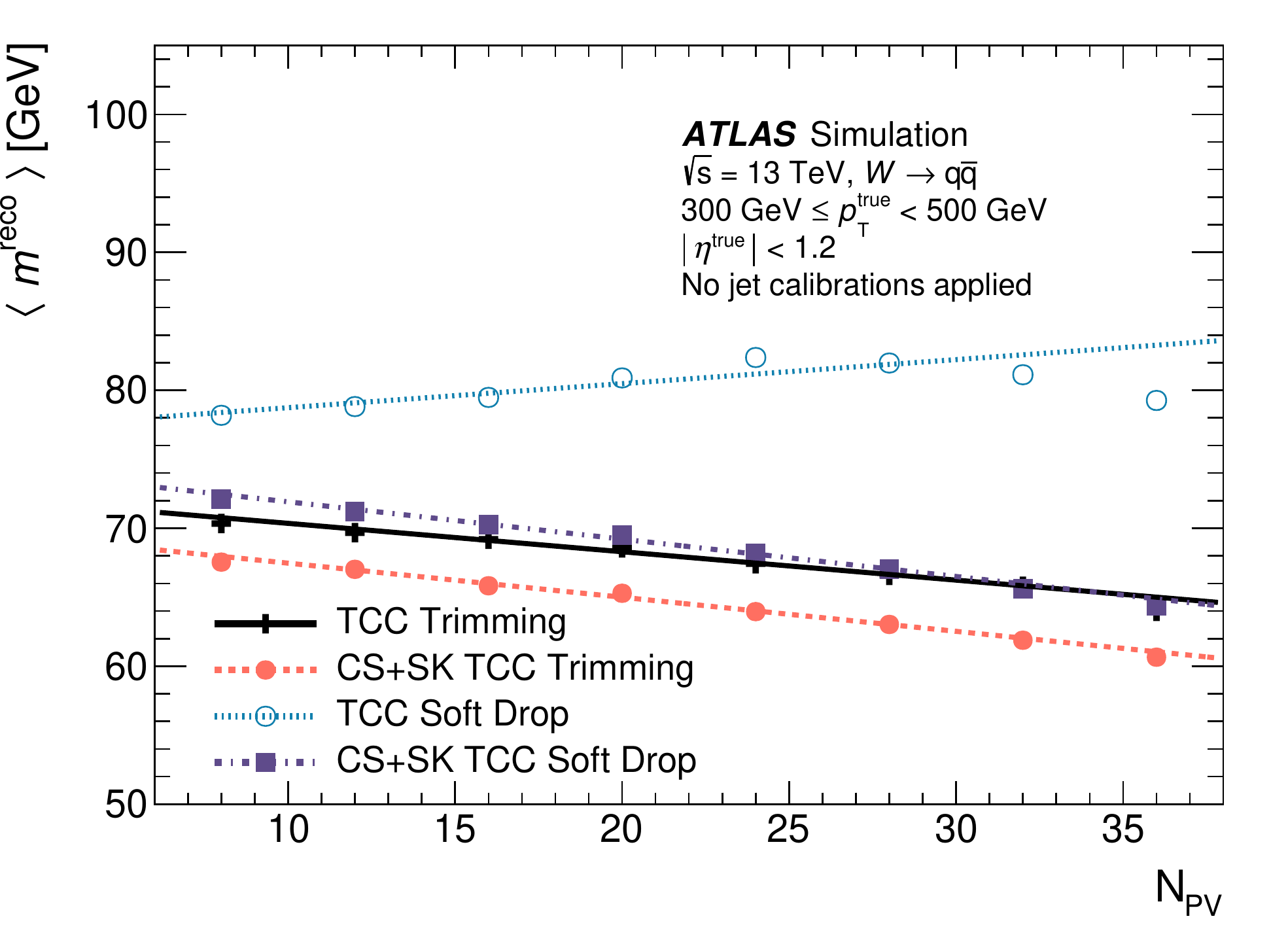}}
\end{center}
\caption{The value of the fitted $W$ boson mass peak as a function of the number of primary vertices, \Npv. Several different jet input object types are shown: (a) topoclusters, (b) particle-flow objects and (c) track-caloclusters. Jet \pt and $\eta$ cuts before tagging are made using the ungroomed particle-level \largeR jet matched to each of the groomed reconstructed \largeR jets. Jets groomed with the trimming ($\Rsub=0.2$, $\fcut=0.05$) and soft-drop ($\beta=1.0$, $\zcut = 0.1$) algorithms are shown.}
\label{nocalib:pileup:wmass:fits}
\end{figure}
 
\subsubsection{Pile-up stability of a simple tagger}
 
The second metric of pile-up stability quantifies the effect of pile-up on the tagging efficiency, which is impacted more by contributions from soft radiation to the tails of jet substructure observables. The $\DTwo$ variable is particularly sensitive to soft radiation, and so a $W$ tagger is defined using the jet mass and $\DTwo$ (Section~\ref{sec:nocalib:tagging}). For a sample of events with \Npv < 15, a mass cut which results in a 68\% signal efficiency is found, and then the $\DTwo$ cut that results in an overall signal efficiency of 50\% is determined. Then, in bins of \Npv, the signal efficiency of applying these cuts is evaluated. These signal efficiencies are plotted as a function of \Npv and the trend is fit with a line. The slope of this line is indicative of pile-up sensitivity in the soft jet substructure of the jet definition. These slopes are shown for the reduced set of jet definitions in Figure~\ref{nocalib:pileup:d2}.
 
As pile-up levels increase, the signal efficiency of the $W$ tagger tends to decrease, although the opposite behaviour is often observed for TCC jets. Similarly to what was found when studying the $W$ mass peak position metric (Section~\ref{sec:nocalib:pumass}), topocluster inputs are the least stable. After pile-up mitigation, the pile-up stability of all inputs, including TCCs, improves. The trends in stability as a function of grooming algorithm are the same as for the $W$ mass peak position.
 
\begin{figure}[htbp]
\begin{center}
\subfigure[]{\includegraphics[width=0.49\textwidth]{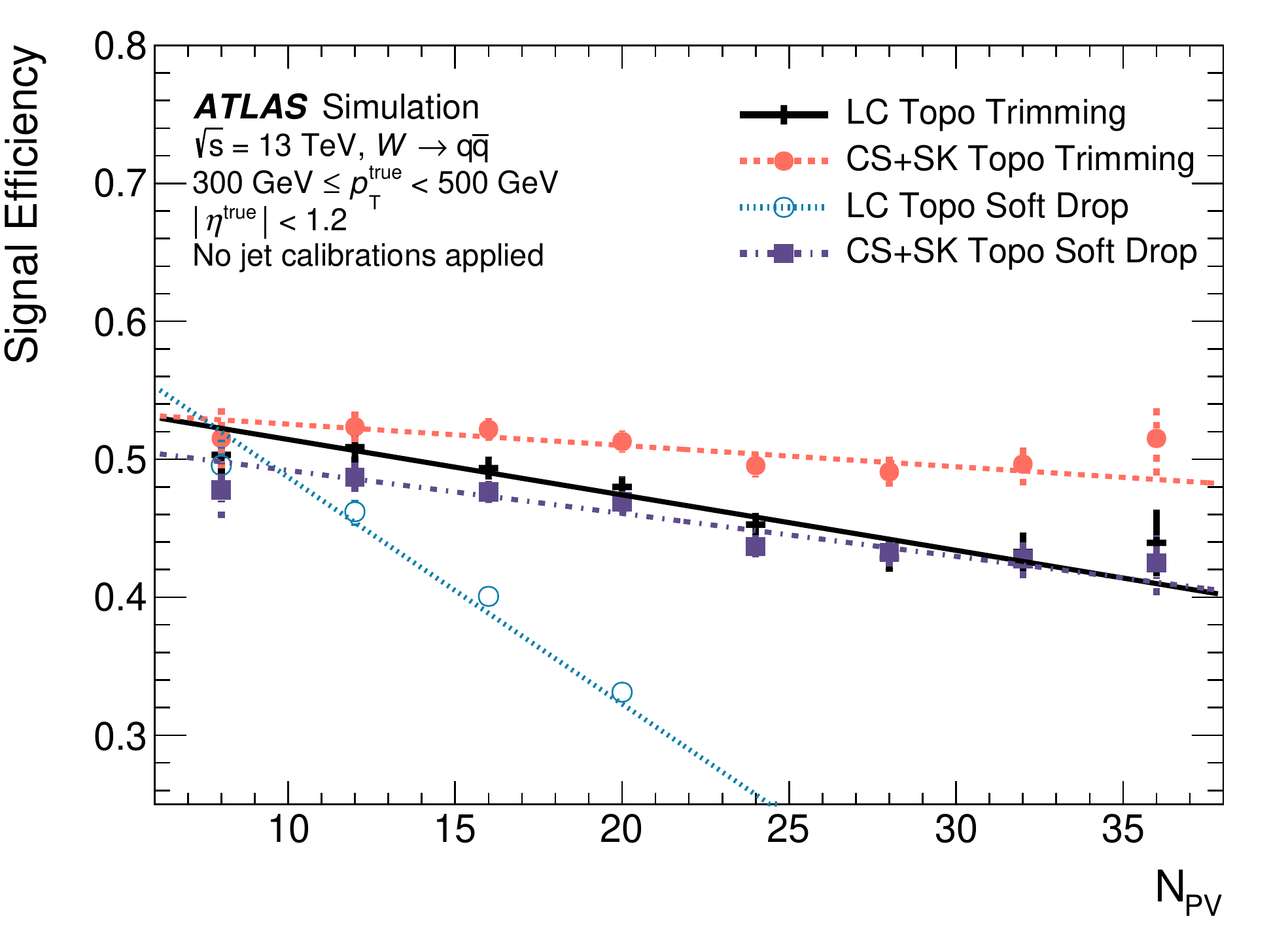}}
\subfigure[]{\includegraphics[width=0.49\textwidth]{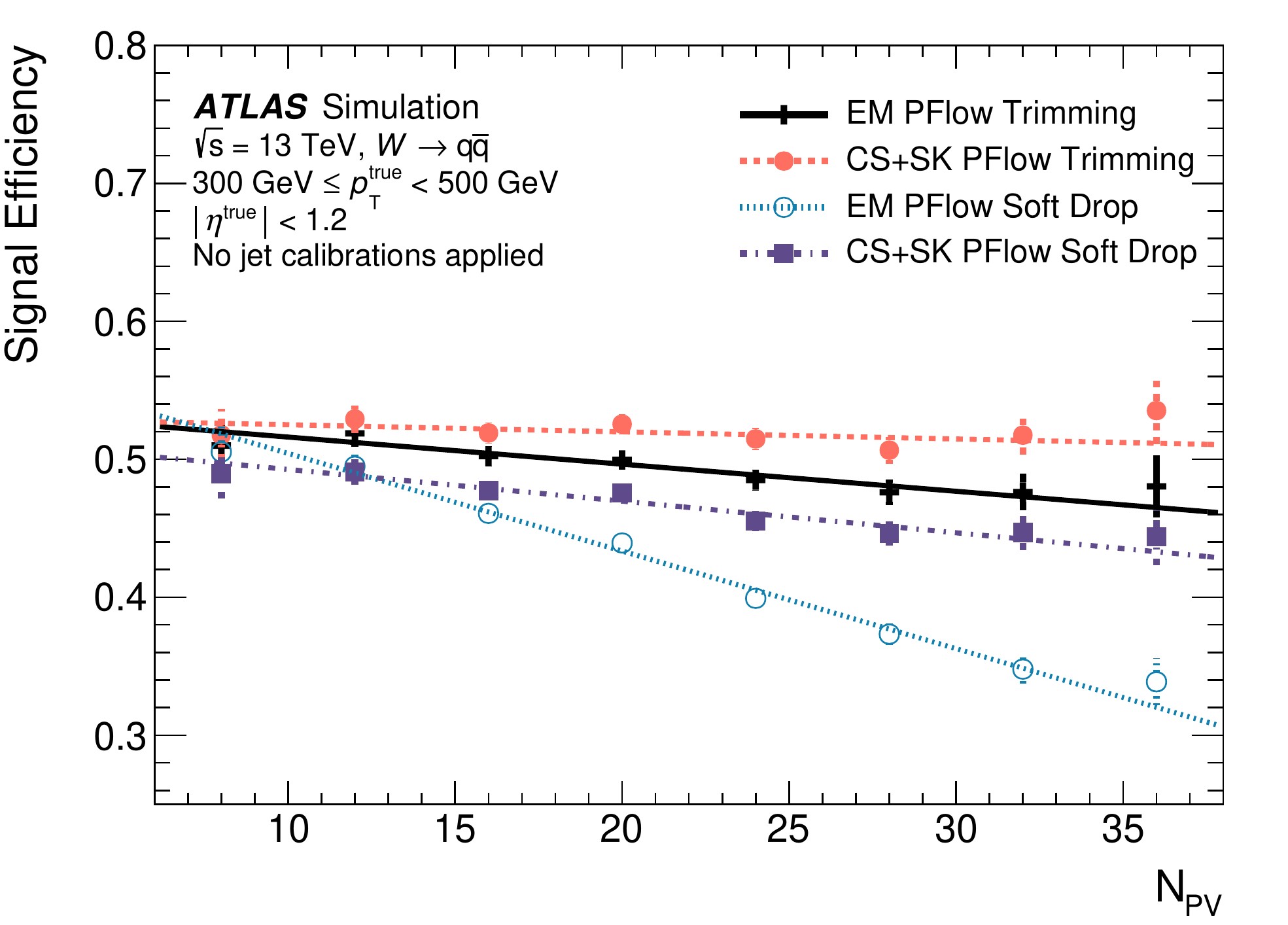}}
\subfigure[]{\includegraphics[width=0.49\textwidth]{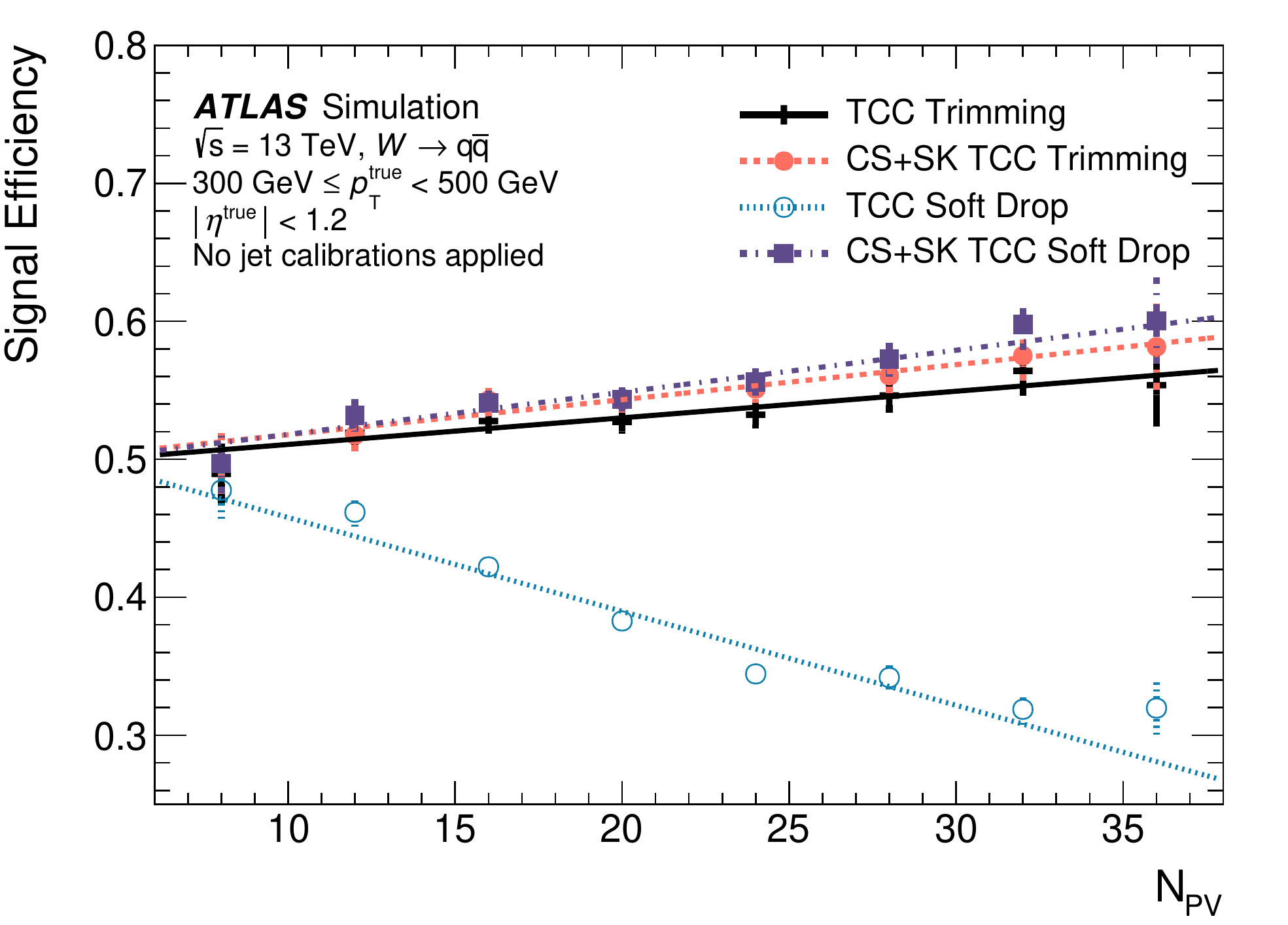}}
\end{center}
\caption{The signal efficiency of a $W$ boson tagger as a function of the number of primary vertices, \Npv. Several different jet input object types are shown: (a) topoclusters, (b) particle-flow objects and (c) track-caloclusters. Jet \pt and $\eta$ cuts before tagging are made using the ungroomed particle-level \largeR jet matched to each of the groomed reconstructed \largeR jets. Jets groomed with the trimming ($\Rsub=0.2$, $\fcut=0.05$) and soft-drop ($\beta=1.0$, $\zcut = 0.1$) algorithms are shown.}
\label{nocalib:pileup:d2}
\end{figure}
 
\FloatBarrier
\subsection{Topological sensitivity}\label{sec:nocalib:topology}
 
ATLAS calibrates \largeR jets using a procedure which involves simulation-based and \emph{in situ} methods~\cite{JETM-2018-02}. For the simulation-based calibration, the average jet energy and mass scale in reconstructed jets are calibrated to the average scale of jets at particle level, using a sample of jets originating from light quarks and gluons (Section~\ref{sec:mcjesjms}). These light-quark- and gluon-derived calibrations are also currently applied to all jets, including to signal jets (e.g.\ $W$/$Z$/$H$/$t$ jets). Dependence of the jet energy and mass scale on the progenitor of the jet is undesirable: if the jet mass scale for signal and background jets with similar kinematics is different, then the signal jets will receive an incorrect calibration factor.
 
In order to examine the topology dependence of the jet mass scale for different jet definitions, the ratio of the mean value of the uncalibrated jet mass response, $R_{m}=m^{\text{reco}}/m^{\text{true}},$ for signal $W$ jets to that of background jets is constructed within a bin of \largeR jet \pt, $\eta$ and mass. Deviations from unity will result in non-closure in the mass response for signal jets following calibration (Section~\ref{sec:mcjesjms}). This effect is relevant at low \pt, where $W$ jets may be contained within an $R=1.0$ jet, but top quarks are not; therefore, only $W$ jets and background jets are considered in this context. The baseline topocluster-based trimmed \largeR jet definition used by the ATLAS experiment exhibits a difference for signal jets of 4\% by this metric; therefore, deviations from unity of 4\% or less have not been found to be problematic at later stages of the calibration workflow~\cite{JETM-2018-02}, given the current level of calibration precision.
 
Figure~\ref{fig:nocalib:topology} shows the jet mass response for signal and background jets built from topological clusters and groomed with either the trimming or soft-drop grooming algorithms. The low-\pt bin, where this topological effect is most pronounced, is shown. A larger sensitivity to the signal- or background-like nature of the jet is observed for soft-drop grooming, which retains more soft radiation. The application of pile-up mitigation can exacerbate topological differences in the jet mass scale by altering the distribution of soft jet constituents differently depending on the jet's signal- or background-like topology.

\begin{figure}[htbp]
\begin{center}
\subfigure[]{\includegraphics[width=0.49\textwidth]{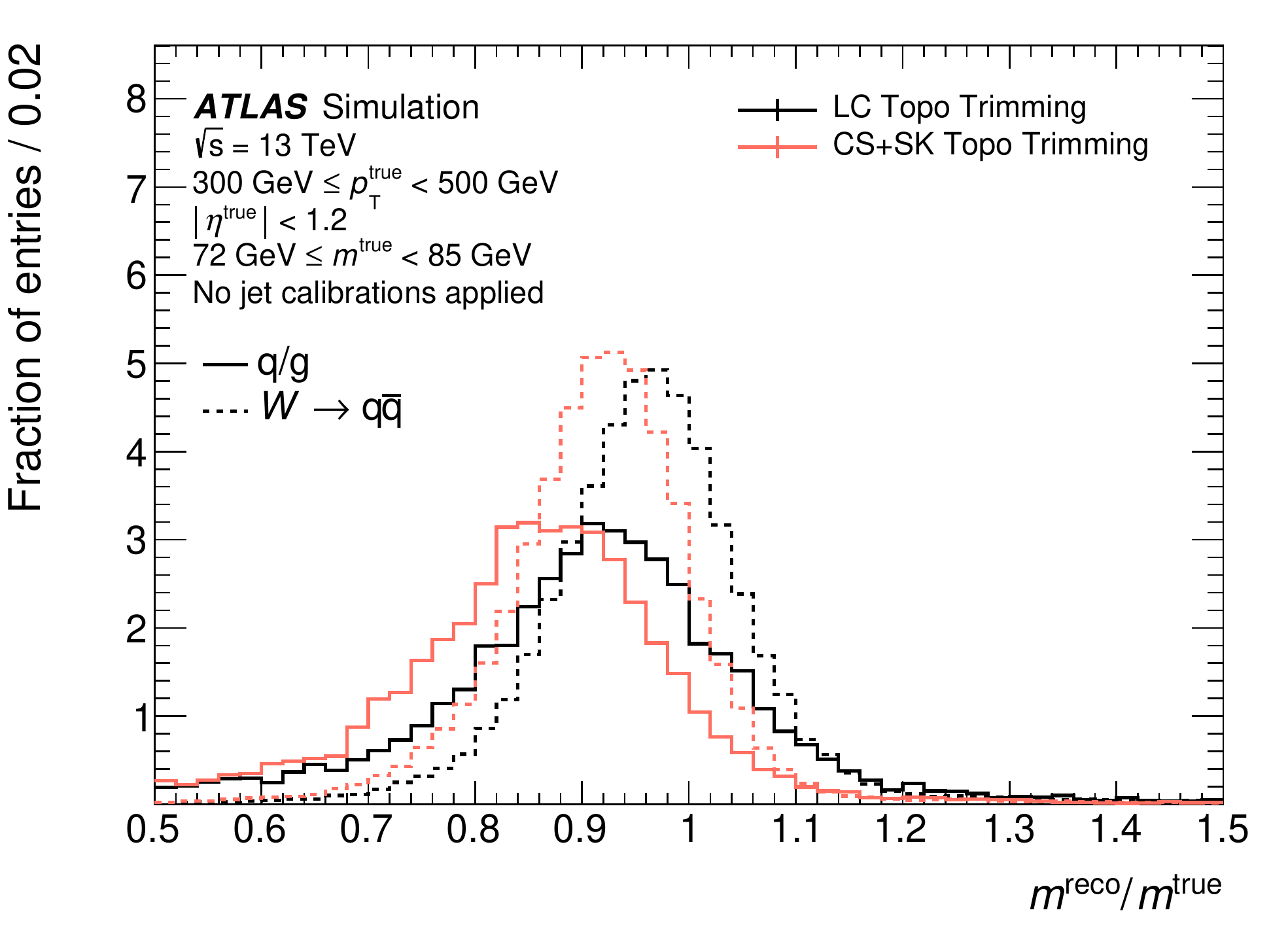}}
\subfigure[]{\includegraphics[width=0.49\textwidth]{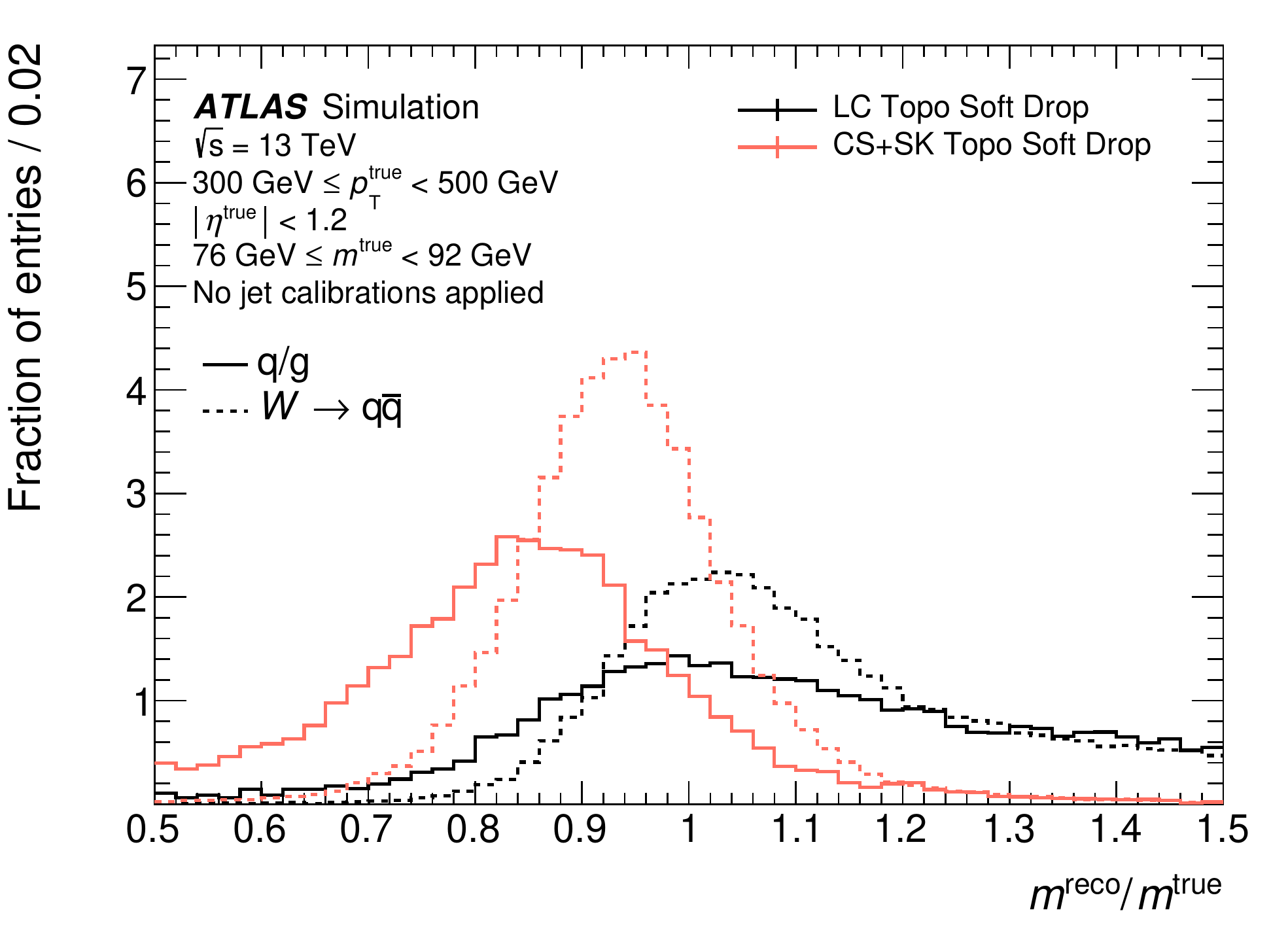}} \\
\end{center}
\caption{Distribution of the jet mass response in $W$ jets and $q/g$ jets reconstructed from topoclusters. The mass response is constructed following application of the (a) trimming ($\Rsub=0.2$, $\fcut=0.05$) or (b) soft-drop ($\beta=1.0$, $\zcut = 0.1$) grooming algorithms at both truth and detector level. Jet \pt and $\eta$ selections are made using the ungroomed particle-level \largeR jet matched to each of the groomed detector-level \largeR jets. The uncertainties from the fits are typically less than 0.005. A particle-level mass-window cut with 68\% signal efficiency is applied to both the groomed signal and background jets.}
\label{fig:nocalib:topology}
\end{figure}
 
\FloatBarrier
\section{Unified Flow Objects (UFOs)}\label{sec:ufos}
After observing the behaviour of the jet input objects currently used by ATLAS in physics analyses (topoclusters, PFOs and TCCs), it is clear even from the reduced set of jet definitions (Section~\ref{sec:nocalib}) that no single jet definition is optimal according to all metrics. While TCCs significantly improve tagging performance at high \pt, their performance is typically worse than the baseline topocluster-based trimmed jet definition at low \pt, and they are more sensitive to pile-up than other definitions. Jets reconstructed from PFOs can improve on the baseline definition for the entire \pt range, but their tagging performance is significantly worse than that of TCC jets at high \pt when given the same grooming algorithm.
 
The relative performance of these jet definitions can be understood by reflecting on how different inputs are reconstructed. For low-\pt particles, PFOs are designed to improve the correspondence between particles and reconstructed objects. However, as the particle \pt increases or the environment close-by to the particle becomes dense, the inner detector's momentum resolution deteriorates, and so the PFlow subtraction algorithm is gradually disabled in order to avoid degradation of the jet energy resolution.
 
The cluster splitting scheme used for TCCs does not utilise a detailed understanding of the correlation between tracks and clusters, and instead is designed to resolve many (charged) particles without double counting their energy. When splitting low-energy topoclusters, this can result in an incorrect redistribution of the cluster's energy, while for high-energy clusters, the ability to resolve many particles increases the relative tagging performance of TCCs over other definitions. TCCs exhibit pile-up instabilities at low \pt, where the mass scale decreases as the number of pile-up interactions increases. This trend is the opposite of what is observed for jets reconstructed from topoclusters and PFOs, and occurs because the TCC algorithm splits clusters into more components when additional tracks from pile-up interactions are present in the reconstruction procedure.
 
These observations motivate the development of a new jet input object, which combines desirable aspects of PFO and TCC reconstruction in order to achieve optimal overall performance across the full kinematic range. These new inputs are called Unified Flow Objects (UFOs).
 
The UFO reconstruction algorithm is illustrated in Figure~\ref{fig:ufos:flowchart}. The process begins by applying the standard ATLAS PFlow algorithm (Section~\ref{sec:objects:pfo}). Charged PFOs which are matched to pile-up vertices are removed. The remaining PFOs are classified into different categories: neutral PFOs, charged PFOs which were used to subtract energy from a topocluster, and charged PFOs for which no subtraction was performed due to their high momentum or being located in a dense environment. Jet-input-level pile-up mitigation algorithms may now be applied to the neutral PFOs if desired. A modified version of the TCC splitting algorithm is then applied to the remaining PFOs: only tracks from the hard-scatter vertex are used as input to the splitting algorithm, in order to avoid pile-up instabilities. Any tracks which have been used for PFlow subtraction are not considered, as they have already been well-matched and their expected contributions have been subtracted from the energy in the calorimeter. The TCC algorithm then proceeds as described in Section~\ref{sec:objects:tccs}, using the modified collection of tracks to split neutral and unsubtracted charged PFOs instead of topoclusters. This approach provides the maximum benefit of PFlow subtraction at lower particle \pt, and cluster splitting where the benefit is maximal at high particle \pt.
 
The performance of UFOs is illustrated in Figures~\ref{fig:nocalib:tag:ufos} and~\ref{fig:nocalib:pileup:wmass:ufos} according to the same metrics as for other jet input objects in Section~\ref{sec:nocalib}. The increased tagging performance of UFOs is demonstrated across both the low and high \pt ranges in Figure~\ref{fig:nocalib:tag:ufos}, where their performance is superior to that of TCC jets at high \pt, and becomes similar to that of PFlow jets as \pt decreases.
 
UFOs are naturally pile-up-stable due to the inclusion of only charged-particle tracks matched to the primary vertex, similar to the ATLAS PFlow algorithm. Figure~\ref{fig:nocalib:pileup:wmass:ufos} demonstrates the additional stability that an input-level pile-up mitigation algorithm such as CS+SK can offer when it is applied to neutral particles (calorimeter deposits), especially at low \pt.
 
The topological dependence of UFOs is not enhanced relative to the other jet definitions previously studied, and options exist with sensitivity equal to or below that of the baseline topocluster-based trimmed definition which improve on other aspects of jet performance.
 
\begin{figure}[htbp]
\begin{center}
\includegraphics[width=0.95\textwidth]{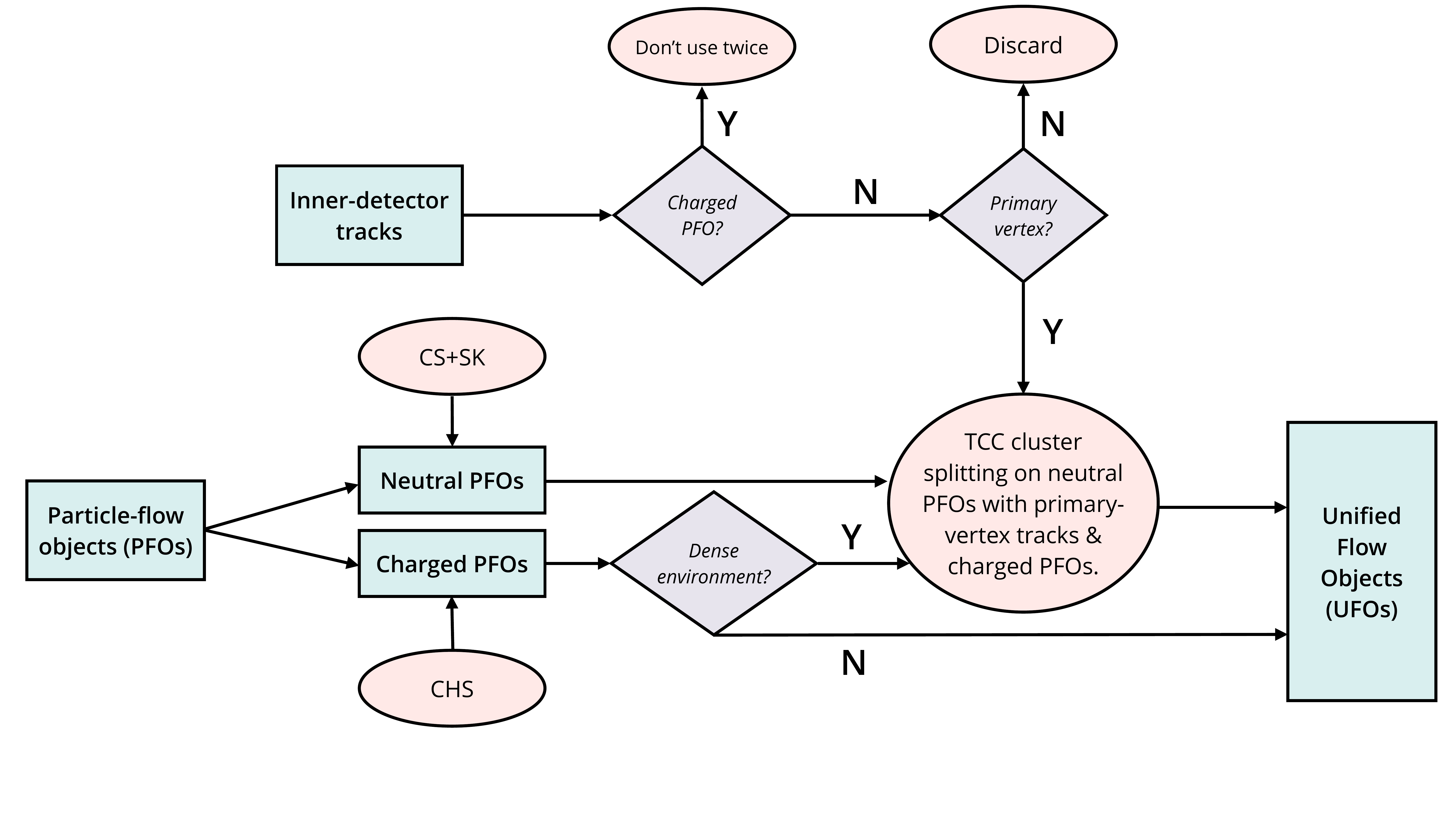}
\end{center}
\caption{An illustration of the unified flow object reconstruction algorithm.}
\label{fig:ufos:flowchart}
\end{figure}
 
\begin{figure}[htbp]
\begin{center}
\subfigure[]{\includegraphics[width=0.49\textwidth]{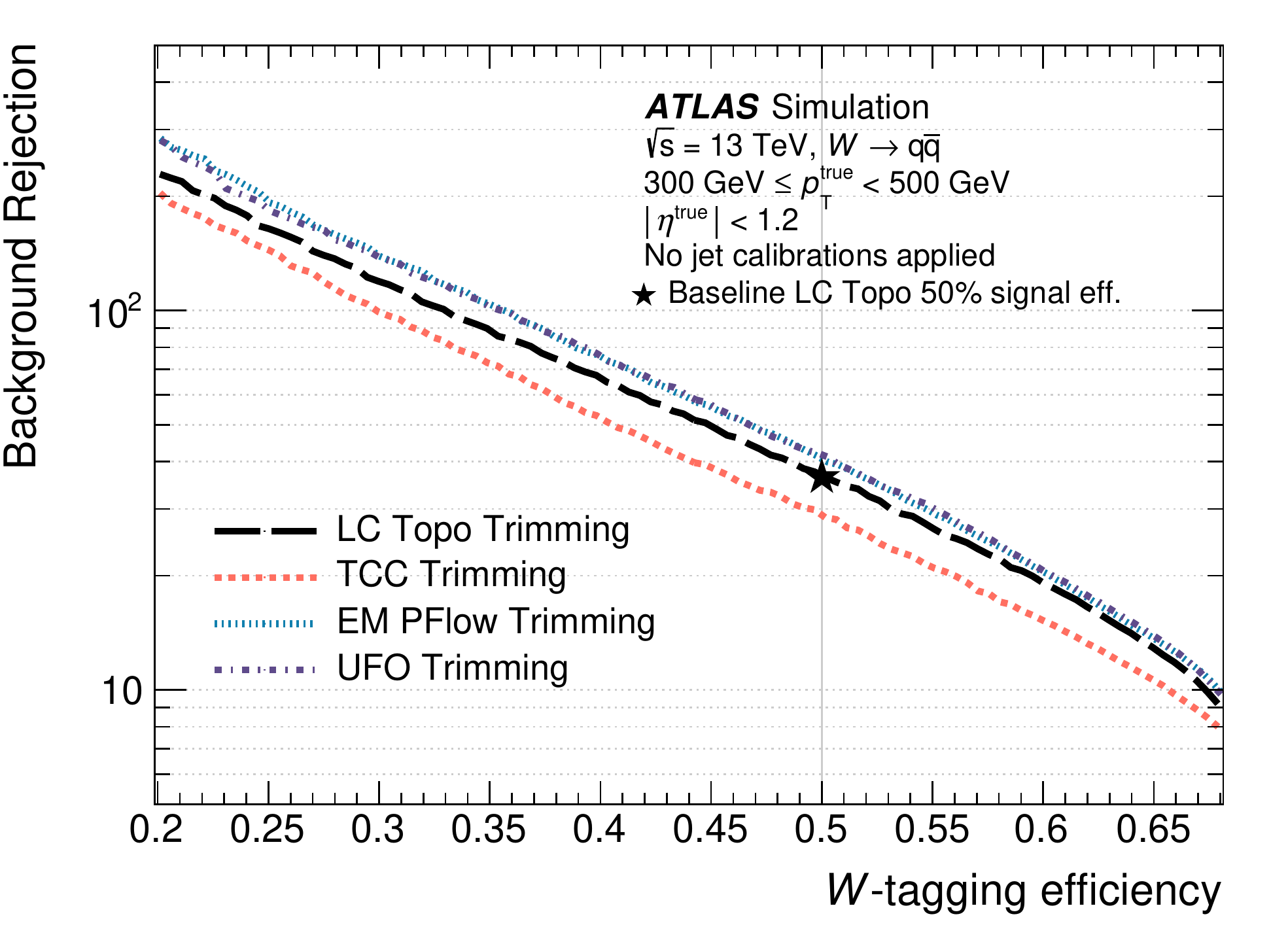}}
\subfigure[]{\includegraphics[width=0.49\textwidth]{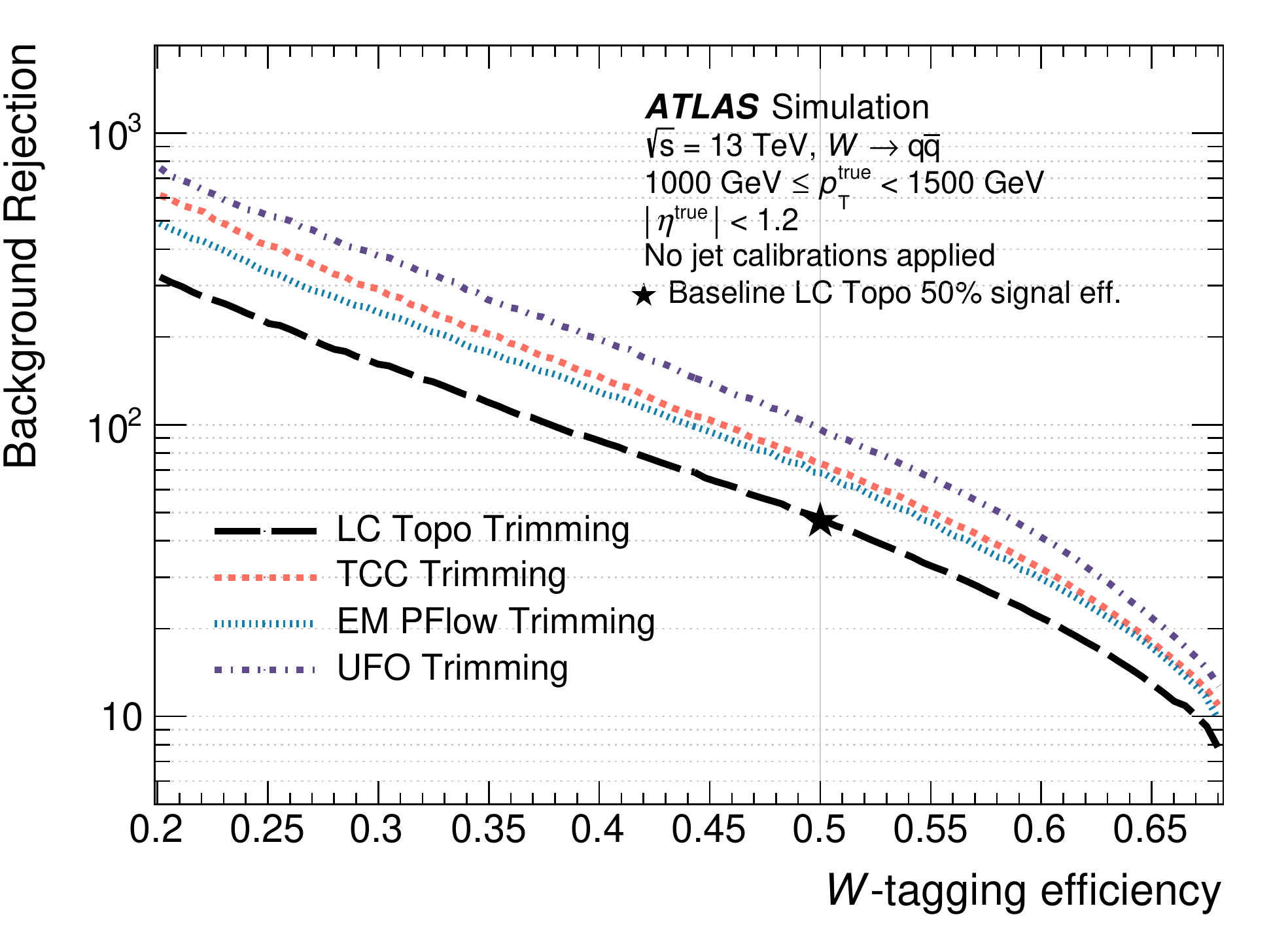}}\\
\subfigure[]{\includegraphics[width=0.49\textwidth]{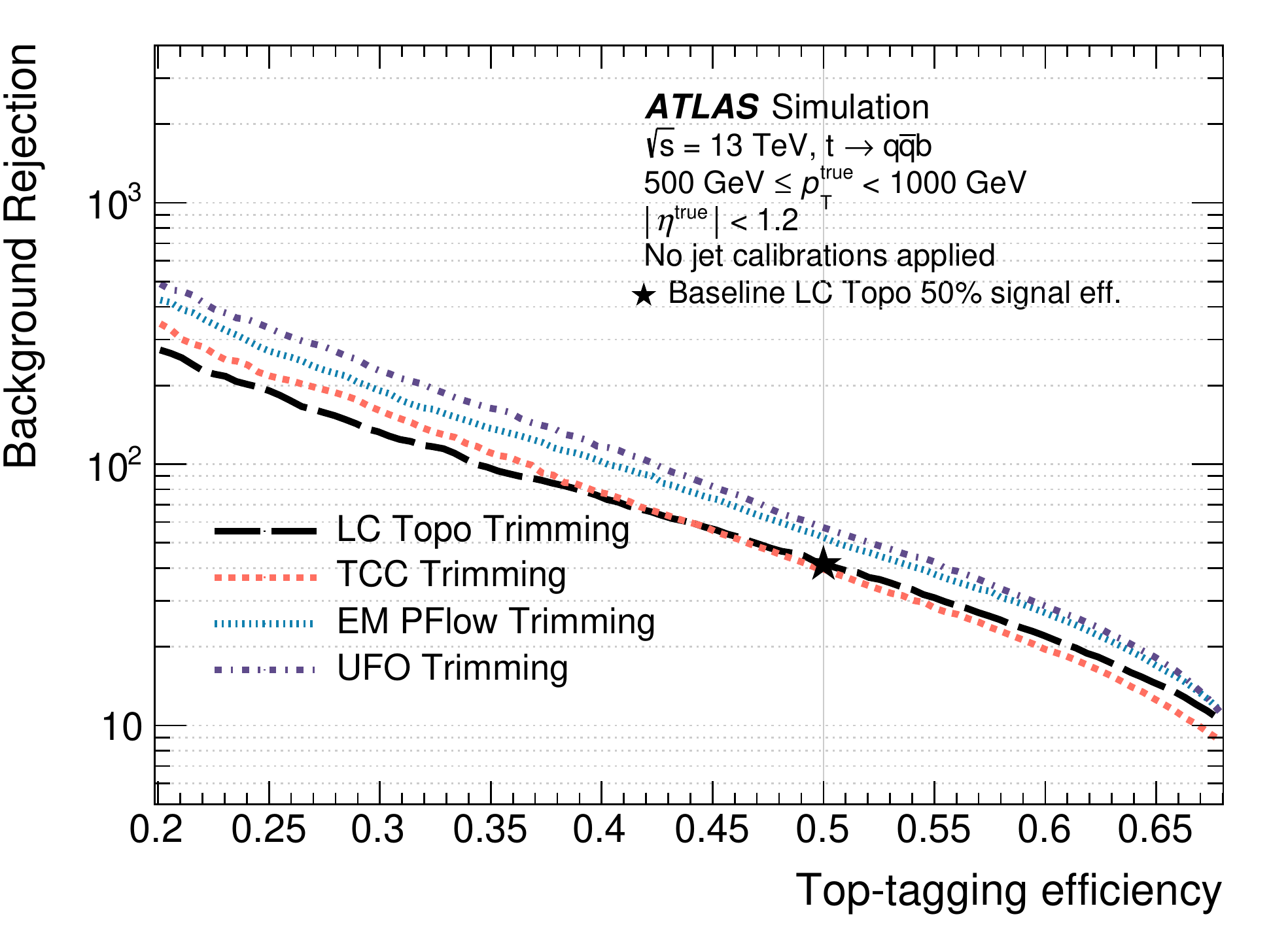}}
\subfigure[]{\includegraphics[width=0.49\textwidth]{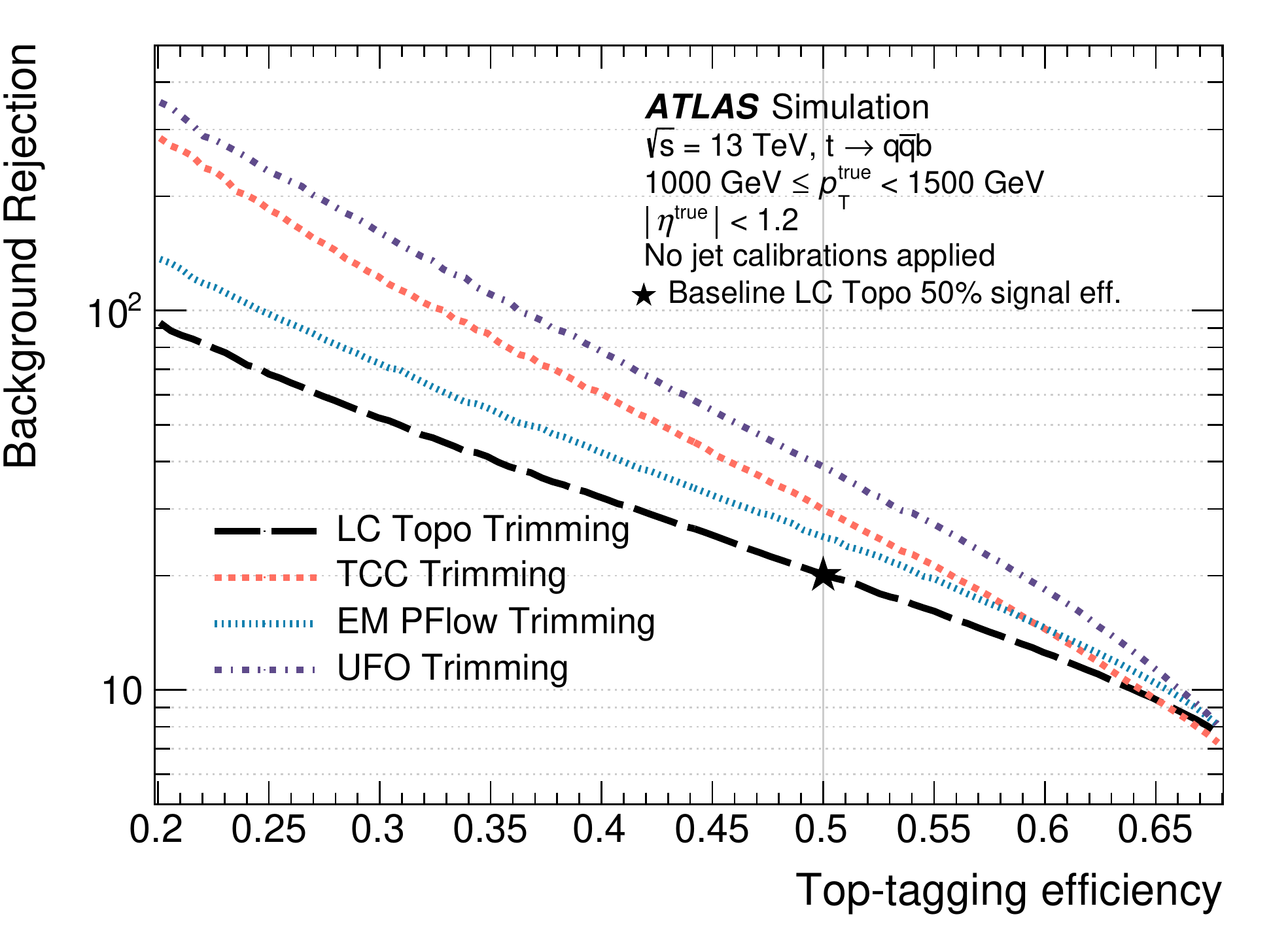}}\\
\end{center}
\caption{Background rejection as a function of signal efficiency for a tagger using (top row) the jet mass and \DTwo for $W$ boson jets, or (bottom row) the jet mass and \tauthrtwo for top quark jets. These results are shown in (left) low-\pt and (right) high-\pt bins, and include a comparison of different jet input object types, including topoclusters, particle-flow objects, track-caloclusters and unified flow objects. The \largeR jets are groomed using the trimming algorithm ($\Rsub=0.2$, $\fcut=0.05$). The background rejection factor of the baseline topocluster-based trimmed collection at a fixed signal tagging efficiency of 50\% is indicated with a $\star$.}
\label{fig:nocalib:tag:ufos}
\end{figure}
 
\begin{figure}[htbp]
\begin{center}
\subfigure[]{\includegraphics[width=0.49\textwidth]{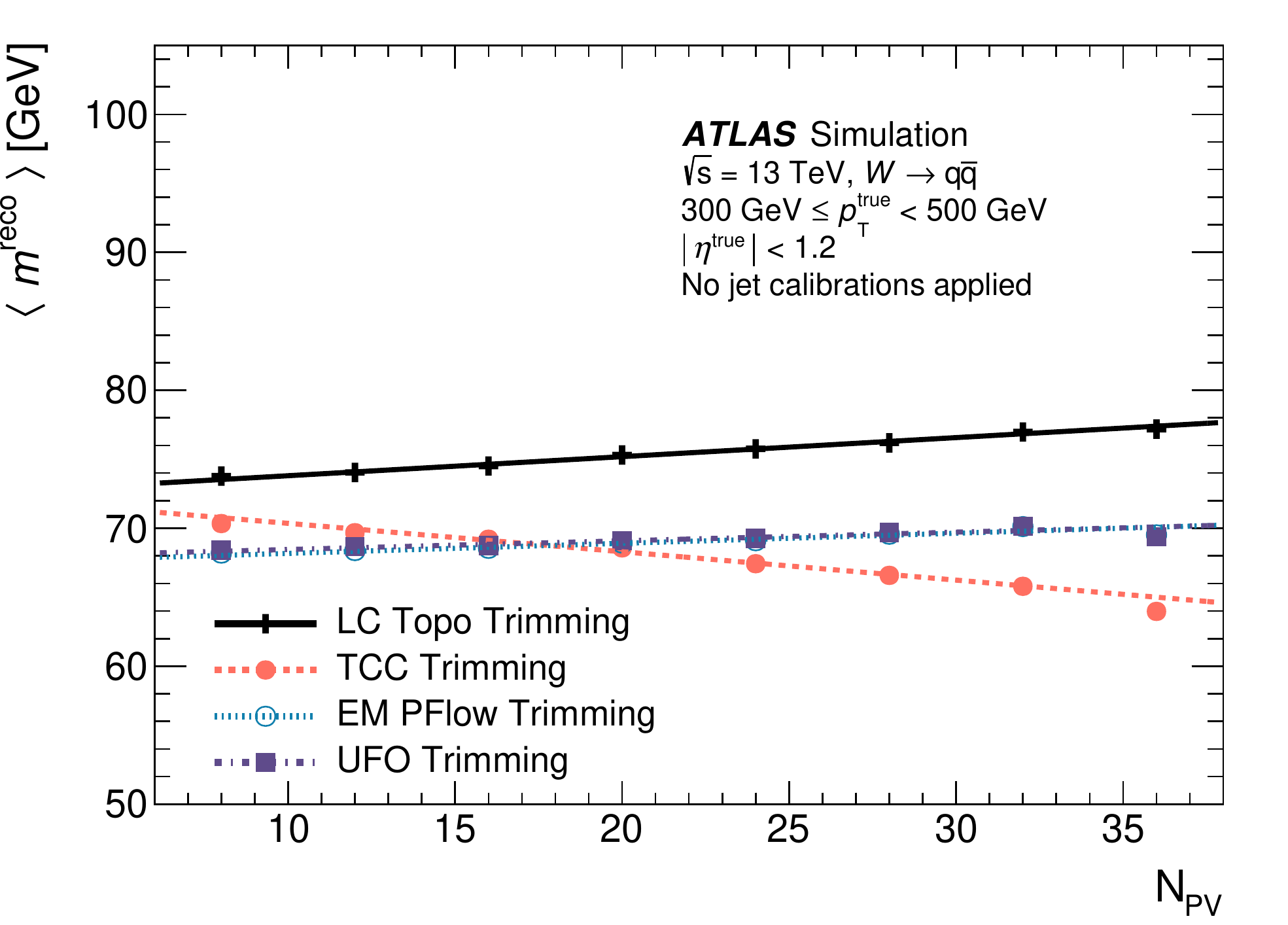}}
\subfigure[]{\includegraphics[width=0.49\textwidth]{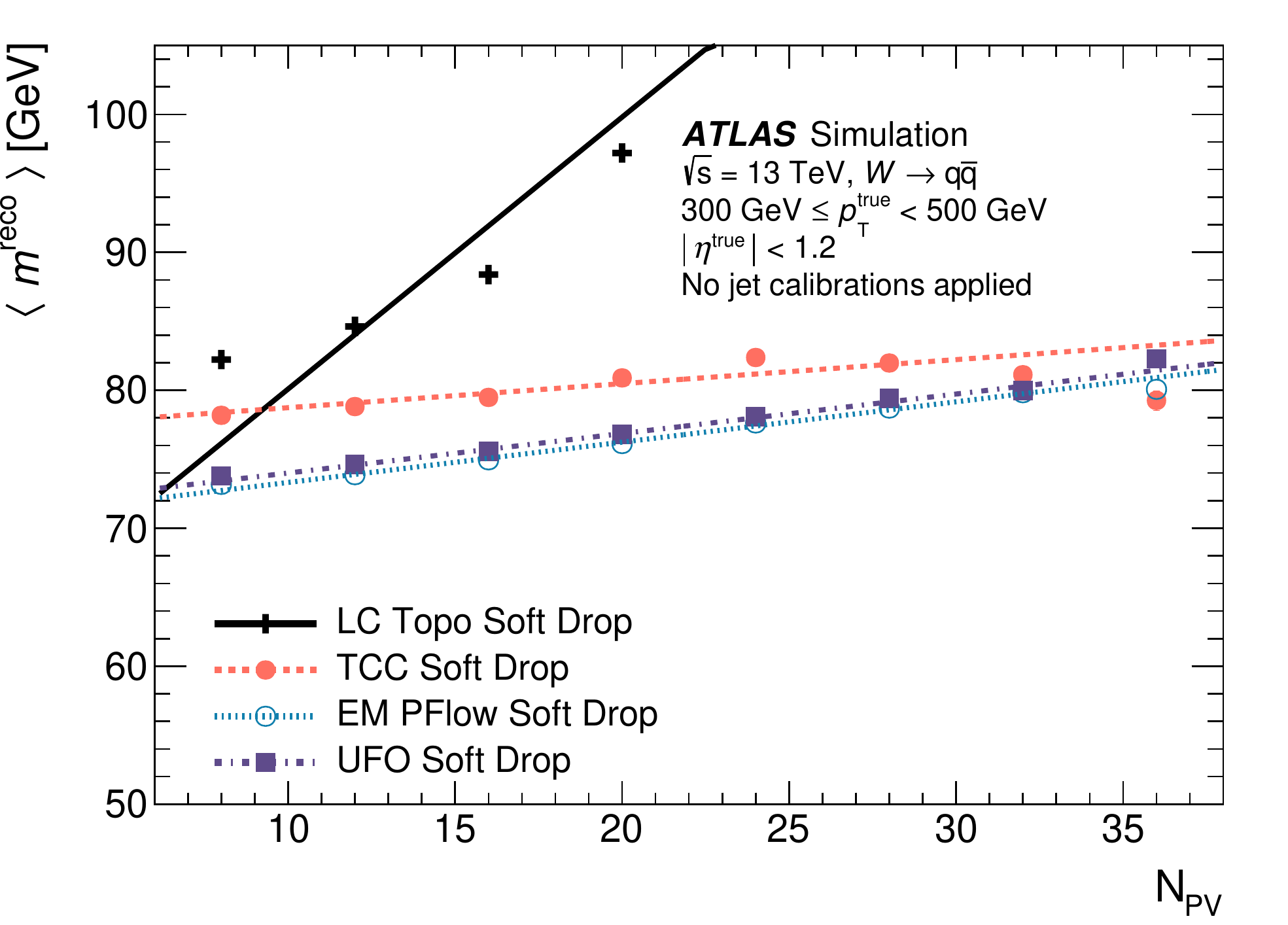}} \\
\subfigure[]{\includegraphics[width=0.49\textwidth]{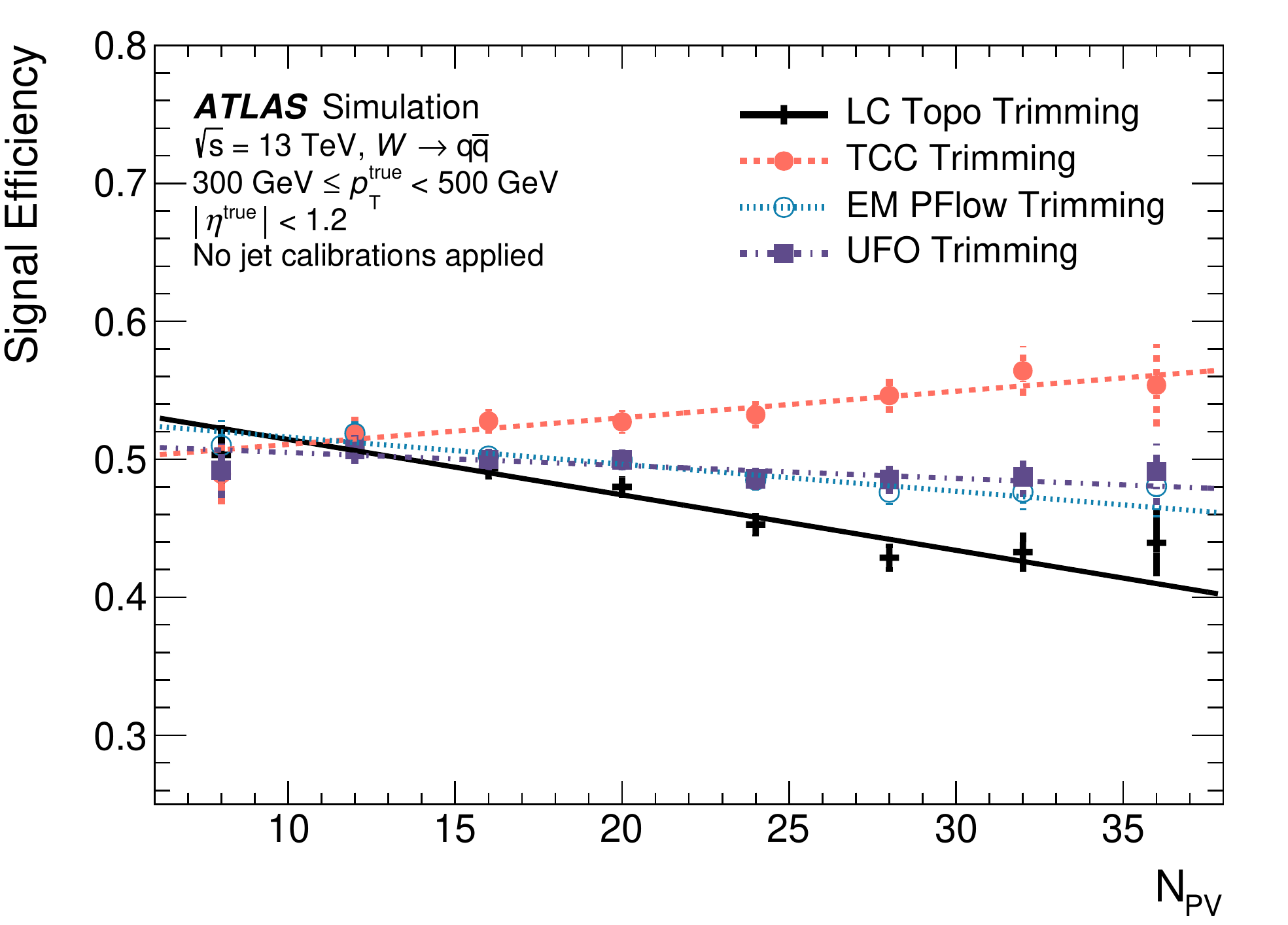}}
\subfigure[]{\includegraphics[width=0.49\textwidth]{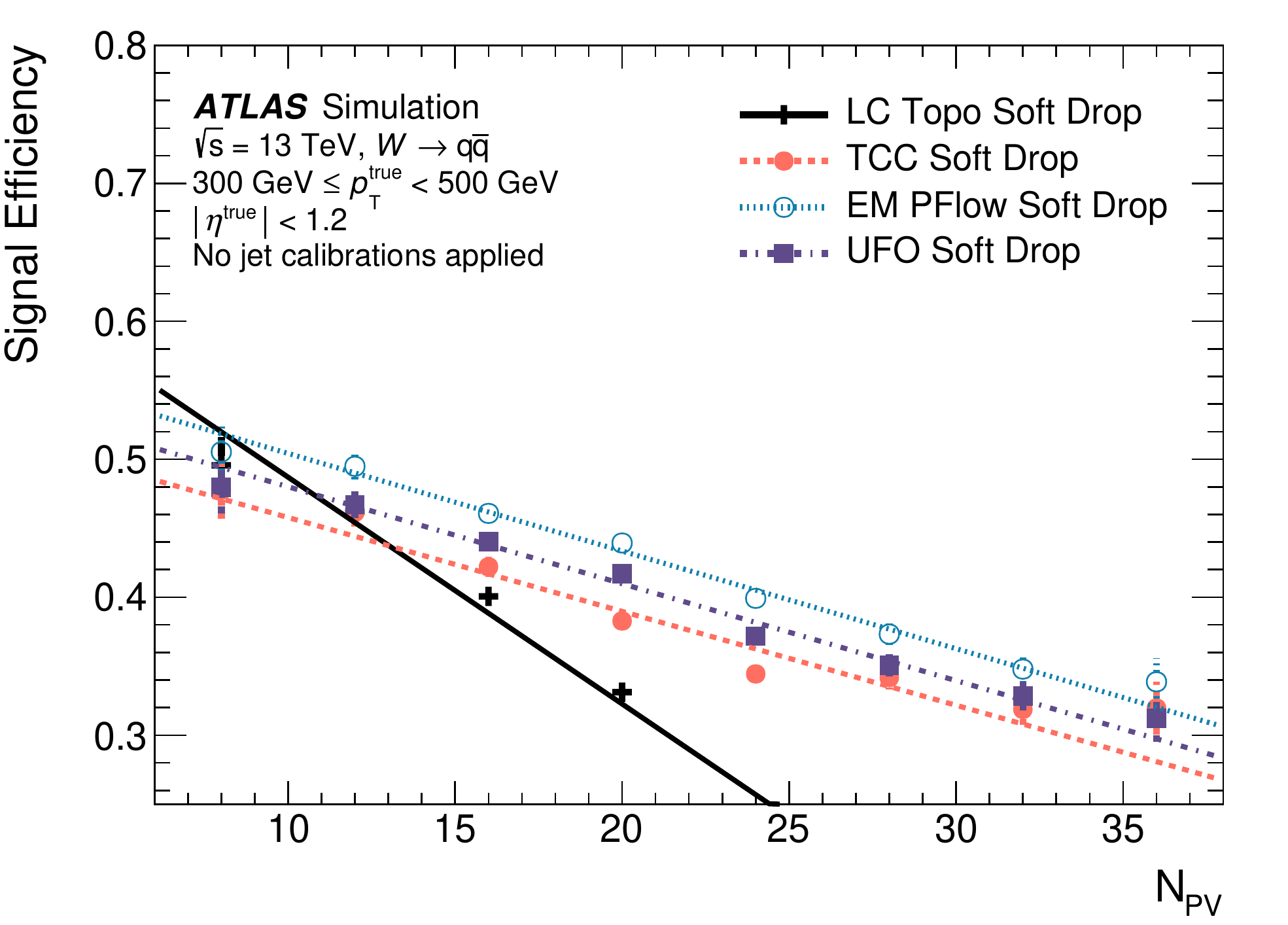}}
\end{center}
\caption{(top row) The value of the fitted $W$ boson mass peak, and (bottom row) the signal efficiency of a $W$ boson tagger as a function of the number of primary vertices, \Npv. These results are shown for \largeR jets groomed with the (left) trimming ($\Rsub=0.2$, $\fcut=0.05$) or (right) soft-drop ($\beta=1.0$, $\zcut = 0.1$) algorithms. A comparison of different jet input object types is made, including topoclusters, particle-flow objects, track-caloclusters and unified flow objects.}
\label{fig:nocalib:pileup:wmass:ufos}
\end{figure}
 
\FloatBarrier
\section{Performance survey}\label{sec:results}
The metrics described in Section~\ref{sec:nocalib} are used to study the performance of all jet definitions listed in Table~\ref{tab:summary:algos}, with the addition of UFOs. This provides a more complete understanding of the interplay between the different aspects of jet reconstruction. The results are summarised in Figures~\ref{fig:heatmap:w}--\ref{fig:heatmap:topology}.
 
\FloatBarrier
\subsection{Tagging performance}
 
A comparison of the background rejection of the $W$ tagger at the 50\% signal tagging efficiency working point is shown in Figure~\ref{fig:heatmap:w} for two \pt bins: a low-\pt bin (300~\GeV~< $\pt^\text{true, ungroomed}$ < 500~\GeV), and a high-\pt bin (1000~\GeV~< $\pt^\text{true, ungroomed}$ < 1500~\GeV).
 
Several trends are apparent from the performance of the taggers. As seen in Section~\ref{sec:nocalib}, for a fixed grooming algorithm, PFO reconstruction improves on topocluster reconstruction for both \pt bins, while TCCs improve background rejection even further at high \pt. In both cases, UFO reconstruction is able to match or improve on the performance of other jet inputs for both \pt bins. In general, pile-up mitigation improves $W$ tagging performance for all input types. The effects of pile-up mitigation are more apparent at low \pt, where soft pile-up radiation has a larger impact on the reconstruction of \DTwo. At high \pt, pile-up mitigation significantly improves the performance of TCC jets. This is related to the greater impact of pile-up mitigation for TCCs on the background mass distribution than the signal distribution, which increases the background rejection.
 
The tagging performance varies significantly among the different grooming algorithms and parameter choices. For trimming algorithms, smaller values of \Rsub or larger values of \fcut result in reduced tagging performance, regardless of the jet input type. These parameter choices correspond to more aggressive grooming, indicating that some of the softer radiation is important for effectively tagging different types of jets. An analogous observation is made for SD jets, where small values of $\beta$, or large values of \zcut generally result in degraded tagging performance.
 
A similar set of results is seen for the top tagger in Figure~\ref{fig:heatmap:top}. In the low-\pt bin, PFlow jets typically outperform both topocluster and TCC jets, while TCC jets outperform the other input object types at high \pt. Again, UFO jets are able to match or improve the performance compared to the other jet input types in both \pt bins. Pile-up mitigation tends to improve results, particularly at low \pt, as observed for $W$ taggers, although in a few cases the background rejection deteriorates. The baseline trimming algorithm works well for all input object types, but at low \pt, the background rejection may be improved by 50\% by instead using a SD algorithm with lighter grooming. The standard SD algorithm with $\beta=1$ and $z_{\mathrm{cut}}=0.1$ works particularly well, although recursive and bottom-up variants can also provide comparable performance.
 
In general, the tagging performance of jets constructed out of UFOs matches or exceeds that of jets reconstructed out of any other input type.
 
\begin{figure}[htbp]
\begin{center}
\subfigure[]{\includegraphics[width=0.95\textwidth]{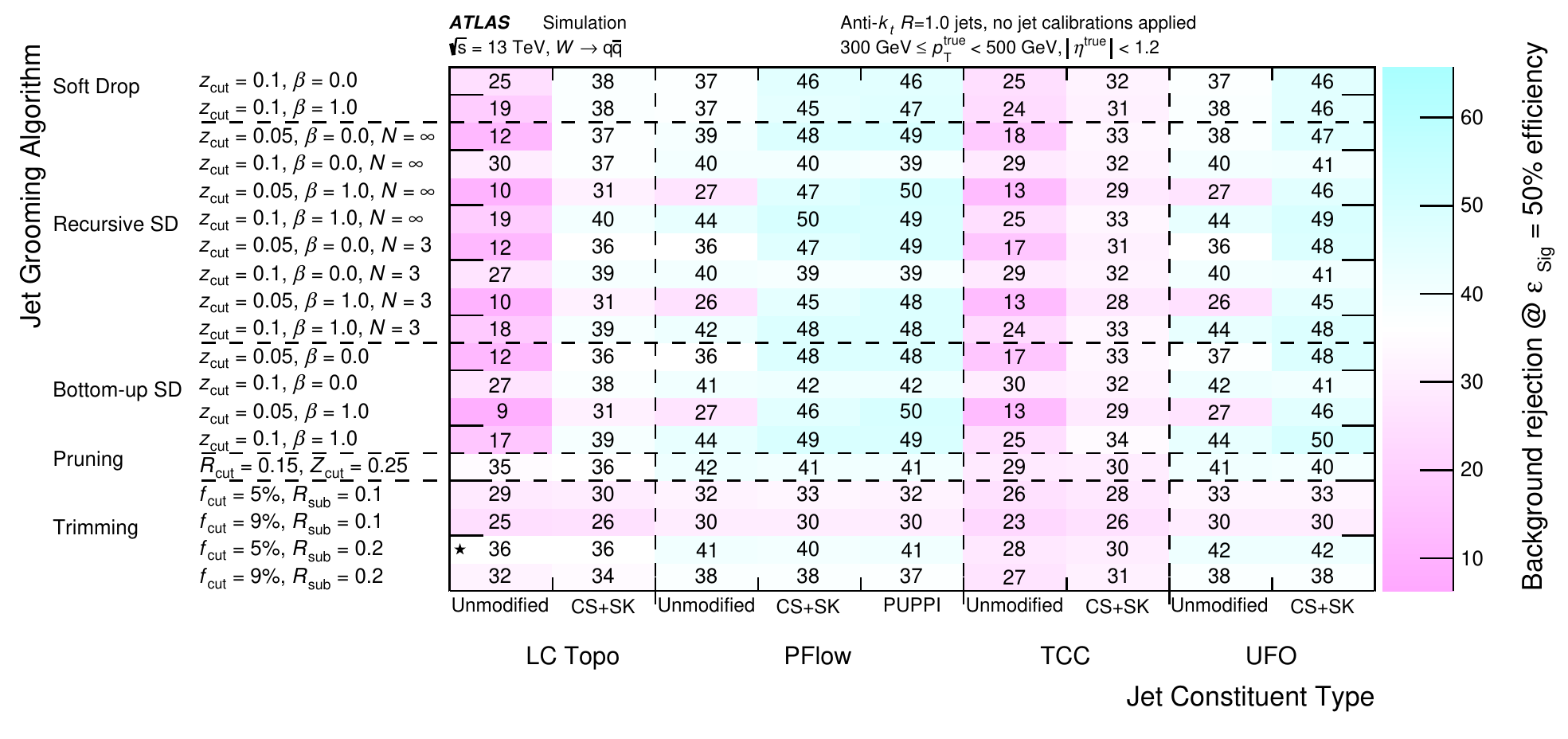}} \\
\subfigure[]{\includegraphics[width=0.95\textwidth]{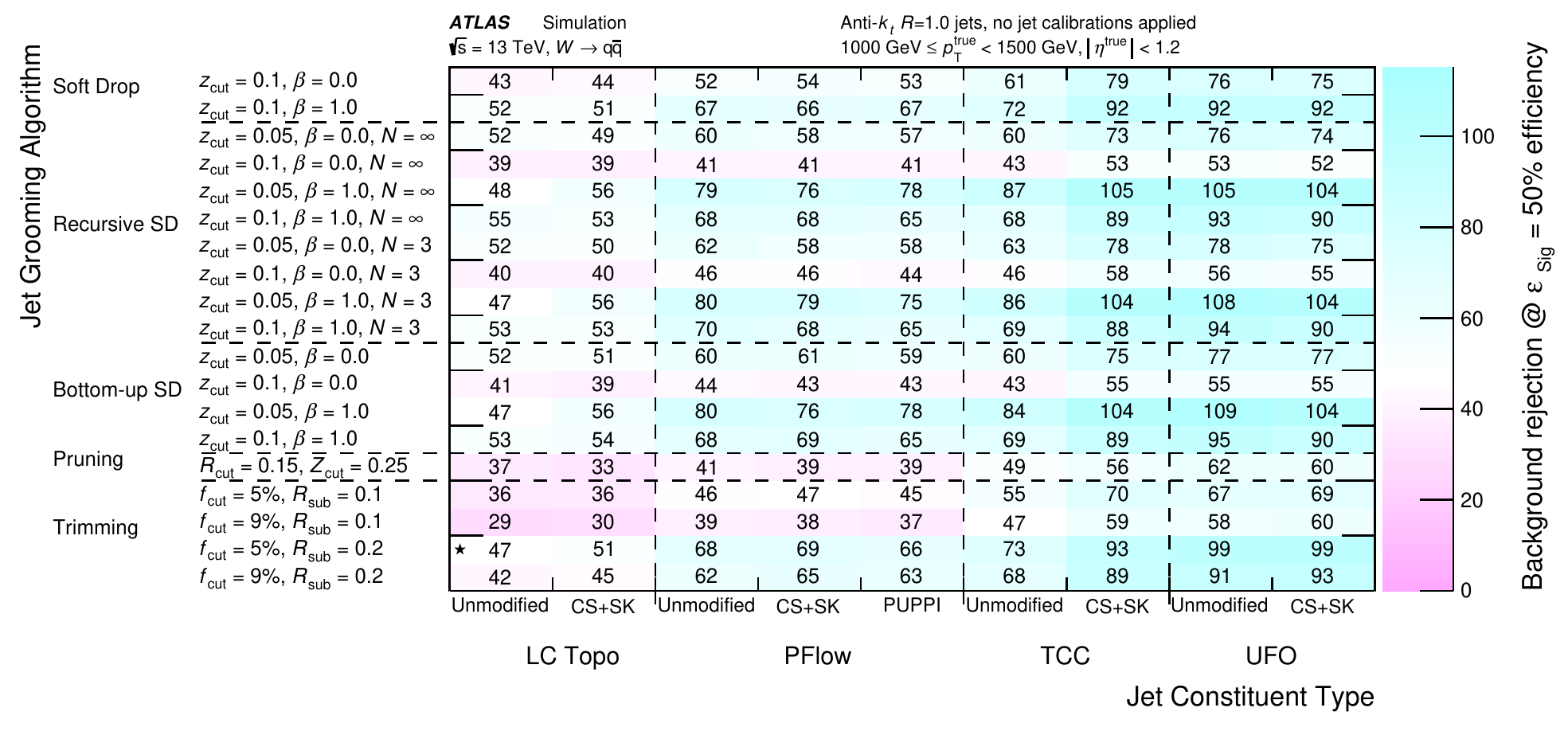}}
\end{center}
\caption{Background rejection at 50\% signal efficiency for a tagger using the jet mass and \DTwo for $W$ boson jets at (a) low \pt, and (b) high \pt. Jet \pt and $\eta$ cuts before tagging are made using the ungroomed particle-level \largeR jet matched to each of the groomed reconstructed \largeR jets. The current baseline topocluster-based trimmed collection is indicated with a $\star$.}
\label{fig:heatmap:w}
\end{figure}
 
\begin{figure}[htbp]
\begin{center}
\subfigure[]{\includegraphics[width=0.95\textwidth]{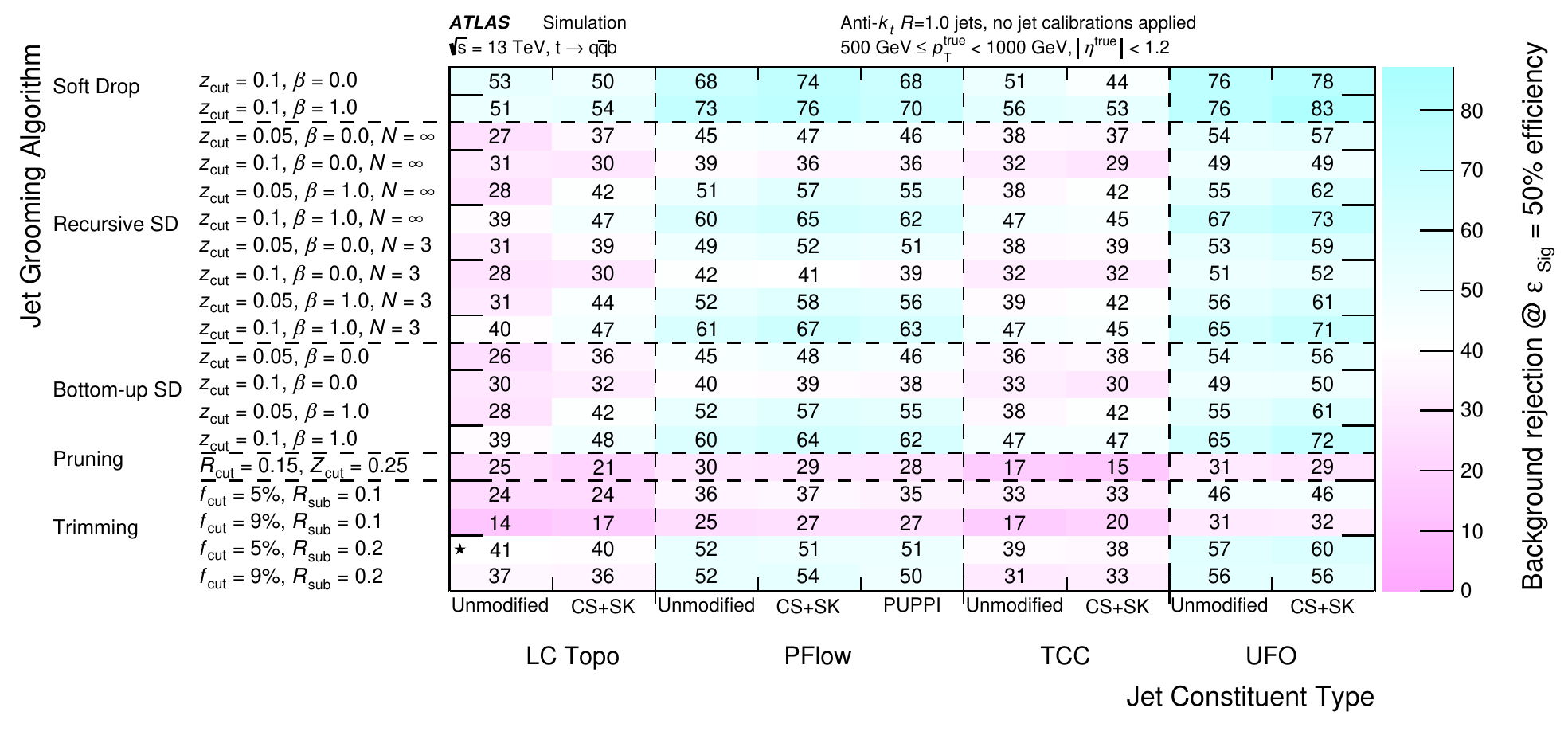}} \\
\subfigure[]{\includegraphics[width=0.95\textwidth]{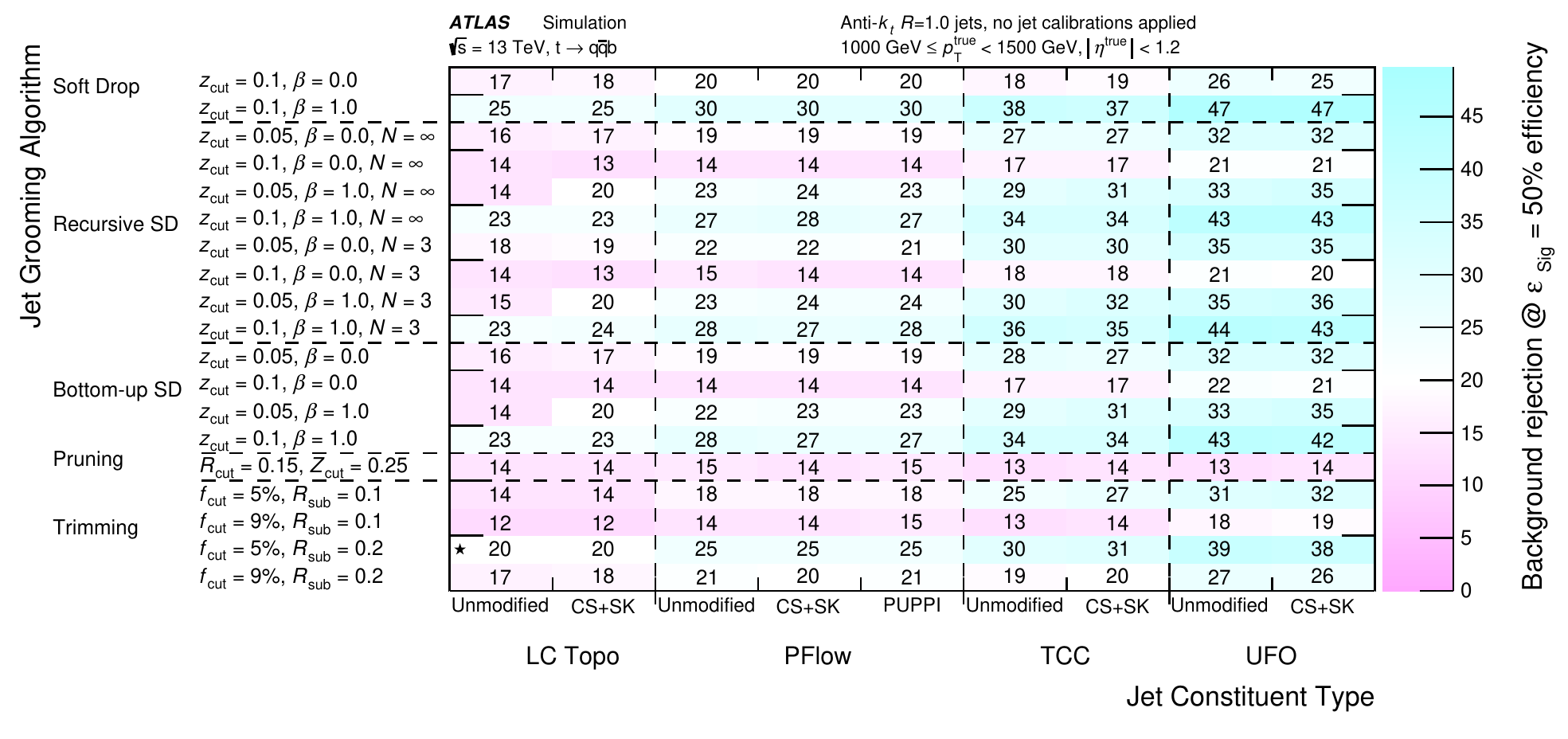}}
\end{center}
\caption{Background rejection at 50\% signal efficiency for a tagger using the jet mass and \tauthrtwo for top quark jets at (a) low \pt, and (b) high \pt. Jet \pt and $\eta$ cuts before tagging are made using the ungroomed particle-level \largeR jet matched to each of the groomed reconstructed \largeR jets. The current baseline topocluster-based trimmed collection is indicated with a $\star$.}
\label{fig:heatmap:top}
\end{figure}
 
\FloatBarrier
\subsection{Pile-up stability}
 
The slopes of the fitted average $W$ boson jet mass as a function of \Npv are shown in Figure~\ref{results:pileup:wmass} for each of the surveyed jet definitions. The uncertainties in the fitted slope values tend to be negligible compared to the differences between reported values. Among jet input types, PFOs and UFOs are the most pile-up-stable. PFOs, TCCs, and UFOs are all more pile-up-stable than topoclusters, due to the ability to easily remove charged particles from pile-up vertices. As discussed in Section~\ref{sec:nocalib}, the fitted value of the TCC $W$ mass peak position decreases as a function of \Npv for most grooming algorithms, although for lighter grooming algorithms which are more affected by pile-up, the slope is sometimes positive. This effect is exacerbated by the use of CS+SK, and for CS+SK TCCs, all of the studied trends are negative.
 
There are significant differences in the pile-up stability of different jet grooming algorithms. In general, all studied configurations of trimming are stable. For SD, RSD and BUSD, stability depends on the parameter choice. Larger values of $\beta$, where more soft and wide-angled radiation is retained, have a larger pile-up dependence. As expected, for the same value of \zcut, RSD and BUSD are more stable than the standard SD definition.
 
For all input types, with the exception of TCCs, jet-input-level pile-up mitigation techniques improve the pile-up stability of the jet definitions. Since too much energy is already subtracted for TCCs because of the inclusion of pile-up tracks in their reconstruction, any additional subtraction further degrades performance. For other jet inputs, the use of pile-up mitigation reduces the pile-up sensitivity so that it is better than or equivalent to the pile-up sensitivity from the baseline trimmed topocluster jet definition. This is true even for lightly groomed algorithms (e.g.\ RSD with $\zcut=0.05$, $\beta=1$, $N=3$), where CS+SK improves stability by a factor of 20. While PUPPI improves the pile-up stability of PFOs, the performance of CS+SK PFOs is better overall, sometimes by more than a factor of two. This improvement is seen for nearly all grooming algorithms. The pile-up stability of UFOs is similar to that of PFOs, which is expected since the modified TCC splitting step does not remove pile-up particles.
 
\begin{figure}[htbp]
\begin{center}
\includegraphics[width=0.95\textwidth]{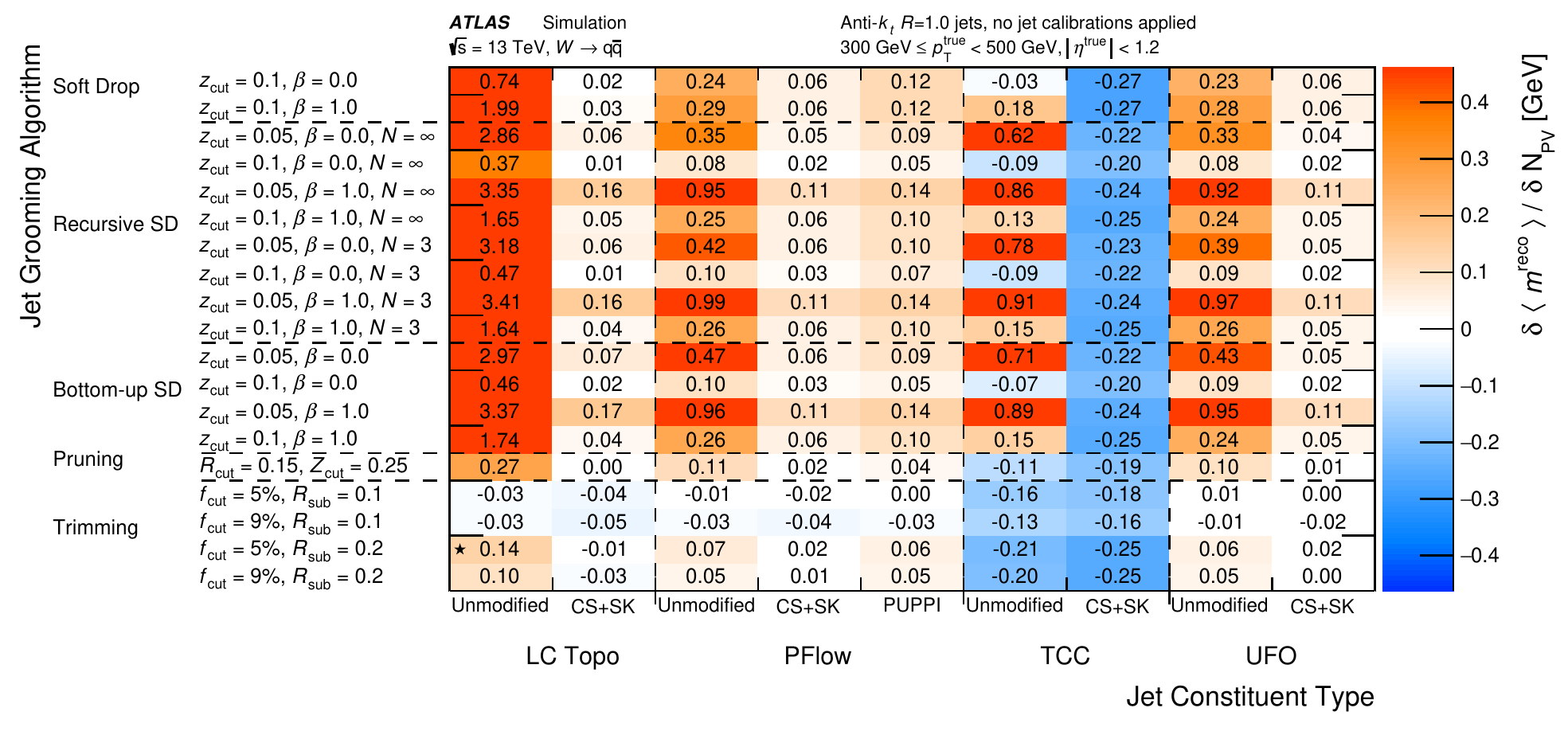}
\end{center}
\caption{Pile-up dependence of the value of the fitted $W$ boson mass peak at low \pt. Jet \pt and $\eta$ cuts before tagging are made using the ungroomed particle-level \largeR jet matched to each of the groomed reconstructed \largeR jets. The current baseline topocluster-based trimmed collection is indicated with a $\star$. The $z$-axis colour range is based on the difference of the baseline collection from a slope of 0. This makes differences between definitions more discernible than those between very unstable collections, which may have values beyond the axis range.}
\label{results:pileup:wmass}
\end{figure}
 
The change in signal efficiency of the \DTwo tagger as a function of \Npv is shown in Figure~\ref{results:pileup:d2}. Uncertainties in the reported values from the fitting procedure tend to be negligible (sub-percent level). As pile-up levels increase, the signal efficiency of the $W$ tagger tends to decrease. As observed when studying the $W$ mass peak position metric, topocluster inputs are the least stable. After pile-up mitigation, the pile-up stability of all inputs, including TCCs, improves by this metric. The trends in stability as a function of grooming algorithm are the same as for the $W$ mass position. While CS+SK is typically still more performant than PUPPI, the degree of improvement is not as large as that observed when studying the pile-up stability of the $W$ jet mass peak-position.
 
\begin{figure}[htbp]
\begin{center}
\includegraphics[width=0.95\textwidth]{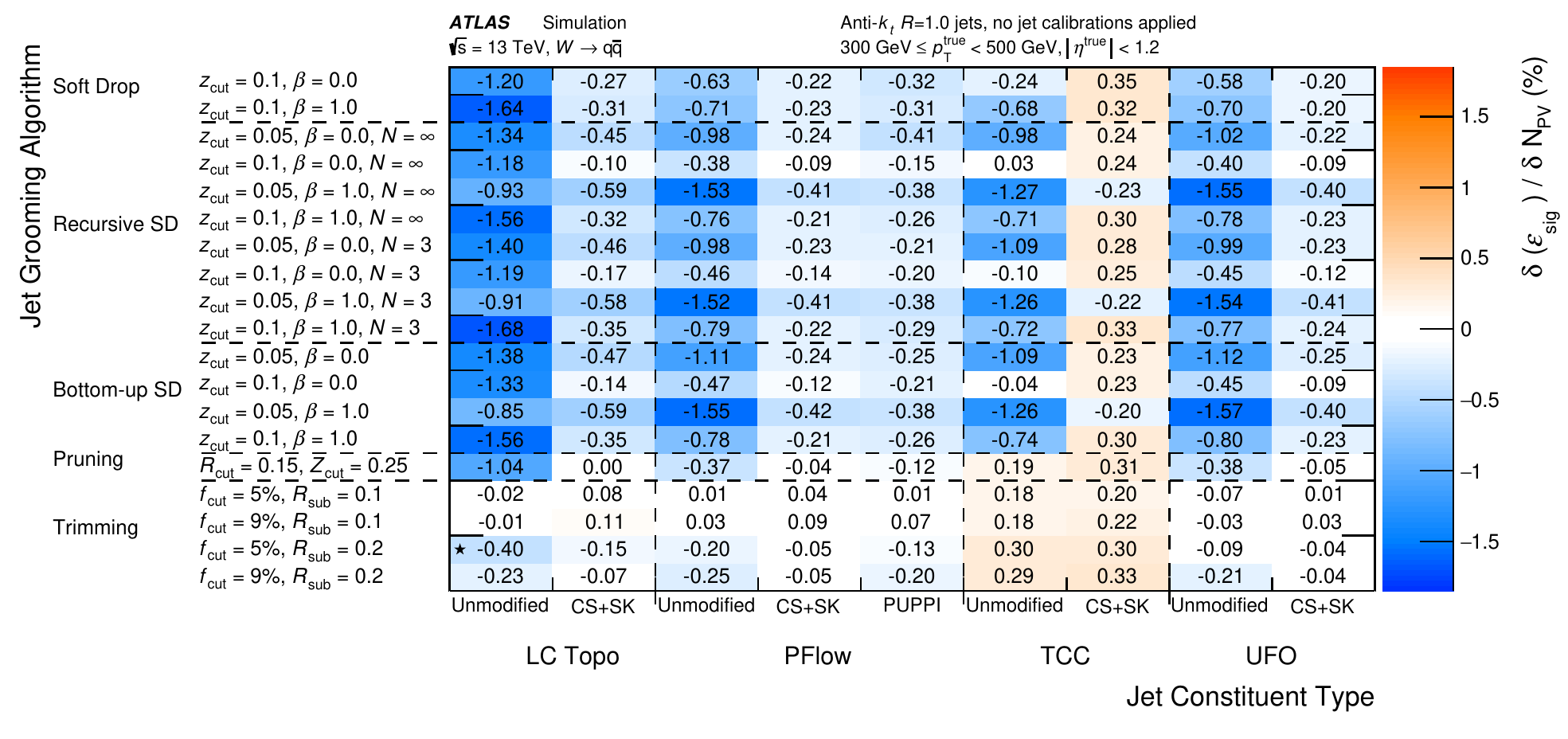}
\end{center}
\caption{Pile-up dependence of a \DTwo cut on the $W$ boson jet selection efficiency at low \pt. Jet \pt and $\eta$ cuts before tagging are made using the ungroomed particle-level \largeR jet matched to each of the groomed reconstructed \largeR jets. The current baseline topocluster-based trimmed collection is indicated with a $\star$. The $z$-axis colour range is based on the difference of the baseline collection from a slope of 0. This makes differences between definitions more discernible than those between very unstable collections, which may have values beyond the axis range.}
\label{results:pileup:d2}
\end{figure}
 
\FloatBarrier
\subsection{Topological sensitivity}
 
In order to examine the topology dependence of the jet energy and mass scale for different jet definitions, the ratio of the mean value of the uncalibrated jet mass response for $W$ jets to that of background jets is constructed. These values can be significantly different, as seen in  Section~\ref{sec:nocalib}. Deviations from unity will result in non-closure in the mass response following calibration. This effect is largest at low \pt, where the reconstruction of $W$ jets is relevant. As seen in Figure~\ref{fig:heatmap:topology}, the baseline topocluster-based trimmed \largeR jet definition used by the ATLAS experiment shows a score of around 4\% in this metric, and so small deviations from unity are not problematic.
 
The topology dependence is increased by the application of jet-input-level pile-up mitigation algorithms. In general, TCCs show the most sensitivity, which can reach 20\% after pile-up mitigation algorithms are applied. The topological sensitivity is increased for all inputs after the application of CS+SK, regardless of the grooming algorithm applied. This effect is generally lower for UFOs than for other jet inputs, even after pile-up mitigation algorithms are applied; the behaviour of PFlow jets is similar.
 
\begin{figure}[htbp]
\begin{center}
\includegraphics[width=0.95\textwidth]{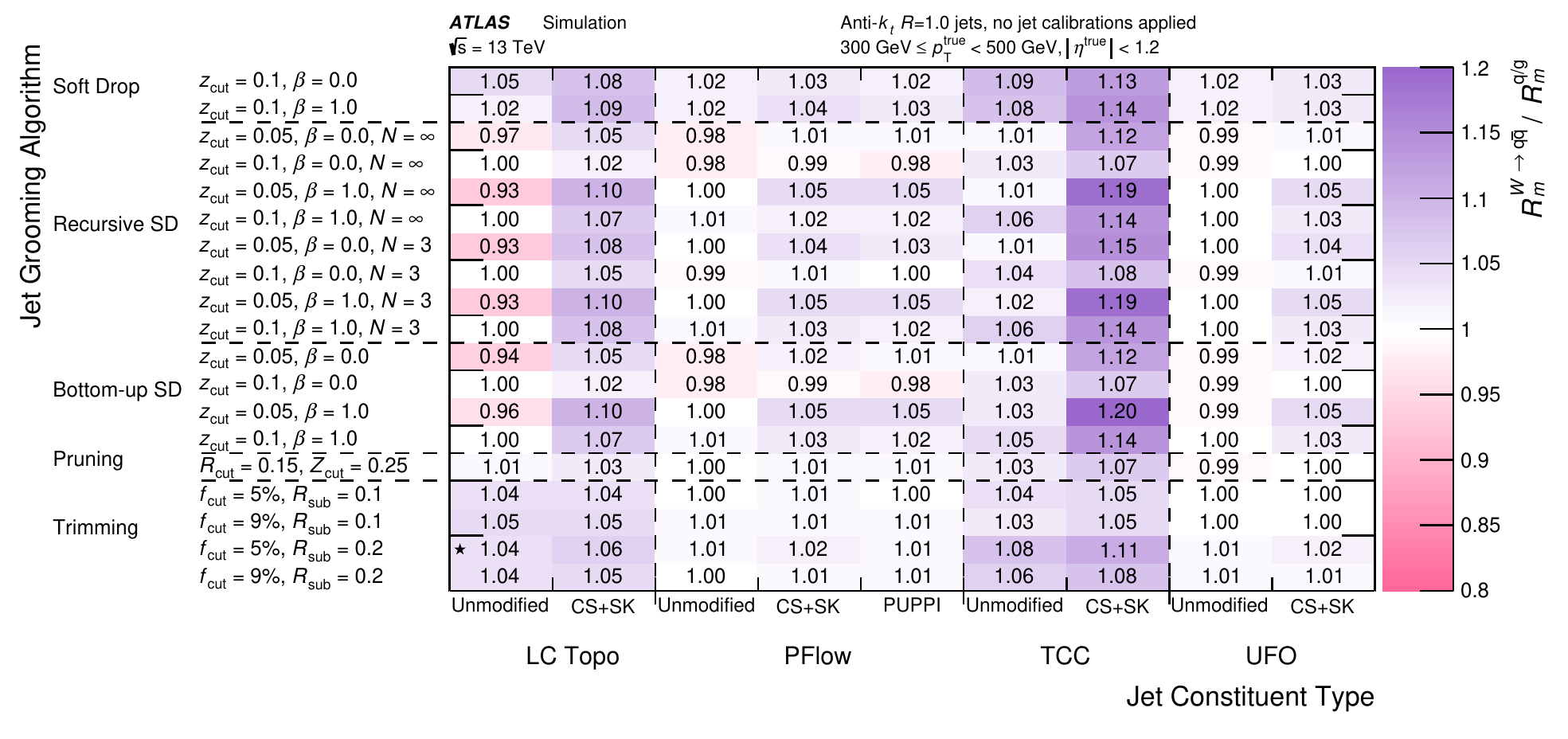}
\end{center}
\caption{Ratio of the mean value of mass response in $W$ jets to that in $q/g$ jets at low \pt. Kinematic selections before tagging are made using the ungroomed particle-level \largeR jet matched to each of the groomed reconstructed \largeR jets. The current baseline topocluster-based trimmed collection is indicated with a $\star$.}
\label{fig:heatmap:topology}
\end{figure}
 
\clearpage
 
\FloatBarrier
\section{Comparison of calibrated jet definitions}\label{sec:calib}
The tagging performance of a jet definition will have the largest impact on the sensitivity of searches for new physics performed by ATLAS, and so it is the primary metric used to determine which definitions are important for further study. The pile-up stability and topological sensitivity of the jet mass scale are also important, but since the performance of the baseline topocluster-based trimmed jet definition is still adequate, they are primarily used to distinguish between otherwise similar jet definitions. The primary motivation for choosing UFO-based definitions for further study is their $W$ boson and top quark tagging performance.
 
Based on their optimal tagging performance over the entire kinematic range of interest, in addition to the increased pile-up stability achieved by utilising tracking information in the jet definition, only jets reconstructed from UFOs are considered further. Several grooming algorithms are promising: soft-drop ($\beta=1.0, \zcut=0.1$) jets perform well when tagging high-\pt top quarks, while the RSD ($\beta=1.0, \zcut=0.05, N=\infty$) and BUSD ($\beta=1.0, \zcut=0.05$) extensions provide further improvements for high-\pt $W$ bosons. Trimmed UFO jets ($\fcut=0.05, \Rsub=0.2$) also provide competitive performance in certain regions. These four UFO jet definitions were selected for calibration and further study, as summarised in Table~\ref{tab:calib:jetchoice} in the category `studied definitions.'
 
\begin{table}[htbp]
\centering
\caption{Summary of the jet reconstruction algorithms, jet-input-level pile-up mitigation algorithms, and grooming algorithms which were determined to merit calibration and further study. Several promising UFO-based definitions are calibrated, as well as other definitions which enable comparisons of the impact of varying different aspects of jet definitions.}
\begin{tabular}{|c|c|c|c|}
\hline
\textbf{Category} & \textbf{Input Objects} & \textbf{Grooming Algorithm} & \textbf{Configuration} \\
\hline \hline
\textbf{Baseline} & LCW Topoclusters & Trimmed & $\Rsub=0.2$, $\fcut=0.05$ \\
\textbf{definitions}& TCCs & Trimmed & $\Rsub=0.2$, $\fcut=0.05$ \\
\hline\hline
& CS+SK UFOs & Trimmed & $\Rsub=0.2$, $\fcut=0.05$ \\
\textbf{Studied} & CS+SK UFOs & SD & $\zcut=0.1$, $\beta=1.0$ \\
\textbf{definitions} & CS+SK UFOs & RSD & $\zcut=0.05$, $\beta=1.0$, $N=\infty$ \\
& CS+SK UFOs & BUSD & $\zcut=0.05$, $\beta=1.0$ \\
\hline \hline
\textbf{Additional}& UFOs & Trimmed & $\Rsub=0.2$, $\fcut=0.05$ \\
\textbf{definitions}& PFOs & Trimmed & $\Rsub=0.2$, $\fcut=0.05$ \\
& UFOs & SD & $\zcut=0.1$, $\beta=1.0$ \\
\hline
\end{tabular}
\label{tab:calib:jetchoice}
\end{table}

\subsection{Simulation-based jet energy and mass scale calibrations}\label{sec:mcjesjms}
 
A simulation-based calibration is derived using \PYTHIA dijet events for each of the UFO collections which were selected for further study, as well as for additional \largeR jet definitions which will permit comparisons of each aspect of the jet definition which is studied. These jet definitions are listed in Table~\ref{tab:calib:jetchoice}. This calibration follows the methodology in Ref.~\cite{JETM-2018-02}, and restores the average reconstructed jet \pt and mass scales (JES, JMS) to those of the particle-level references. For each jet definition, a reference set of particle-level jets are reconstructed as described in Section~\ref{sec:objects:truth}, and the same grooming algorithm is applied as that used for the detector-level jet definition.
 
Detector-level jets are matched to particle-level jets using a procedure which minimises the distance $\DeltaR = \sqrt{(\Delta\phi)^2+(\Delta\eta)^2}$. The \pt and mass responses are defined respectively as $R_{\pt} = \langle \pt^{\text{reco}}/\pt^{\text{true}} \rangle$ and $R_{m} = \langle m^{\text{reco}}/m^{\text{true}} \rangle$, where the `reco' quantities correspond to the value of the jet energy or mass before any calibration has been applied. The truth quantities are defined using particle-level jets, reconstructed following the procedure described in Section~\ref{sec:objects:truth}. The average response is determined using a Gaussian fit to the core of each response distribution.
 
For the JES calibration, these fits are performed in bins of jet energy and detector pseudorapidity $\eta^{\mathrm{det}}$, defined as the jet pseudorapidity calculated relative to the geometrical centre of the ATLAS detector. This parameterisation yields a more accurate representation of the active calorimeter cells than that obtained when using the pseudorapidity calculated relative to the PV, and results in an improved evaluation of the calorimeter response. The JES correction factor, $c_\text{JES} = 1/R_{\pt}$ is smoothed in energy and $\eta^\text{det}$, and is applied to the four-momentum of the reconstructed jet as a multiplicative scale factor. A correction to the jet $\eta$ (`$\Delta\eta$' below) is also applied to correct for biases with respect to the particle-level reference in certain detector regions~\cite{ATLAS-CONF-2015-037}. The JES correction is similar for each of the four CS+SK UFO jet definitions which are calibrated, regardless of the grooming algorithm which is applied.
 
After the JES correction has been applied, the jet mass scale calibration is derived using the same procedure in bins of $E_{\text{reco}}$, $\eta^{\text{det}}$, and log($m_{\text{reco}}/E_{\text{reco}}$). The jet mass calibration factor $c_\text{JMS} = 1/R_{m}$ is applied only to the mass of the jet, keeping the  jet energy fixed and thus allowing the $\pt$ to vary. This factor is also a smooth function of the \largeR jet kinematics. The reconstructed \largeR jet kinematics are thus given by:
\begin{eqnarray*}
E_{\mathrm{reco}} = c_\text{JES}\,E_0,~~~m_{\mathrm{reco}} = c_\text{JES}\,c_\text{JMS}\,m_0,~~~\eta_{\mathrm{reco}} = \eta_0+\Delta\eta,~~~\pt^{\mathrm{reco}} = c_\text{JES} \frac{\sqrt{E_0^2-c^2_\text{JMS}\,m_0^2}}{\cosh{(\eta_0+\Delta\eta)}},
\end{eqnarray*}
where the quantities $E_0$, $m_0$ and $\eta_0$ refer to the jet properties prior to any calibration, but following the jet grooming procedure. The JMS correction is mostly similar for each of the four CS+SK UFO jet definitions which are studied, but differences in the size of the correction become largest for massive jets at high \pt. Figure~\ref{fig:calib:JMS:W} presents the average jet mass response $R_{m}$ for jets with a particle-level jet mass equal to that of the $W$ boson, for the four CS+SK UFO jet definitions which are calibrated. The response for \largeR jets with this mass is obtained by directly taking a profile through the smoothed response maps. High-\pt trimmed jets require a smaller calibration factor than jets which are groomed using the SD, RSD or BUSD algorithms. This indicates that there are differences in the high-\pt behaviour of grooming algorithms: trimming removes more pile-up from jets at high \pt, bringing the average JMS of these jets closer to particle level before calibration.
 
\begin{figure}[htbp]
\begin{center}
\subfigure[]{\includegraphics[width=0.47\textwidth]{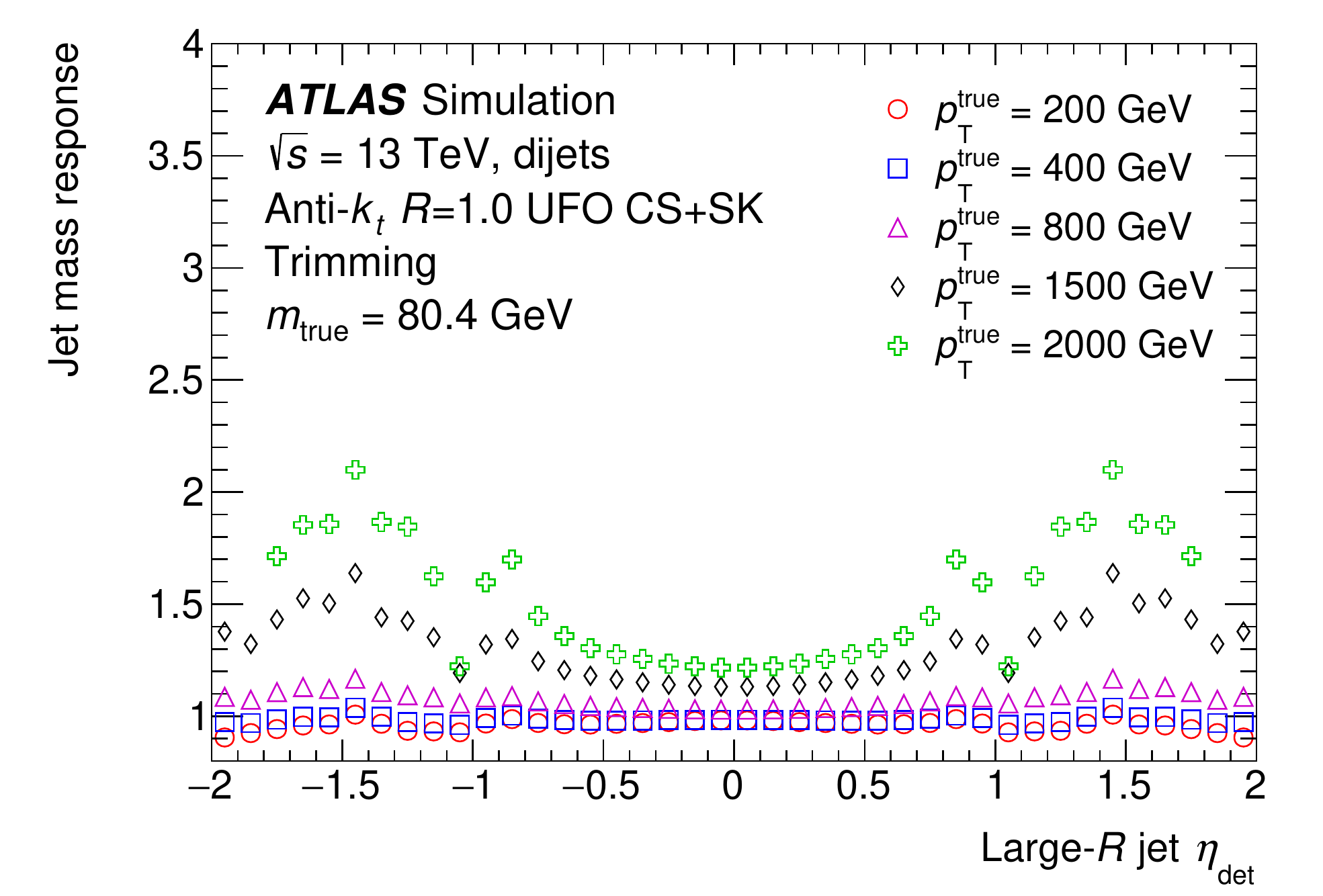}}
\subfigure[]{\includegraphics[width=0.47\textwidth]{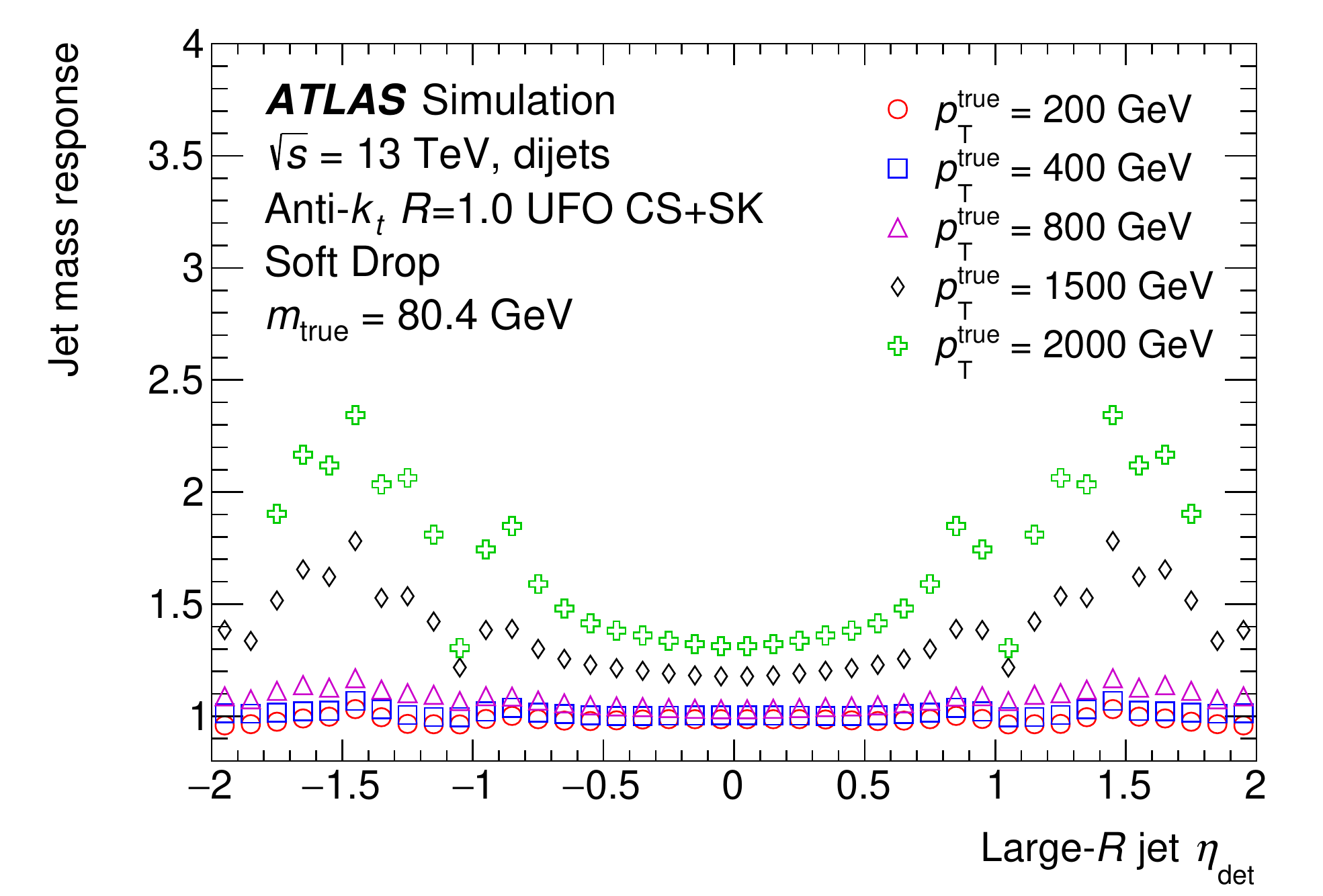}}
\subfigure[]{\includegraphics[width=0.47\textwidth]{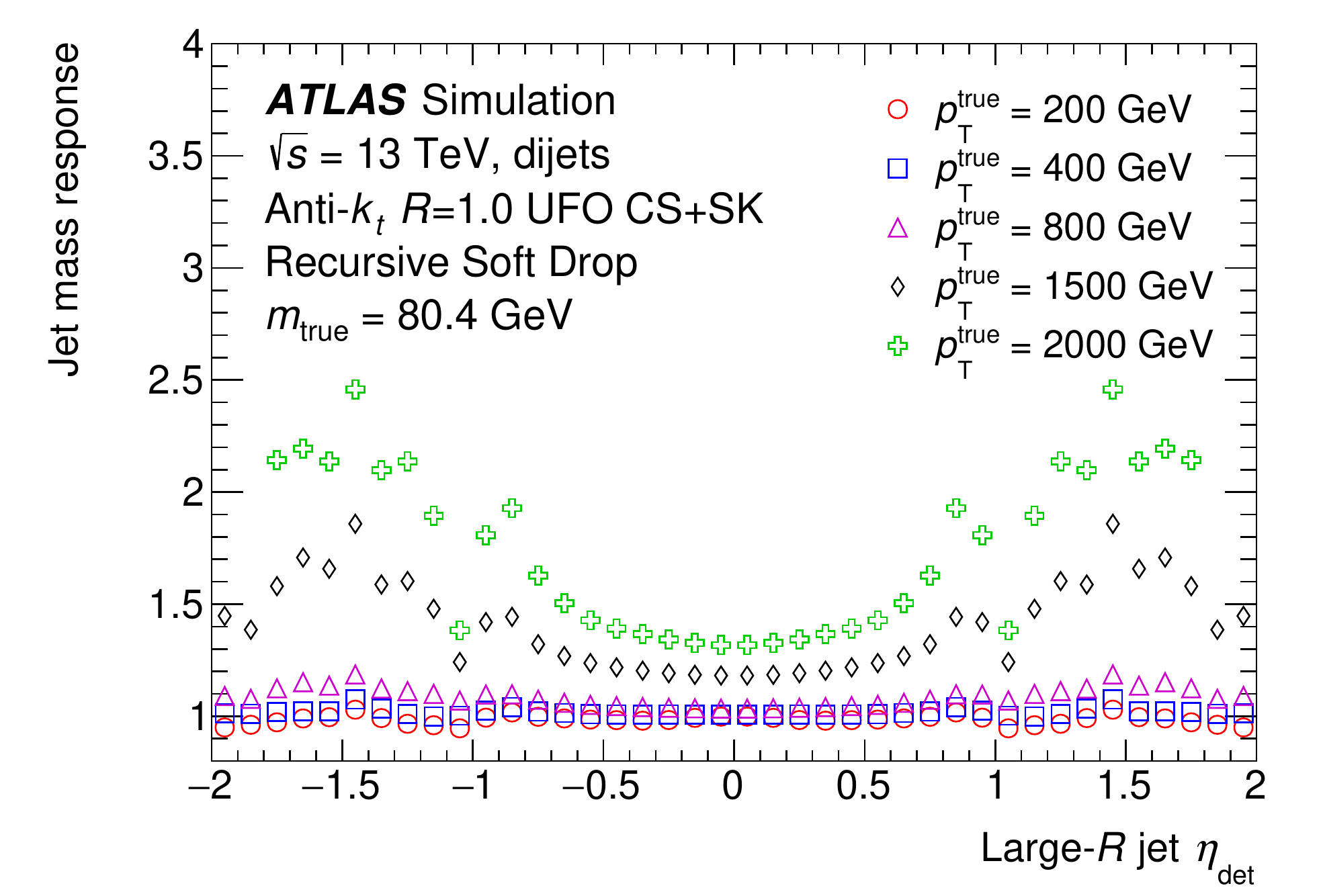}}
\subfigure[]{\includegraphics[width=0.47\textwidth]{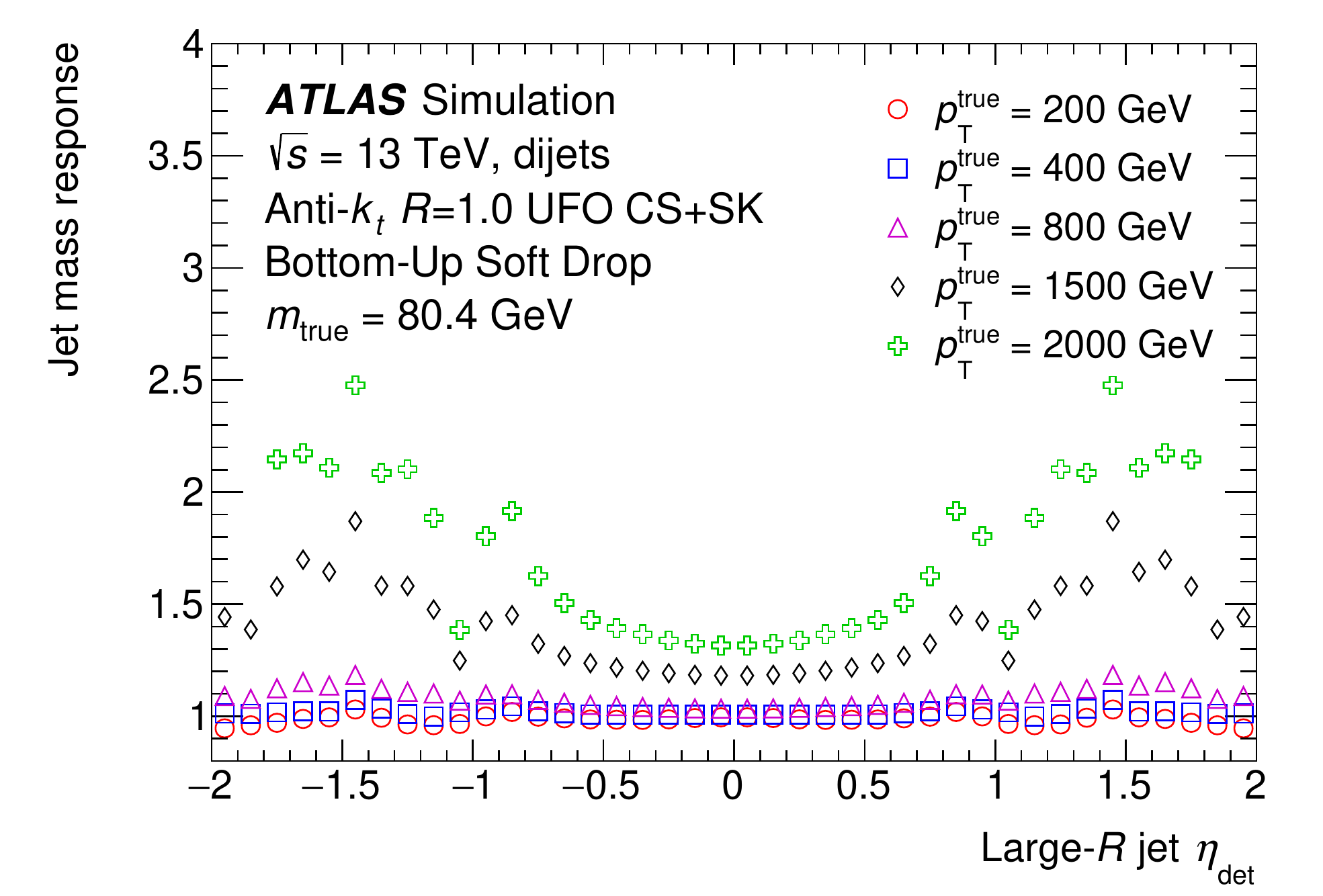}}
\end{center}
\caption{The jet mass response for UFO CS+SK \largeR jets which have been groomed with (a) trimming, (b) soft-drop, (c) recursive soft-drop and (d) bottom-up soft-drop. The jet mass response is presented as a function of jet pseudorapidity for several values of the jet transverse momentum from 200~\gev{} to 2~\tev{}, for jets with a particle-level mass equal to the $W$ boson mass. The mass responses for \largeR jets with this mass are obtained by directly taking a profile through the smoothed response maps.}
\label{fig:calib:JMS:W}
\end{figure}
 
All figures where JES+JMS calibrations have been applied to the \largeR jet four-vector are labelled `JES+JMS'.
 
\subsection{Comparison of calibrated jet definition performance}
 
\subsubsection{Jet mass and \pt resolution}
 
The expected \largeR jet mass resolution, defined to be the 68\% interquantile range divided by twice the median of the distribution, is shown in Figure~\ref{fig:calib:w:mResponse} for samples of signal jets. For these studies (as for all studies in this document), the baseline trimmed topocluster mass is used directly, rather than the combined mass~\cite{JETM-2018-02} (which incorporates additional measurements from the inner tracking detector), allowing a direct comparison of the unmodified performance of the different jet definitions. In Figures~\ref{fig:calib:w:mResponse:ufosGoodWs} and~\ref{fig:calib:w:mResponse:ufosGoodTops}, the resolution for all UFO jet definitions is shown to be better than for the baseline trimmed topocluster definition, particularly at high \pt. The expected mass resolution of UFO jets is stable across the entire \pt spectrum. In the low-\pt region	the mass resolution of UFO jets is typically similar to that of topocluster jets, while in the high-\pt region, it more closely follows the behaviour of TCC jets. For hadronically decaying high-\pt top quarks, UFOs improve the jet mass resolution relative to topocluster-based jets by 26\%, and by 40\% for high-\pt hadronically decaying $W$ bosons.
 
In order to help factorise the performance gains from various sources, comparisons of the jet mass resolution are also provided for several other calibrated jet definitions. Figures~\ref{fig:calib:w:mResponse:trimmedWs} and~\ref{fig:calib:w:mResponse:trimmedTops} show a comparison of the four unmodified input object types using the trimming algorithm. In general, at high-\pt the mass resolution of top quarks is better than that of $W$ bosons due to the fact that $W$ bosons are lighter, and their decay products are typically more collimated, making the calorimeter granularity relevant at lower values of \pt.
UFO jets outperform topocluster and TCC jets for both $W$ boson and top quark jets. PFlow jets are also found to be more performant than topocluster and TCC jets for top quark jets, although their performance deteriorates for highly boosted $W$ bosons. The trimming and soft-drop algorithms are compared for UFO jets with and without CS+SK pile-up mitigation in Figures~\ref{fig:calib:w:mResponse:CSSKWs} and~\ref{fig:calib:w:mResponse:CSSKTops}. The application of CS+SK does not significantly alter the mass resolution of trimmed UFO jets; however, it is found to improve the mass resolution for soft-drop jets at low \pt by nearly 40\%.
 
\begin{figure}[htbp]
\begin{center}
\subfigure[]{\includegraphics[width=0.47\textwidth]{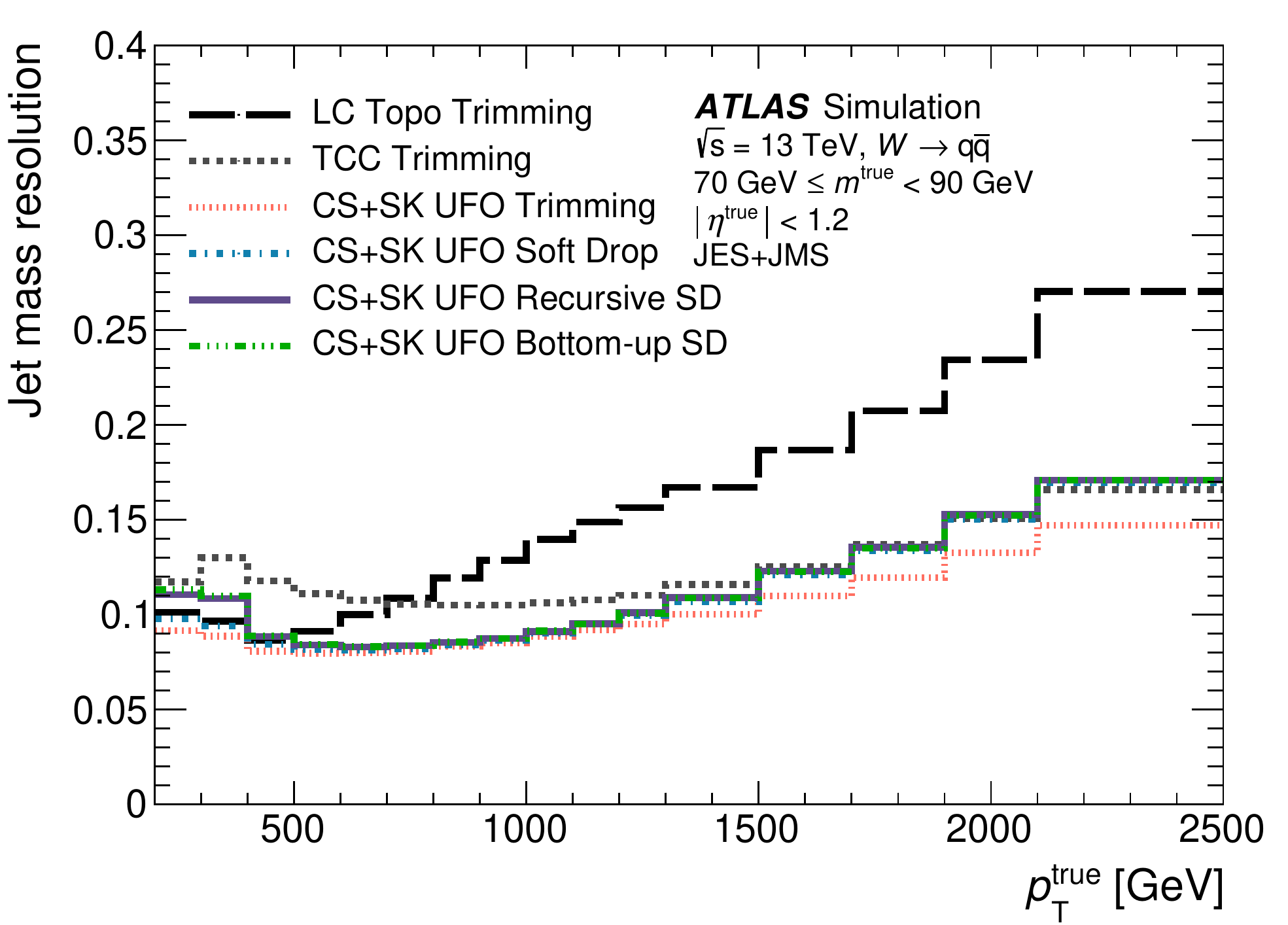}\label{fig:calib:w:mResponse:ufosGoodWs}}
\subfigure[]{\includegraphics[width=0.47\textwidth]{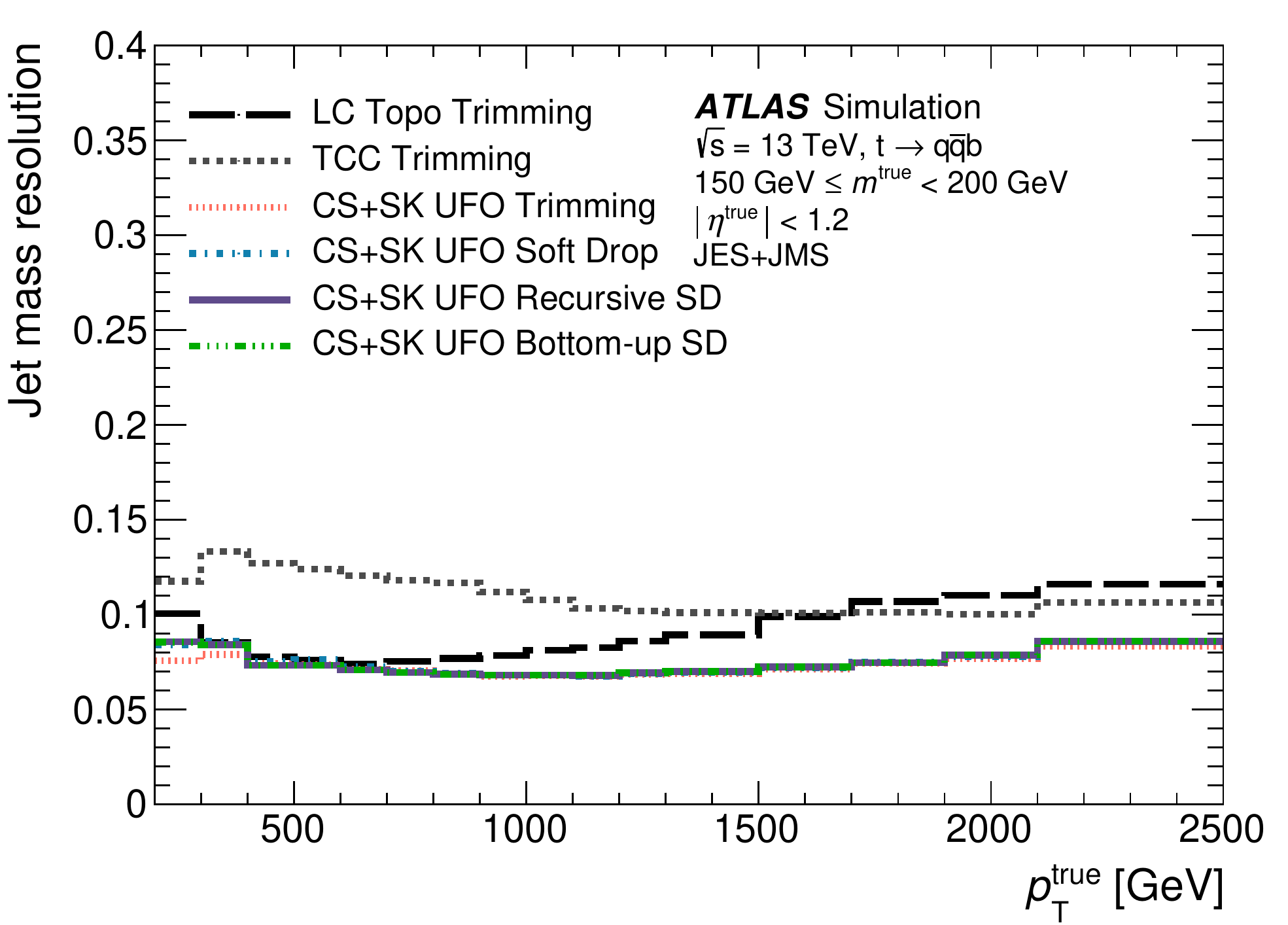}\label{fig:calib:w:mResponse:ufosGoodTops}} \\
 
\subfigure[]{\includegraphics[width=0.47\textwidth]{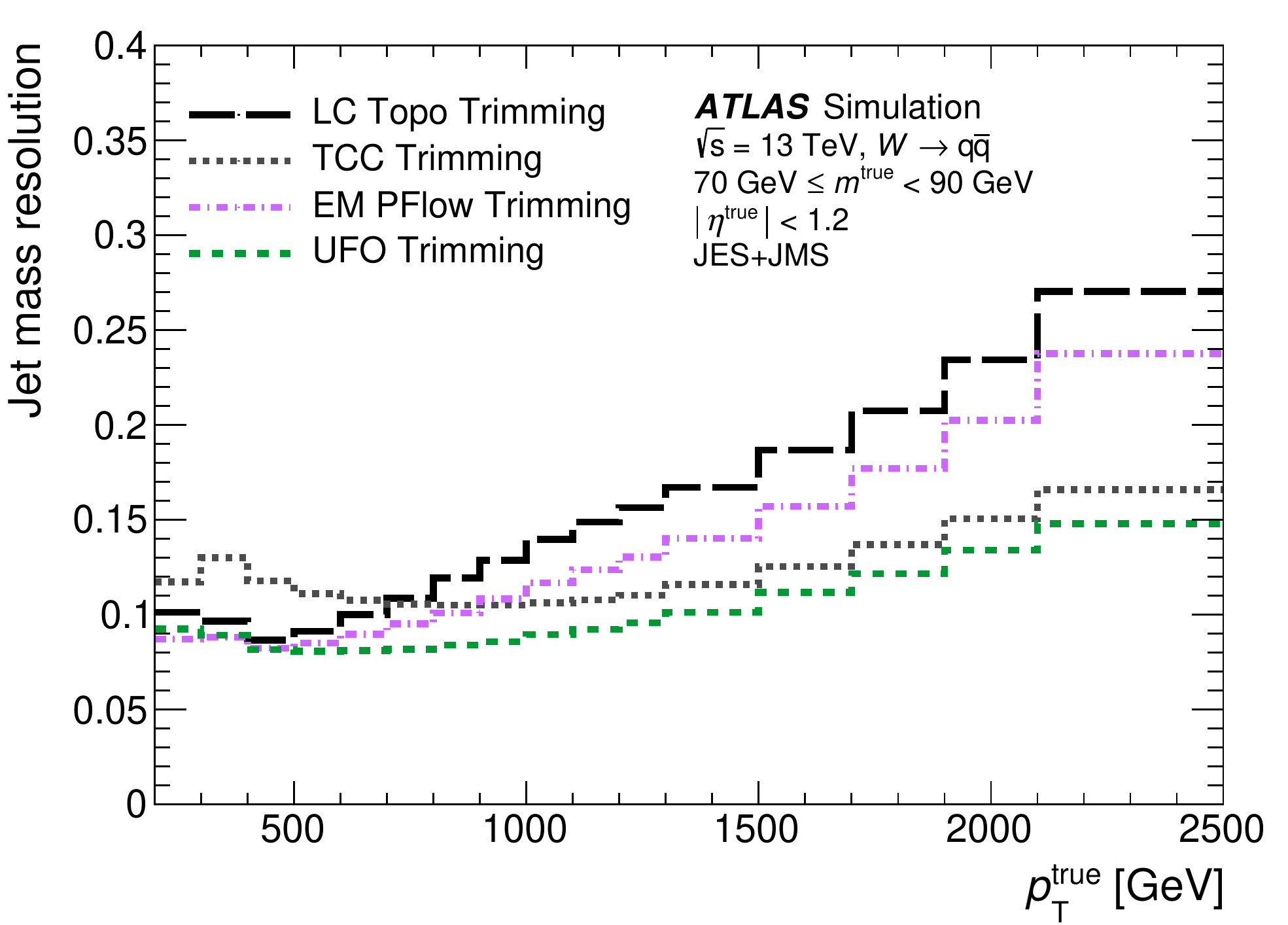}\label{fig:calib:w:mResponse:trimmedWs}}
\subfigure[]{\includegraphics[width=0.47\textwidth]{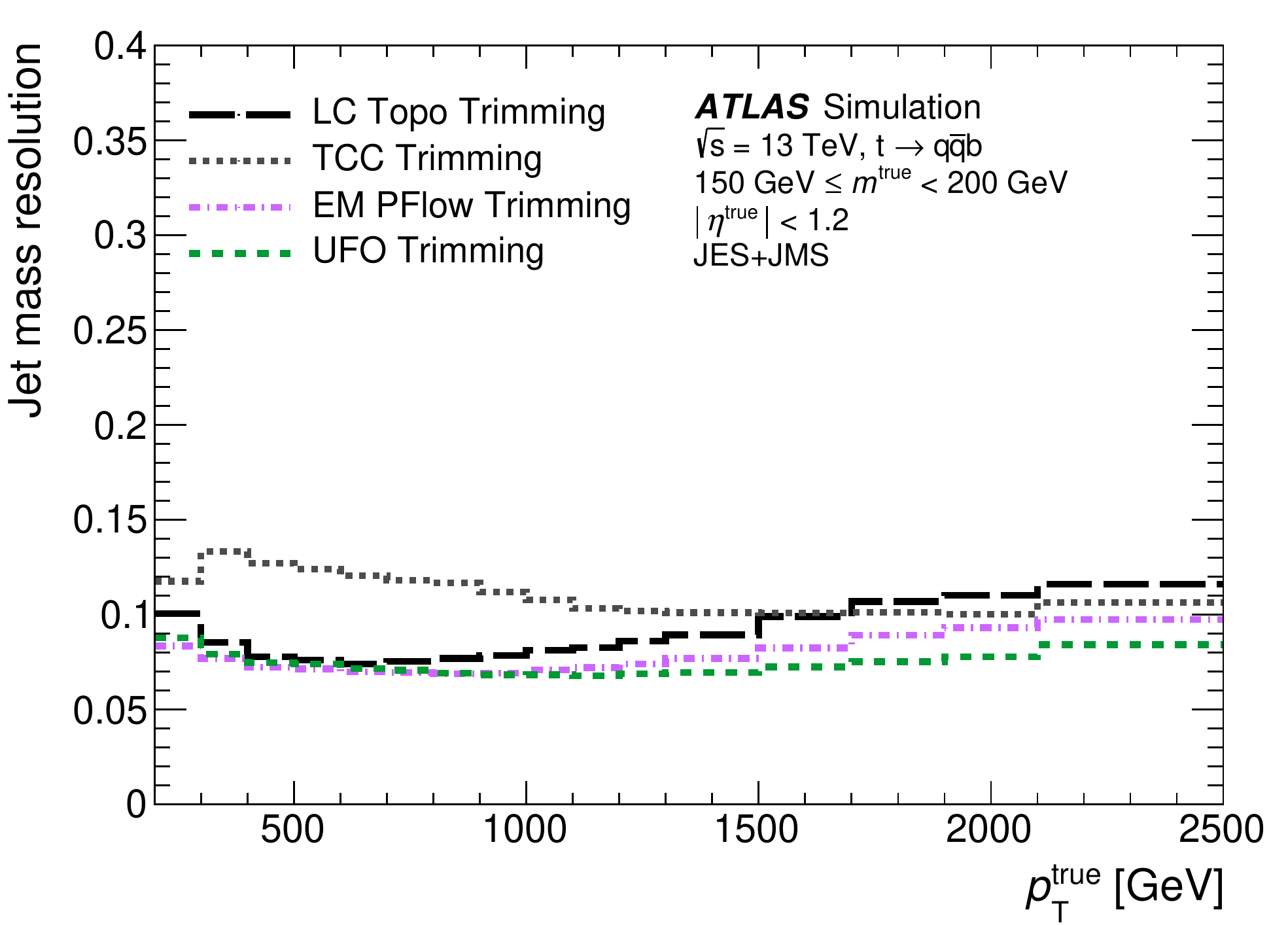}\label{fig:calib:w:mResponse:trimmedTops}} \\
 
\subfigure[]{\includegraphics[width=0.47\textwidth]{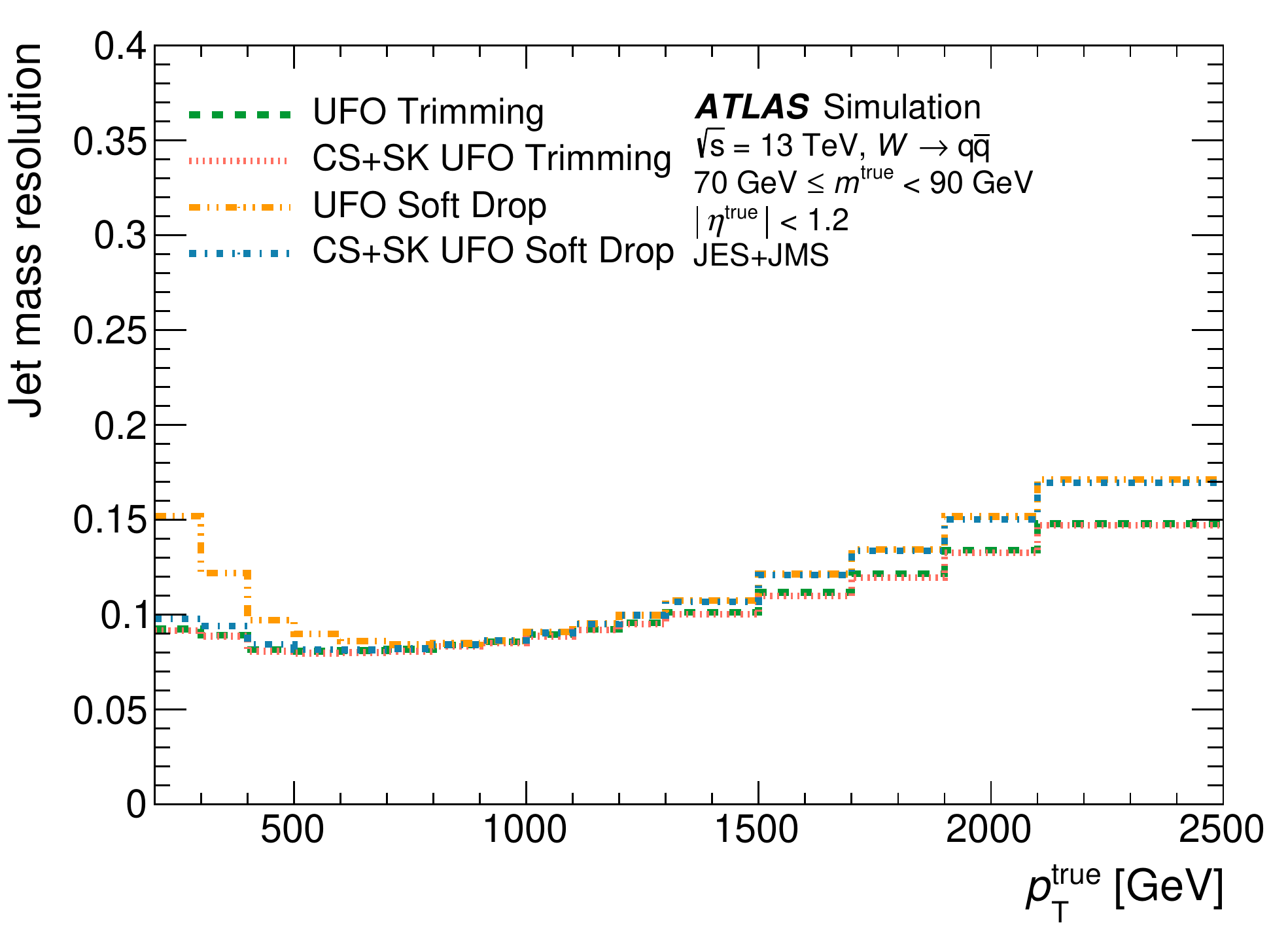}\label{fig:calib:w:mResponse:CSSKWs}}
\subfigure[]{\includegraphics[width=0.47\textwidth]{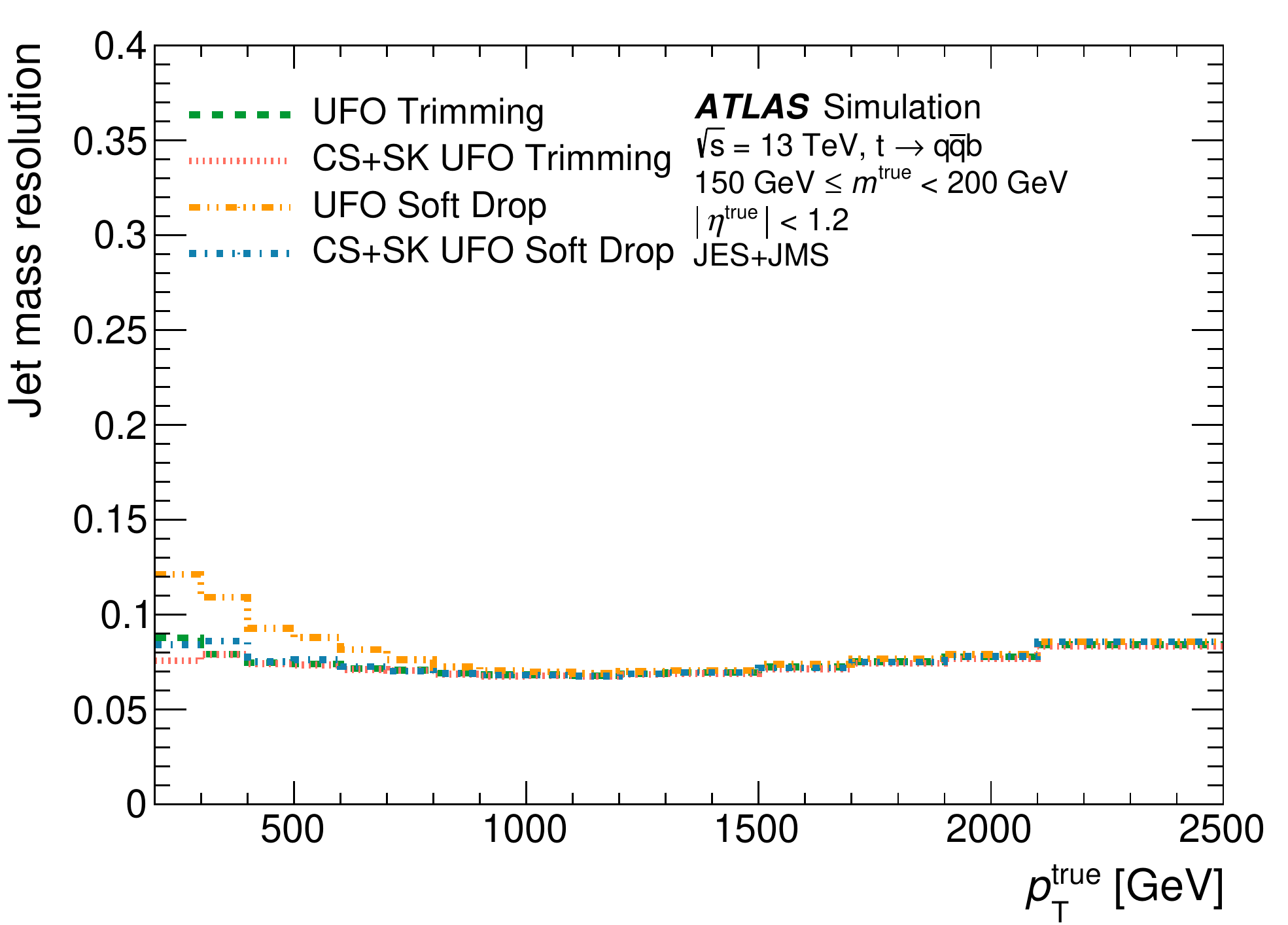}\label{fig:calib:w:mResponse:CSSKTops}} \\
\end{center}
\caption{The jet mass resolution for (a,c,e) $W$ boson jets, and (b,d,f) top quark jets as a function of \pt. In (a,b) the relative performance of the studied UFO definitions is compared with the current ATLAS baseline topocluster and TCC jets, while in (c,d) only jet input object types are compared, and in (e,f) the impact of pile-up mitigation is highlighted.}
\label{fig:calib:w:mResponse}
\end{figure}
 
The \largeR jet \pt resolution for background jets is shown in Figure~\ref{fig:calib:qcd:ptResolution}, determined as the one-standard-deviation width of Gaussian fits to the $R_{\pt}$ distributions divided by their fitted mean. The \pt resolution of trimmed topocluster jets is superior to that of either TCC trimmed jets or any of the UFO jet definitions studied. UFO jets do not use the LC correction because PFOs are reconstructed using topoclusters at the EM scale, which results in a degraded correlation between the particle-level and detector-level \largeR jet \pt. While TCC jets take topoclusters calibrated to the LC scale as input, the energy resolution of TCC trimmed jets is worse than for topocluster trimmed jets, while the UFO trimmed jet resolution is almost identical to the resolution of PFlow trimmed jets. This indicates that the energy resolution degradation of TCC is due to the inclusion of pile-up tracks in the energy sharing, since these are not included in the UFO implementation.
 
\begin{figure}[htbp]
\begin{center}
\subfigure[]{\includegraphics[width=0.48\textwidth]{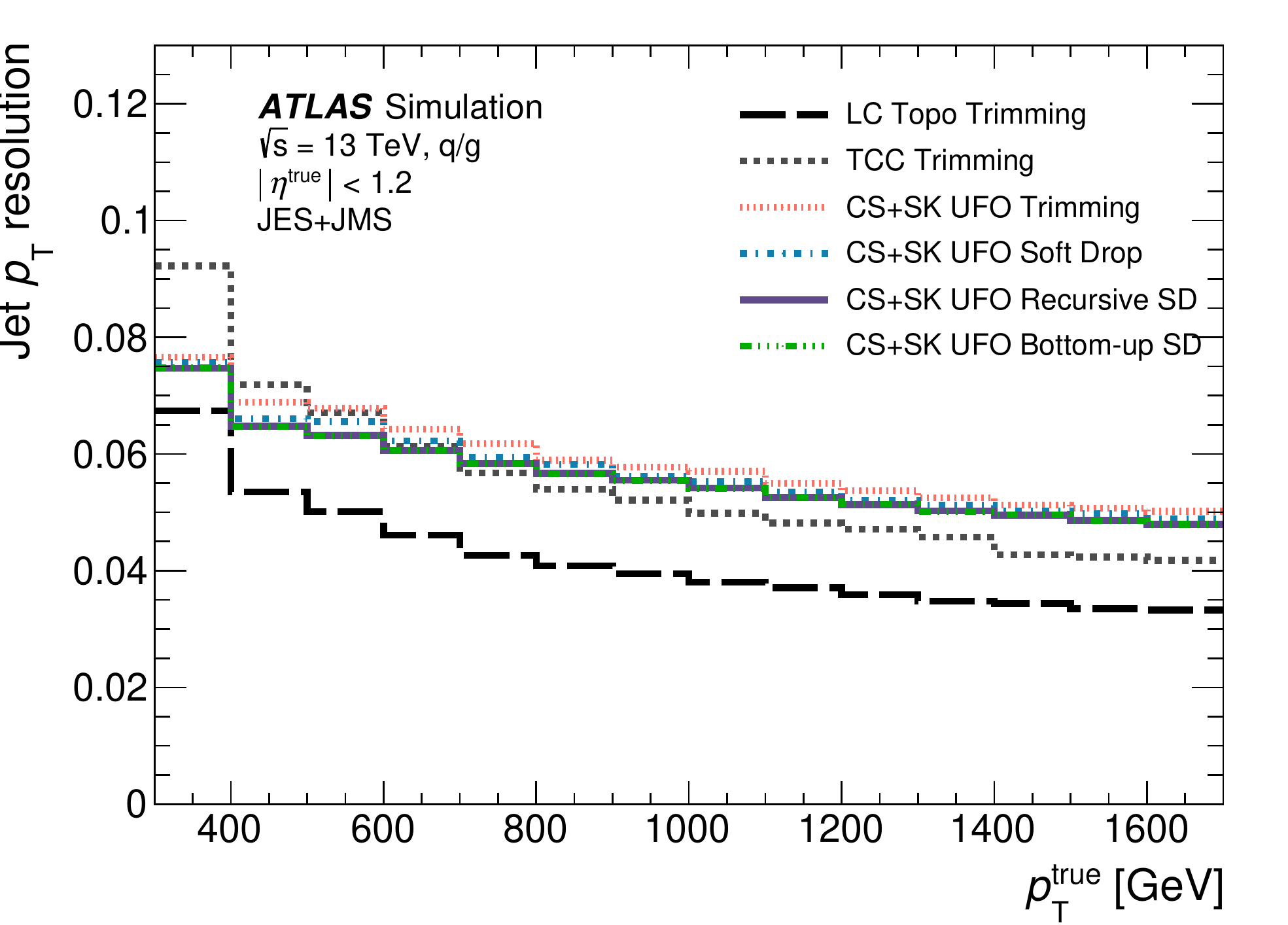}\label{fig:calib:w:ptResponse:shortlist}}
\subfigure[]{\includegraphics[width=0.48\textwidth]{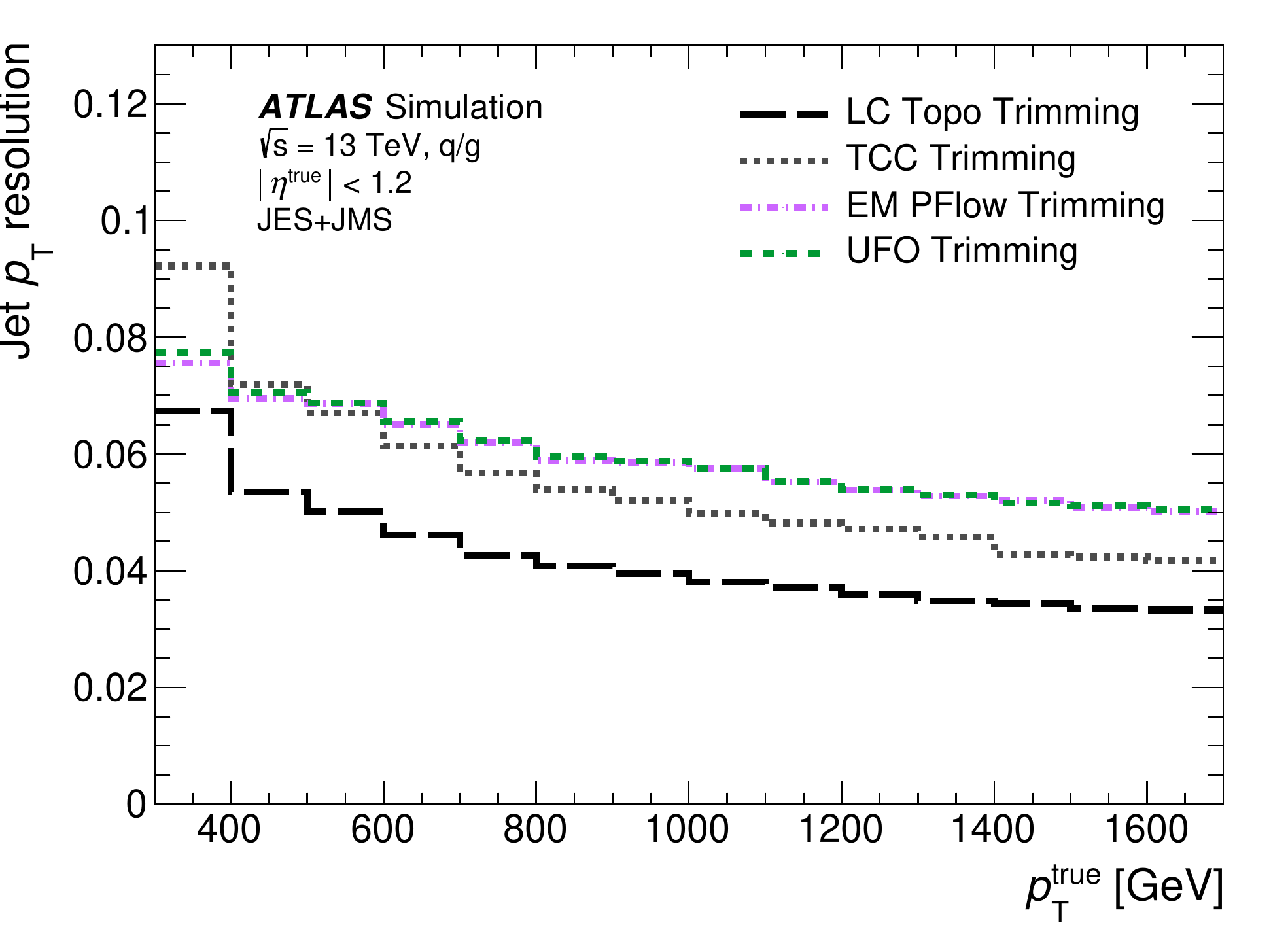}\label{fig:calib:w:ptResponse:trimmed}}\\
\subfigure[]{\includegraphics[width=0.48\textwidth]{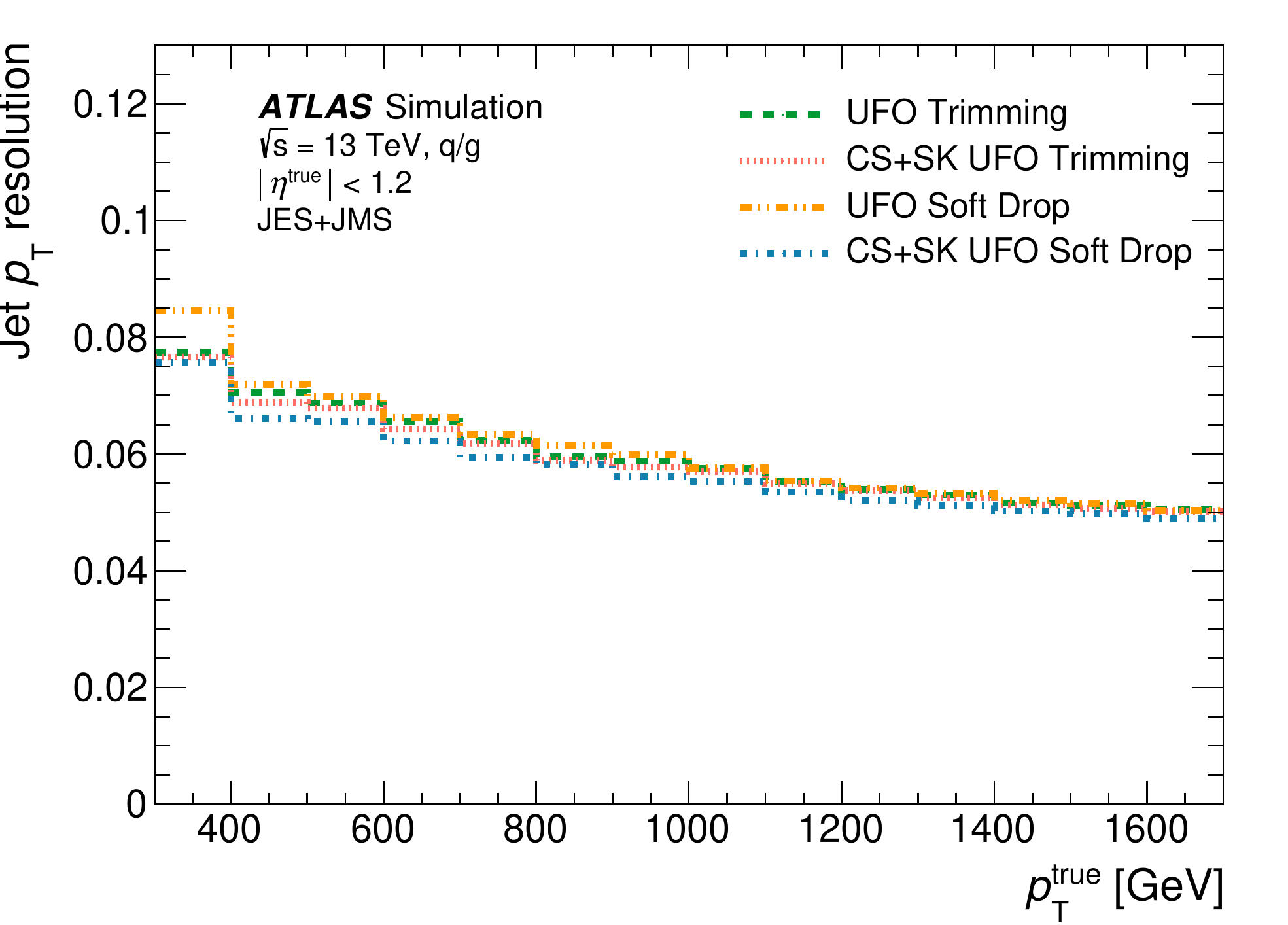}\label{fig:calib:w:ptResponse:cssk}}
\end{center}
\caption{The jet \pt resolution in dijet events.  In (a) the relative performance of the studied UFO definitions is compared with the current ATLAS baseline topocluster and TCC jets, while in (b) only jet input object types are compared, and in (c) the impact of pile-up mitigation is highlighted.}
\label{fig:calib:qcd:ptResolution}
\end{figure}
 
\subsubsection{Jet mass+JSS tagging performance}
 
In this section, a comparison of the tagging performance of the calibrated jet definitions is reported. Instead of considering a single efficiency working point (Section~\ref{sec:nocalib:tagging}), the tagging performance is studied using ROC curves. Figures~\ref{fig:calib:w:2var} and~\ref{fig:calib:top:2var} show the tagger background rejection as a function of the tagger signal efficiency, using the same jet mass $+$ jet substructure taggers discussed in Section~\ref{sec:nocalib:tagging}: a fixed mass-window cut with 68\% signal efficiency is applied, and then a one-sided \DTwo or \tauthrtwo cut is made to obtain the desired signal efficiency.
 
When tagging high-\pt, hadronically decaying $W$ bosons (Figure~\ref{fig:calib:w:2var}), the considered UFO definitions bring significant improvement over the LCTopo and TCC definitions. At high \pt, UFOs outperform the baseline topocluster-based jet definition in terms of their background rejection by about 120\% at a fixed signal-tagging efficiency of 50\%. For high-\pt, hadronically decaying top quarks (Figure~\ref{fig:calib:top:2var}), UFO definitions outperform all other choices, improving the background rejection by 135\% when compared with the baseline topocluster-based jet definition at a fixed signal-tagging efficiency of 50\%. Use of the recursive or bottom-up soft-drop grooming algorithm is noted to further improve performance over the trimmed UFO definition by an additional 10\% for a signal efficiency of 50\%, and the application of CS+SK pile-up mitigation is also found to increase performance by roughly 10\% when it is applied in conjunction with the soft-drop grooming algorithm.
 
\begin{figure}[htbp]
\begin{center}
\subfigure[]{\includegraphics[width=0.47\textwidth]{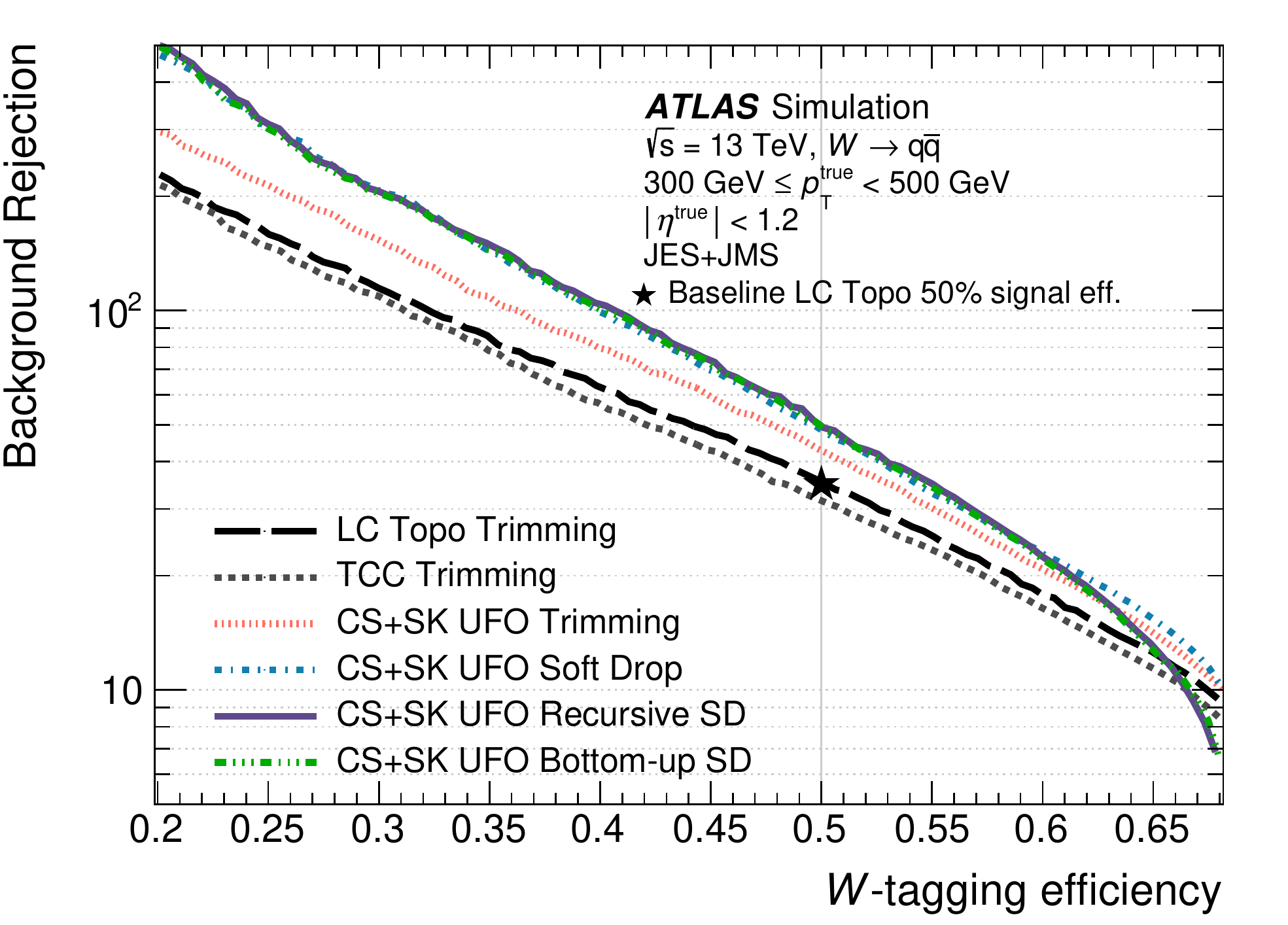}}
\subfigure[]{\includegraphics[width=0.47\textwidth]{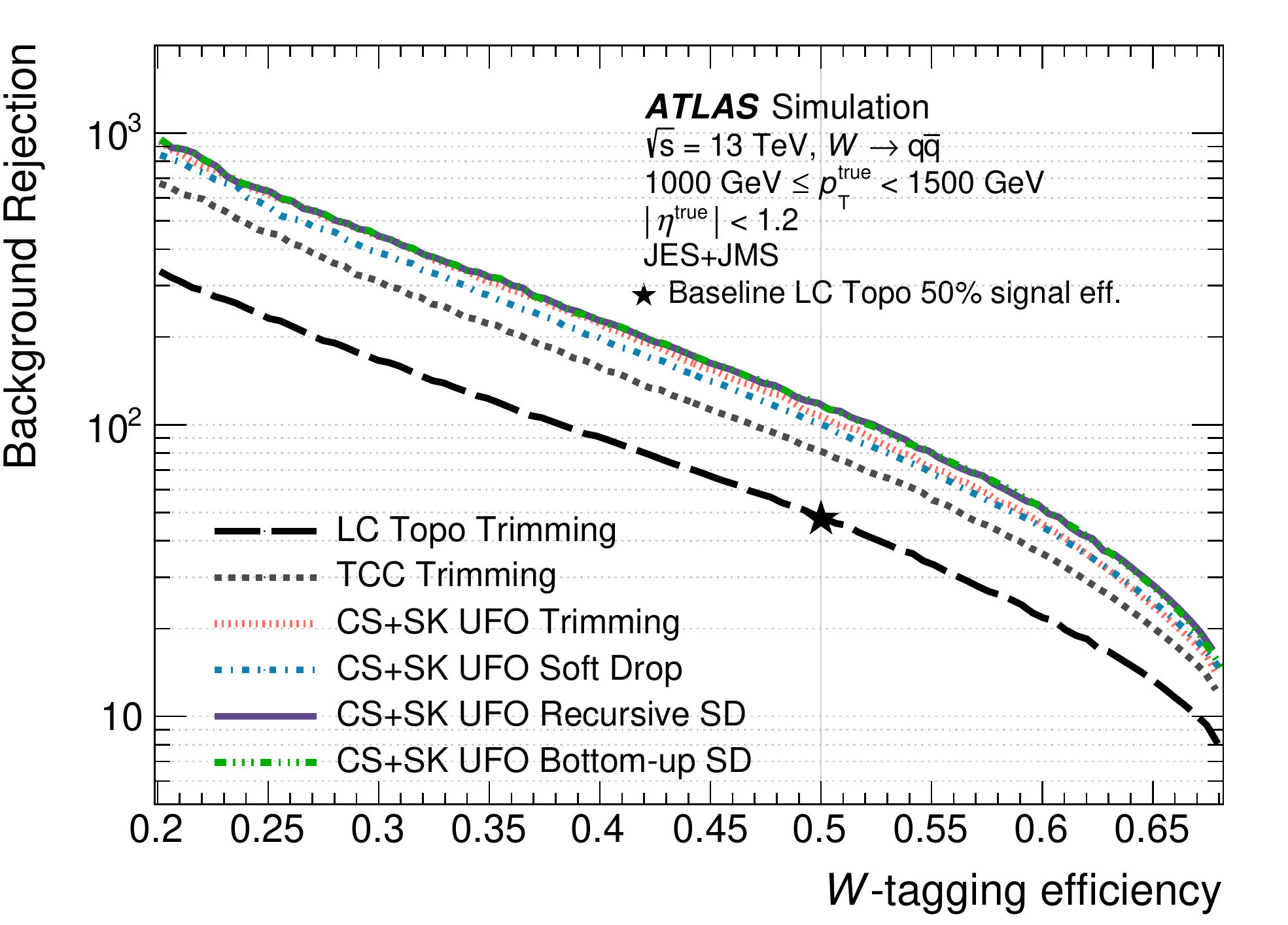}}\\
 
\subfigure[]{\includegraphics[width=0.47\textwidth]{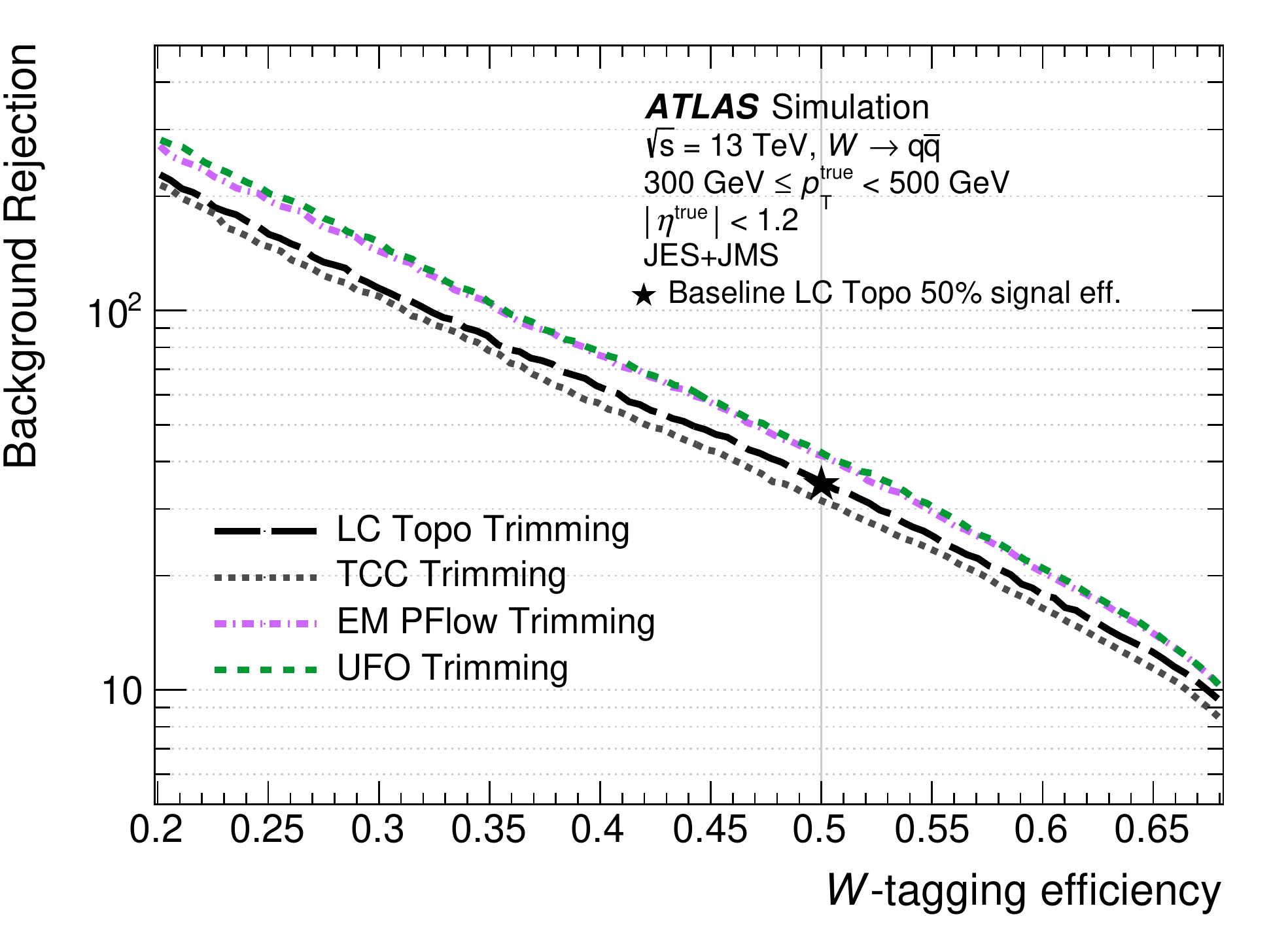}}
\subfigure[]{\includegraphics[width=0.47\textwidth]{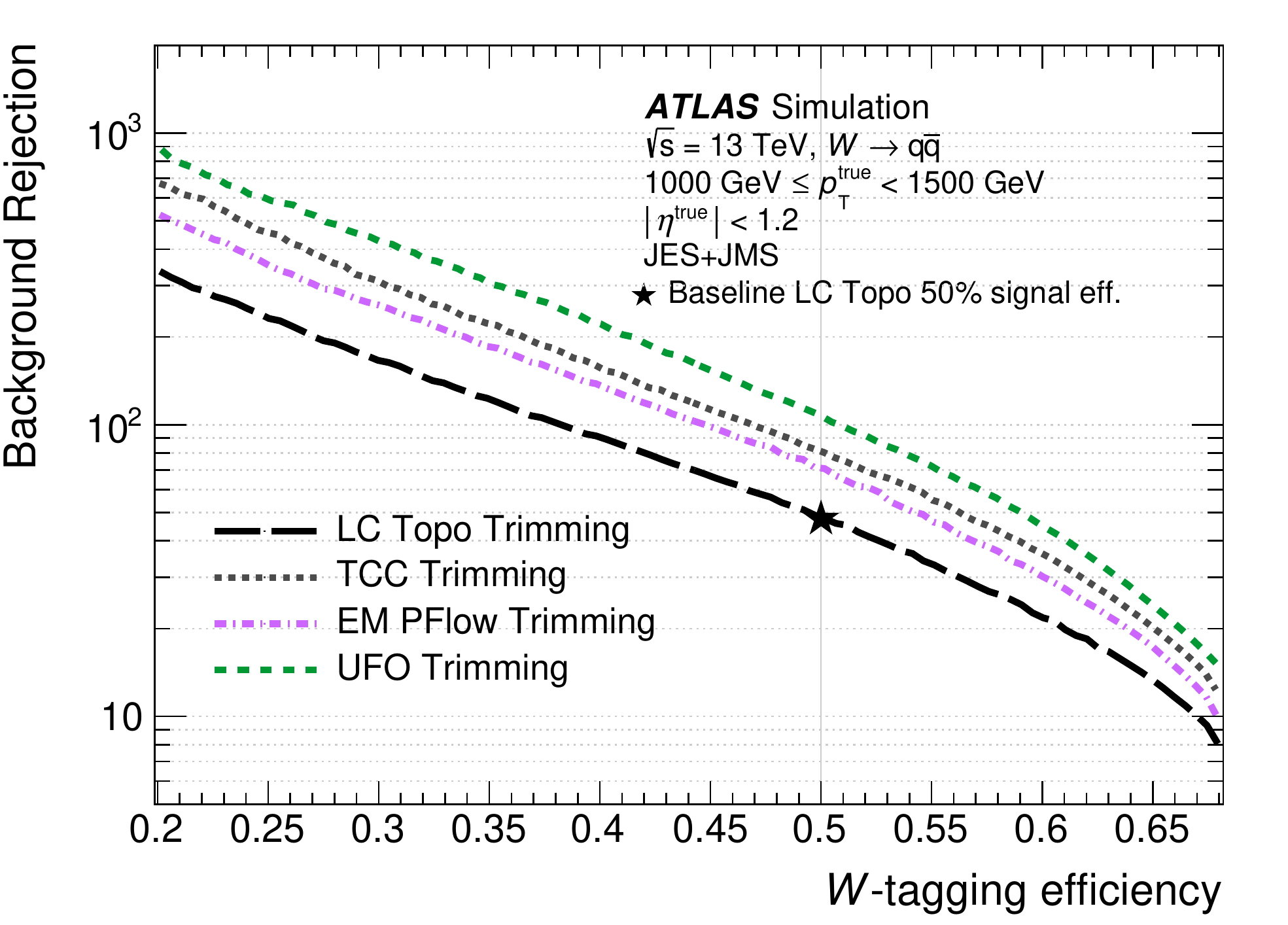}} \\
 
\subfigure[]{\includegraphics[width=0.47\textwidth]{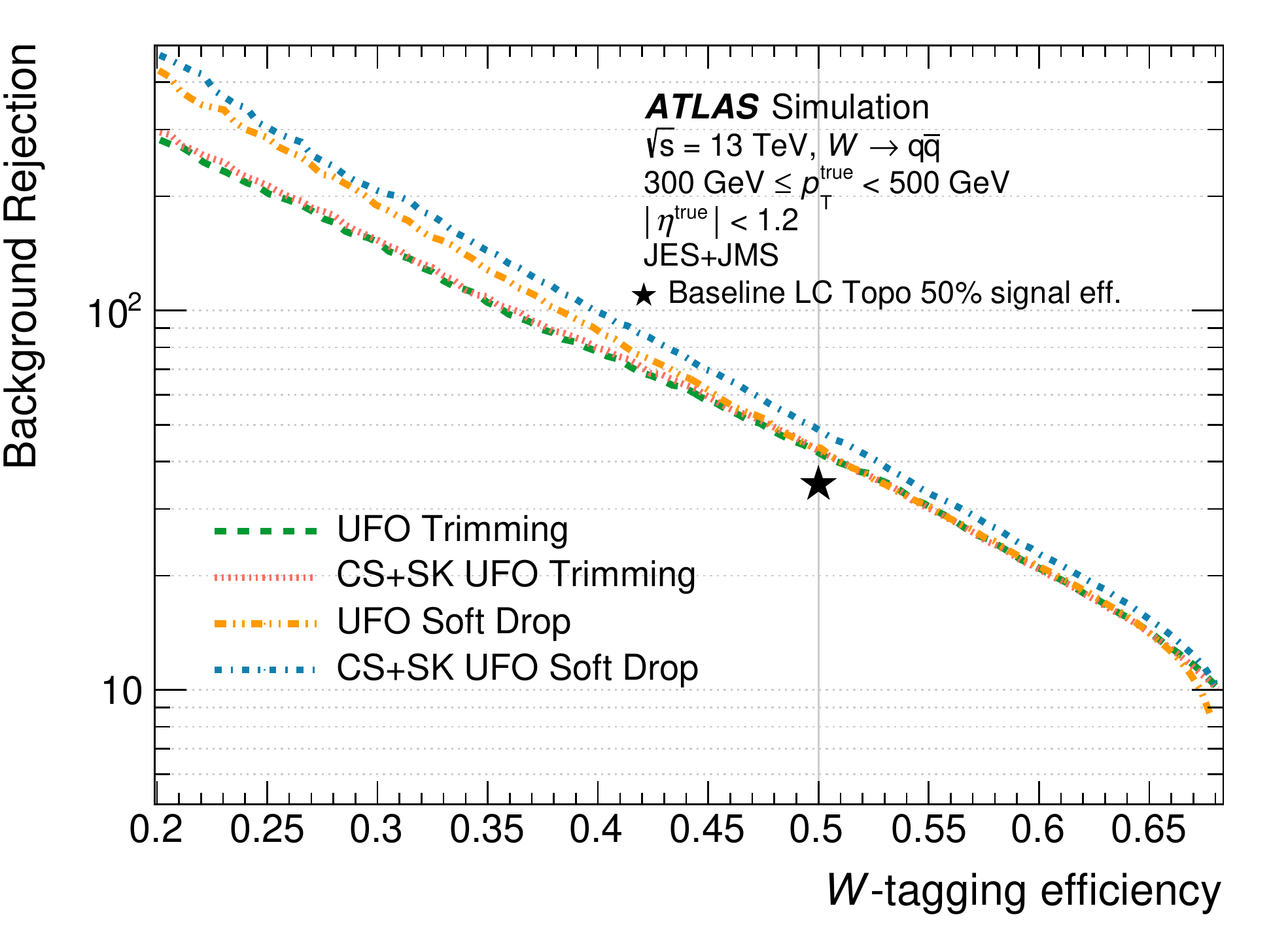}}
\subfigure[]{\includegraphics[width=0.47\textwidth]{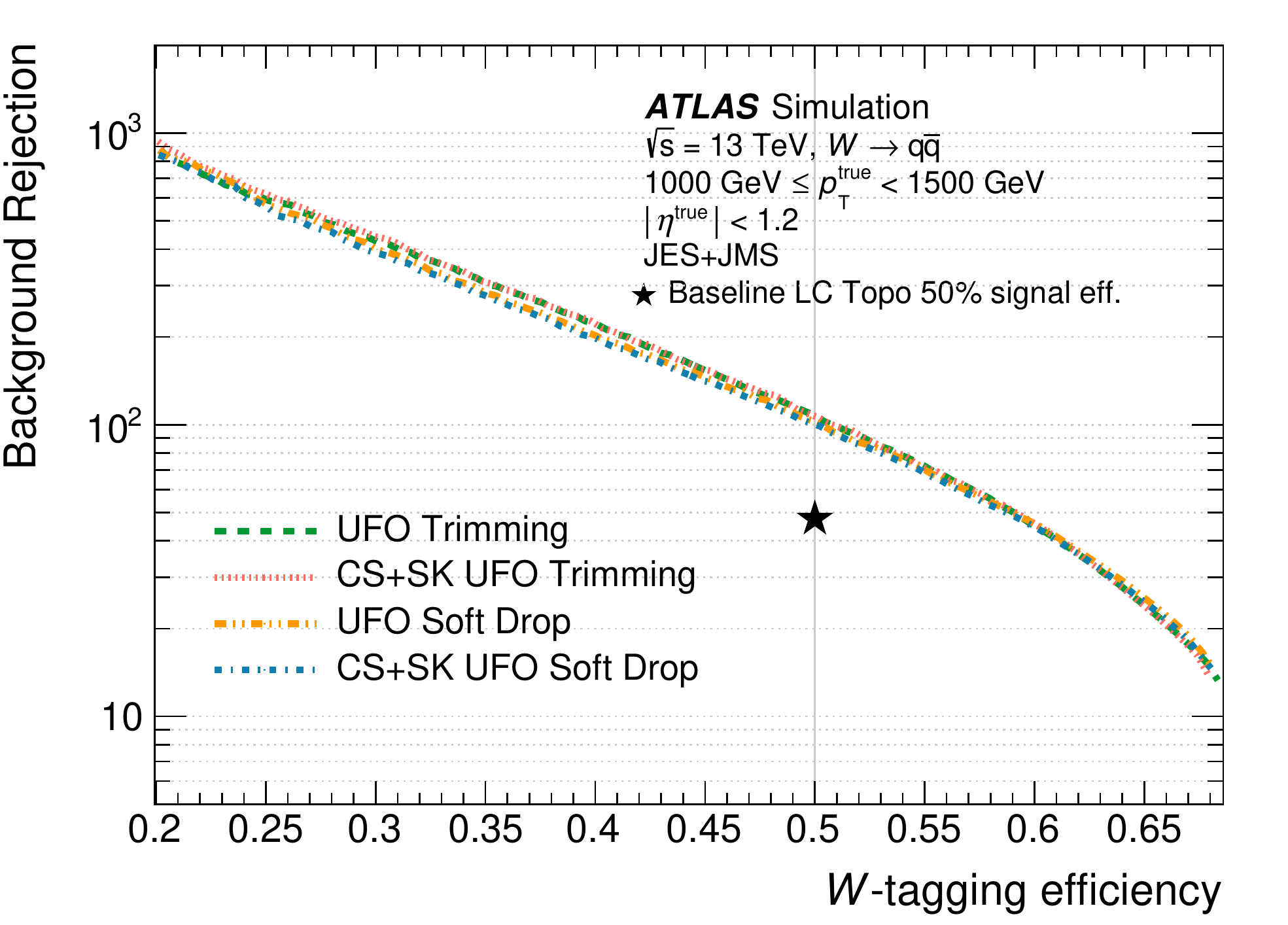}} \\
 
\end{center}
\caption{Background rejection as a function of signal efficiency for a tagger using the jet mass and \DTwo for $W$ boson jets at (left) low \pt, and (right) high \pt.  In (a,b) the relative performance of the studied UFO definitions is compared with the current ATLAS baseline topocluster and TCC jets, while in (c,d) only jet input object types are compared, and in (e,f) the impact of pile-up mitigation is highlighted. The background rejection factor of the baseline topocluster-based trimmed collection at a fixed signal tagging efficiency of 50\% is indicated with a $\star$.}
\label{fig:calib:w:2var}
\end{figure}
 
\begin{figure}[htbp]
\begin{center}
 
\subfigure[]{\includegraphics[width=0.47\textwidth]{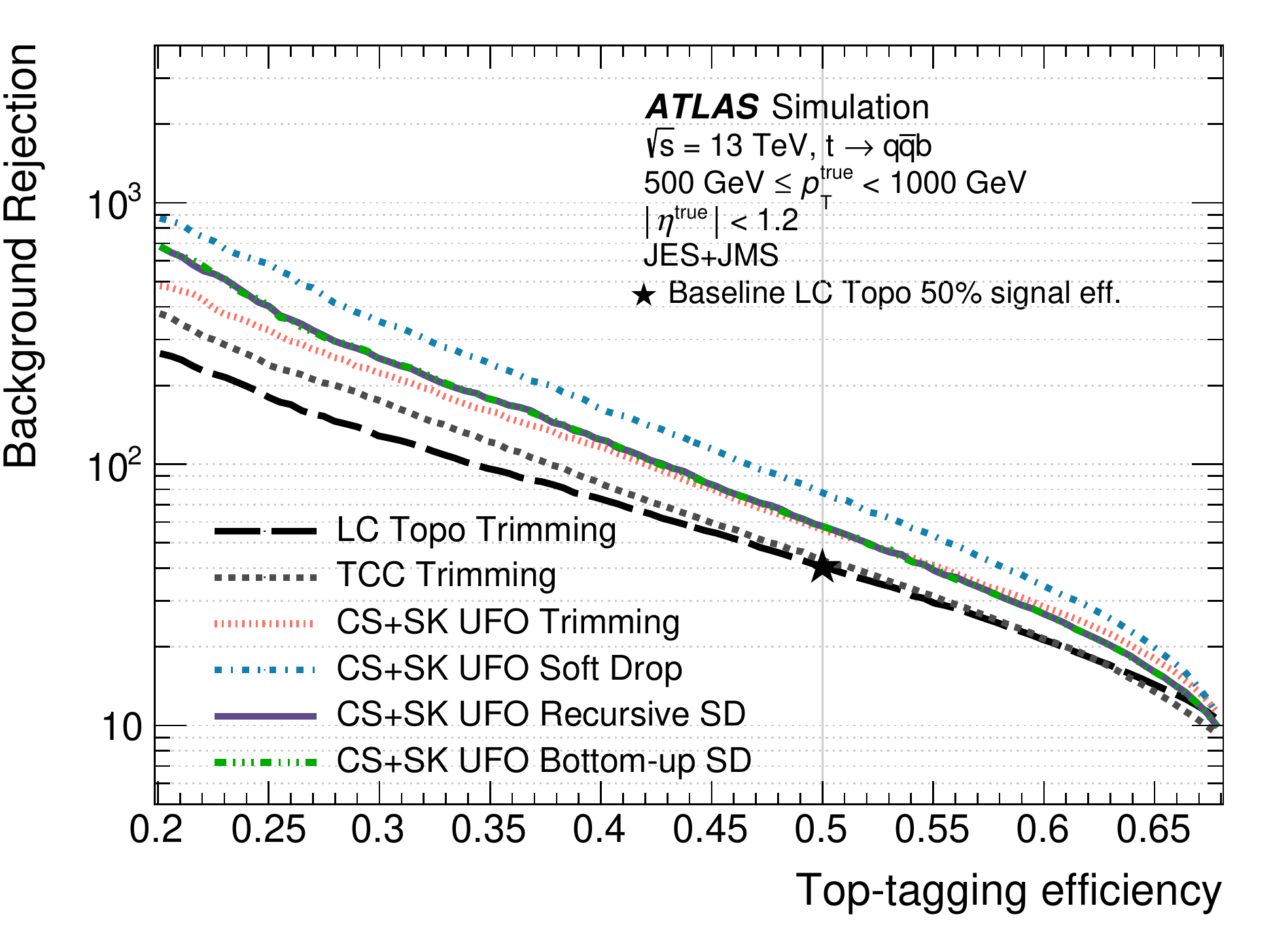}}
\subfigure[]{\includegraphics[width=0.47\textwidth]{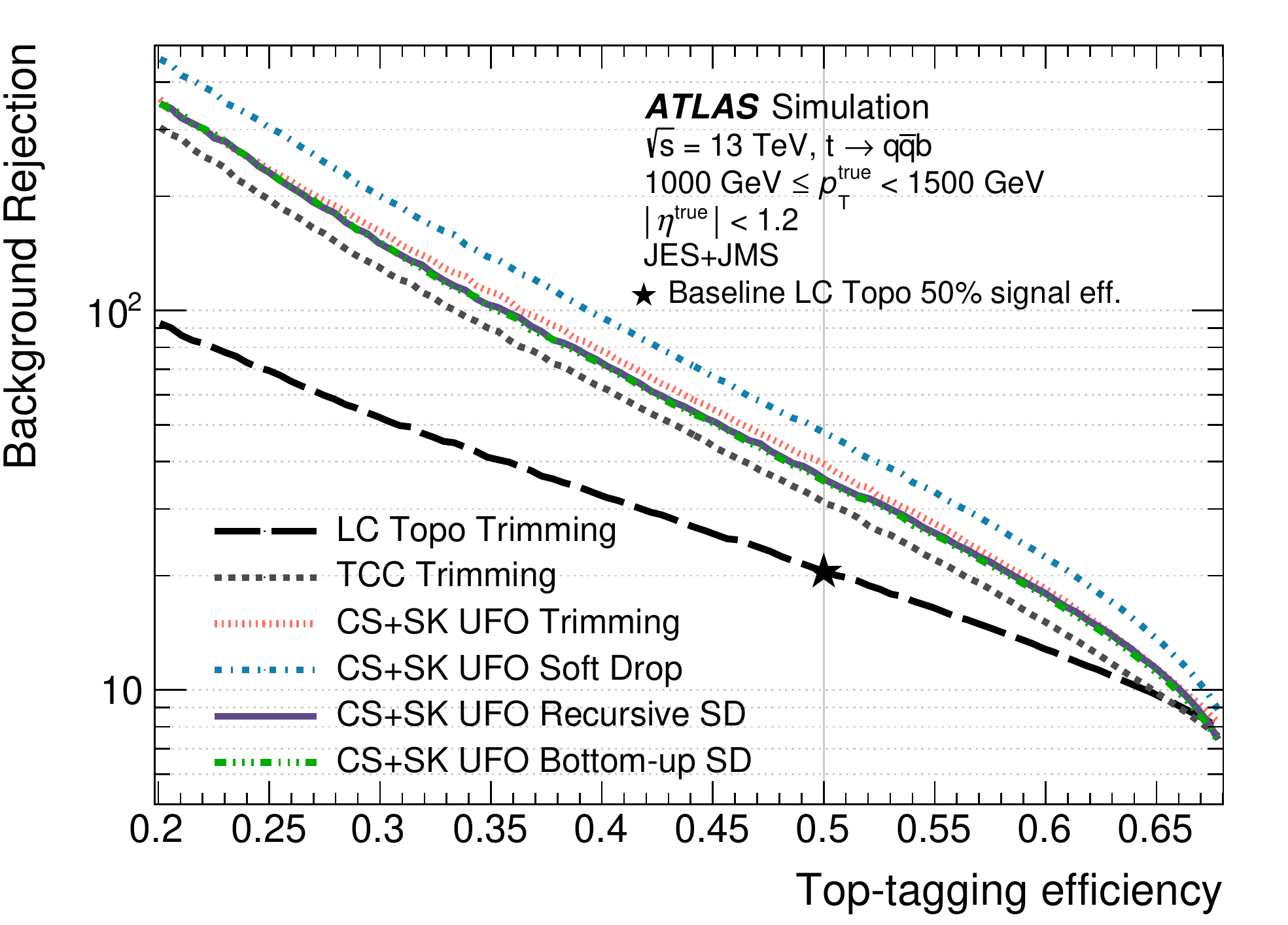}}
 
\subfigure[]{\includegraphics[width=0.47\textwidth]{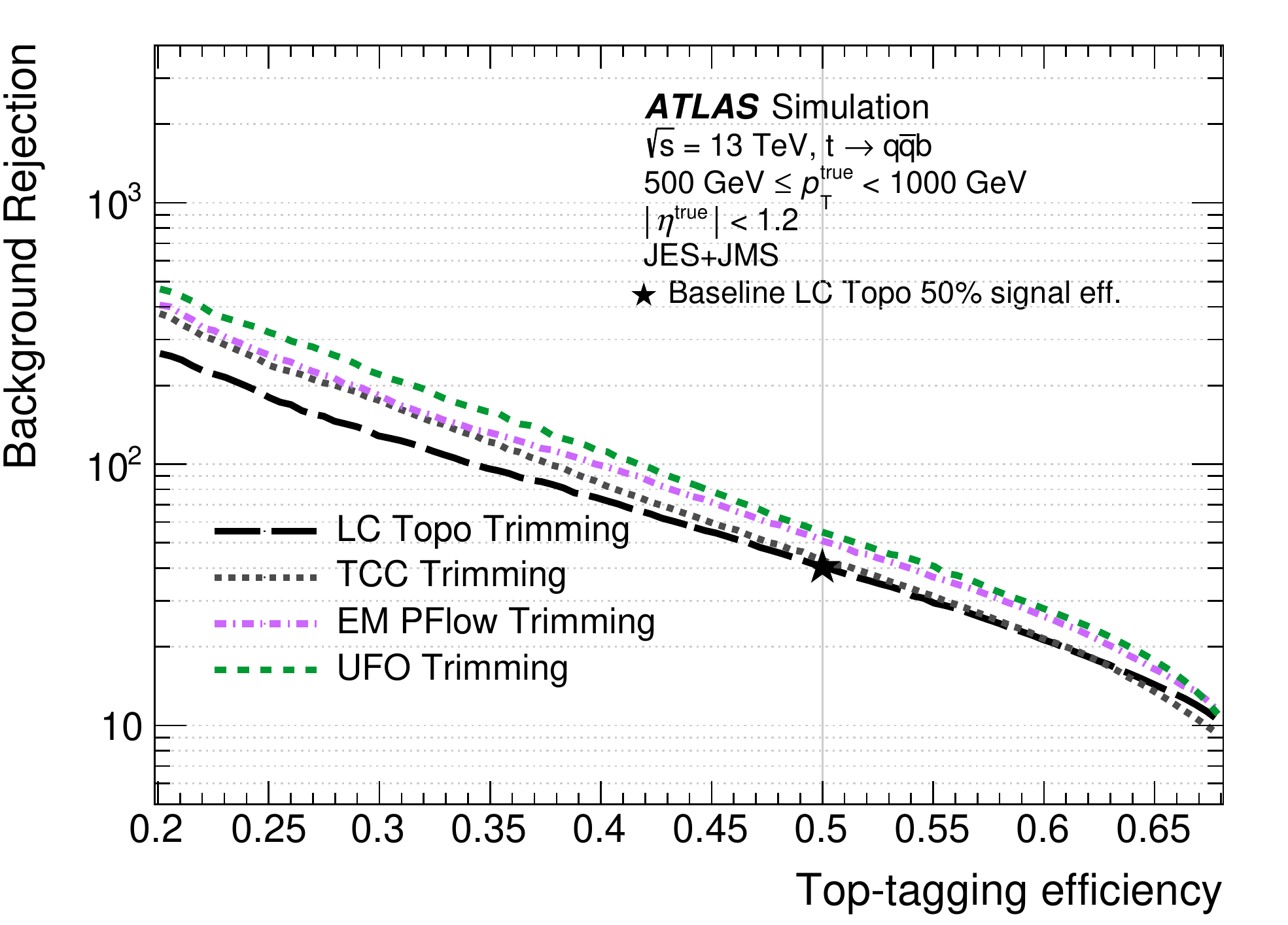}}
\subfigure[]{\includegraphics[width=0.47\textwidth]{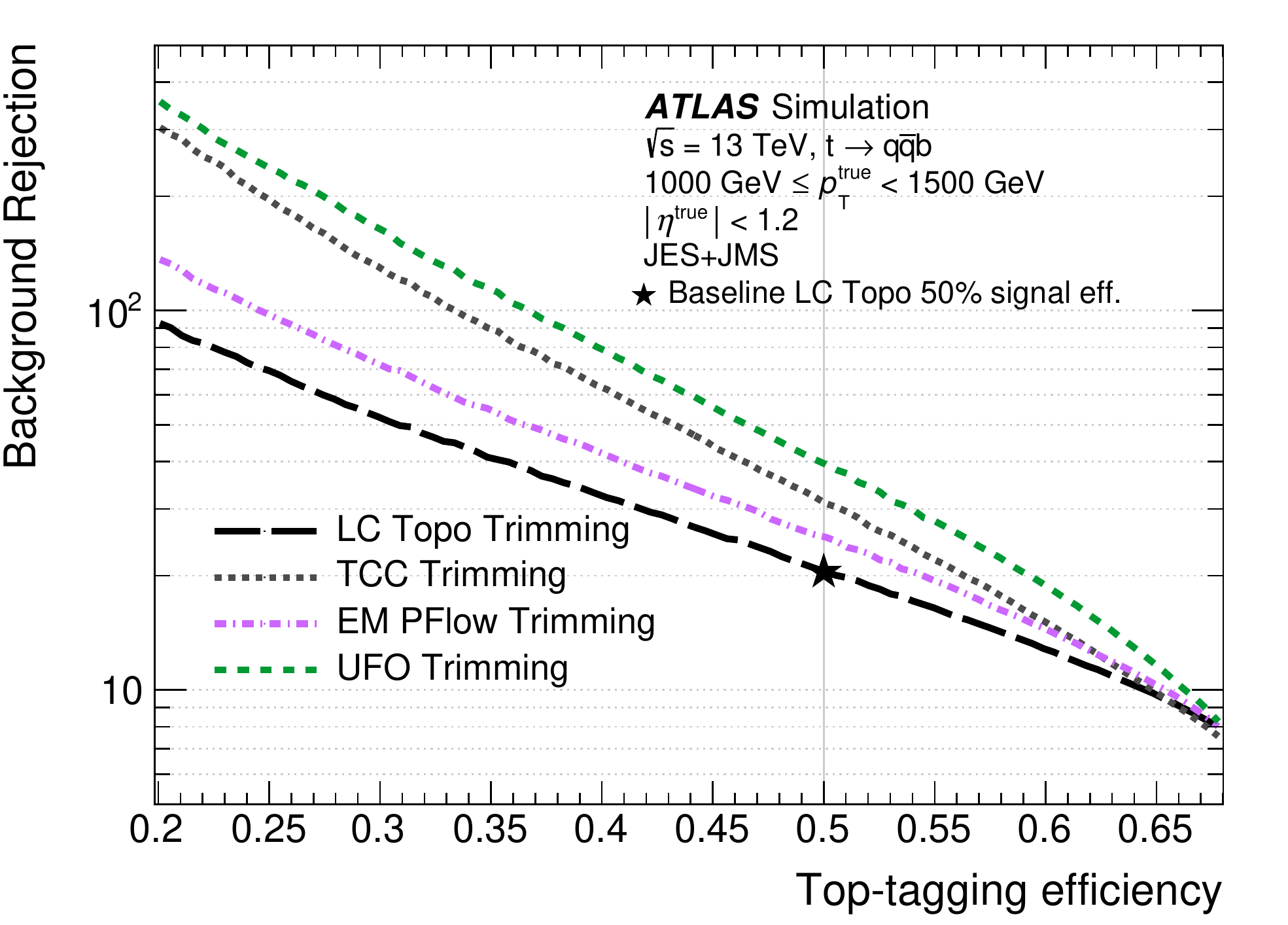}}\\
 
\subfigure[]{\includegraphics[width=0.47\textwidth]{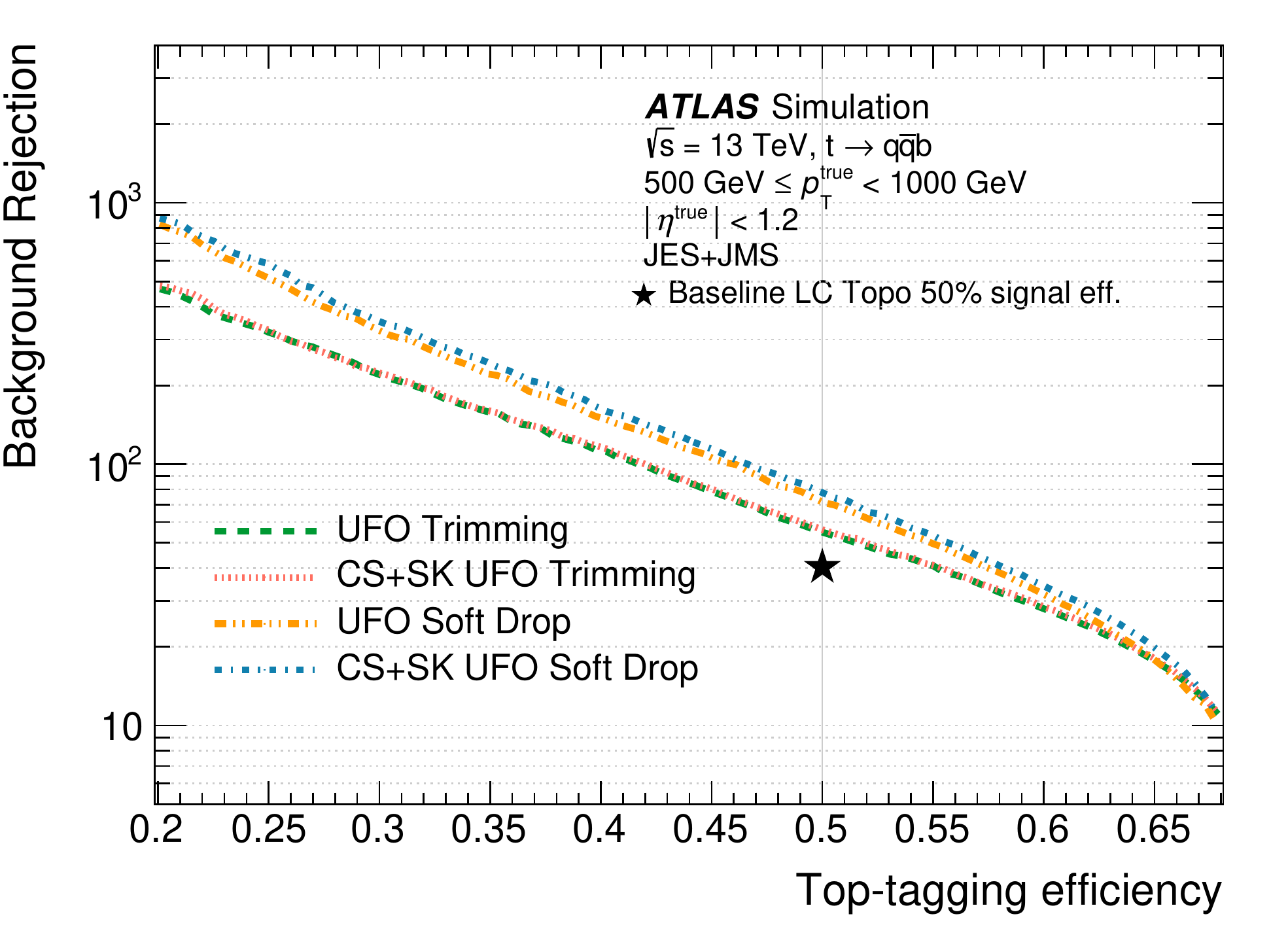}}
\subfigure[]{\includegraphics[width=0.47\textwidth]{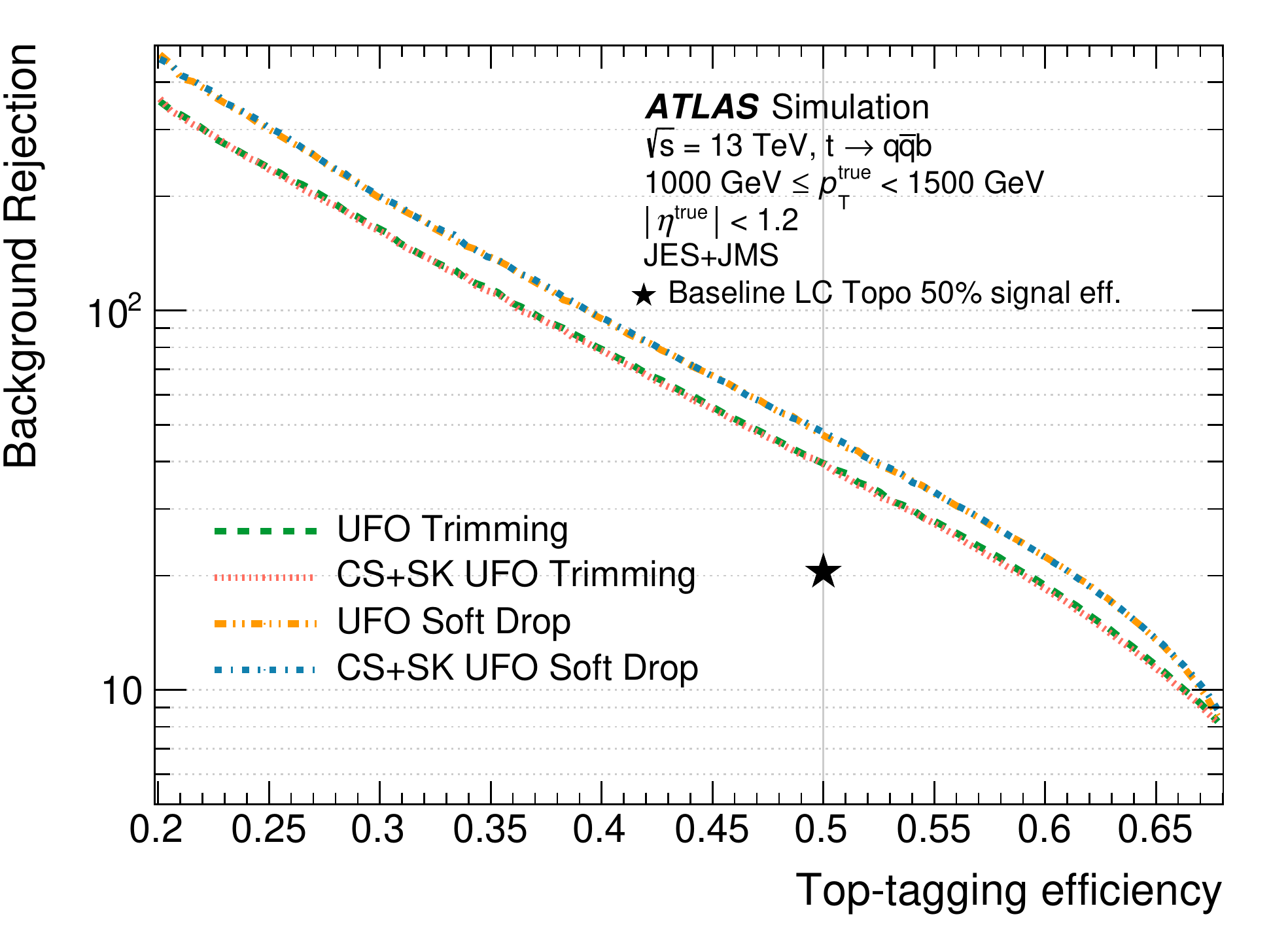}}\\
 
\end{center}
\caption{Background rejection as a function of signal efficiency for a tagger using the jet mass and \tauthrtwo for top quark jets at (left) low \pt, and (right) high \pt.  In (a,b) the relative performance of the studied UFO definitions is compared with the current ATLAS baseline topocluster and TCC jets, while in (c,d) only jet input object types are compared, and in (e,f) the impact of pile-up mitigation is highlighted.}
\label{fig:calib:top:2var}
\end{figure}
 
\subsection{Data-to-simulation comparisons}
 
Robust modelling of jet substructure is crucial to reduce uncertainties related to Monte Carlo modelling of parton showers in physics analyses that rely on jet-substructure-based techniques. To verify the accuracy of the simulation, predictions were generated at the detector level for several jet substructure observables in high-\pt dijet events using \PYTHIA and reconstructed using the full ATLAS detector simulation~\cite{SOFT-2010-01} based on $\GEANT4$~\cite{Agostinelli:2002hh}. The results are compared with the distributions observed in data collected during 2017. Events are selected using the lowest unprescaled single \largeR jet trigger. This trigger is fully efficient for ungroomed \largeR jets with \pt > 600~\GeV. Data are required to pass a series of quality requirements and cleaning cuts. In addition, overlap removal and pile-up reweighting are applied. Events are required to have at least one jet with a groomed jet \pt above 600~\GeV, and all jets are required to have $\pt > 600$~\GeV\ and $|\eta| < 1.2$. When studying the behaviour of $\tau_{32}$ and $D_{2}$, the jet mass is required to be greater than 40~\GeV. Data and simulated events are required to pass the same event selection.
 
The observed data are compared with simulated dijet events in Figure~\ref{fig:dataStudies}. The jet mass, number of jet constituents, \DTwo, and \tauthrtwo are studied. Only statistical uncertainties are displayed, and the statistical uncertainty of the simulation is negligible compared to that of the data. In general, the level of agreement between data and simulation for the UFO jets is similar to that of topocluster trimmed jets, indicating that this level of agreement is tolerable for general use on ATLAS. The exception to this is the number of constituents, which is known to be modelled poorly~\cite{PERF-2014-07}. The modelling is improved for UFO jets relative to topocluster-based trimmed jets, particularly at large constituent multiplicities. 
 
\begin{figure}[htbp]
\centering
\subfigure[]{\includegraphics[width=0.47\textwidth]{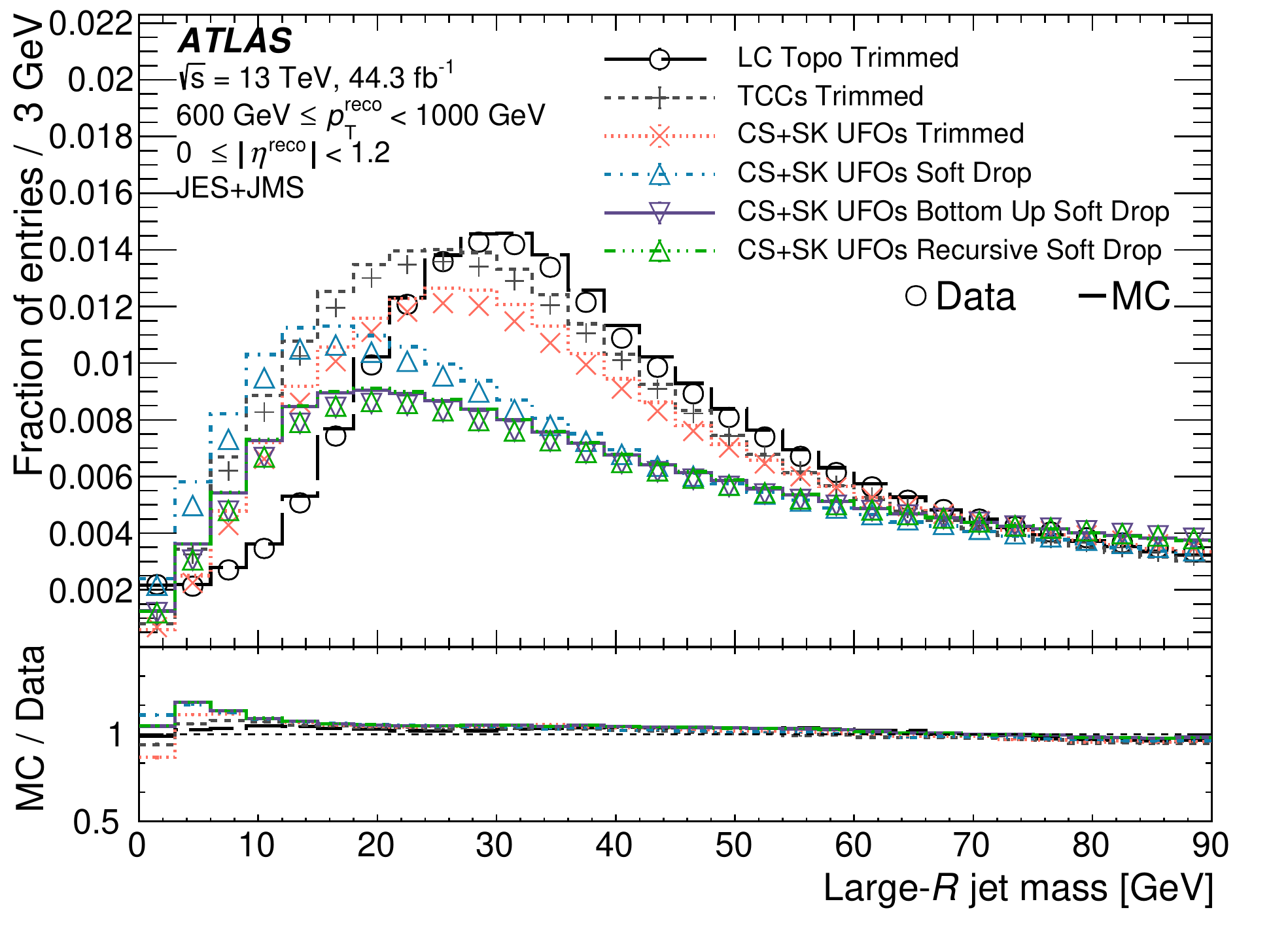}}
\subfigure[]{\includegraphics[width=0.47\textwidth]{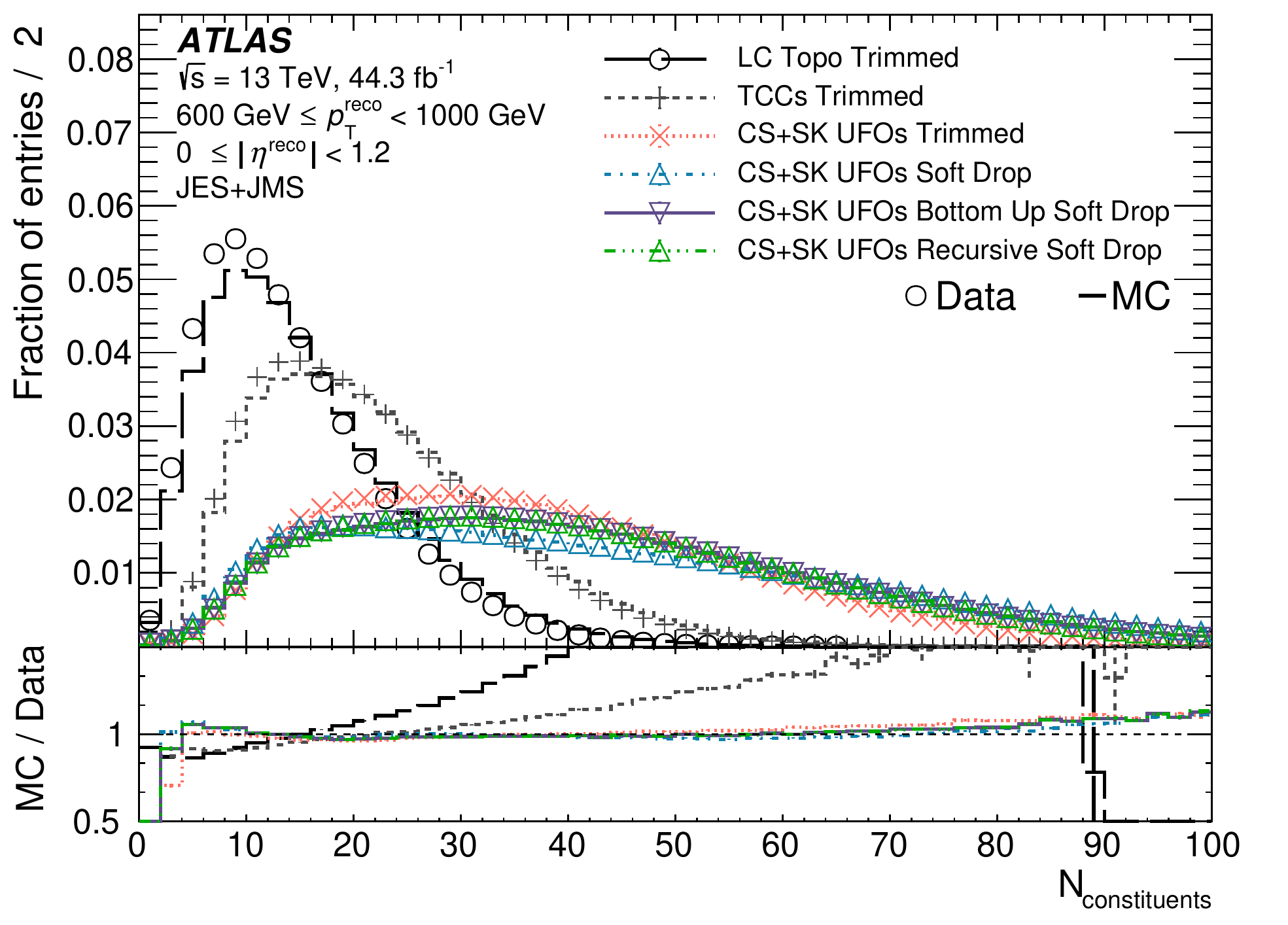}}\\
\subfigure[]{\includegraphics[width=0.47\textwidth]{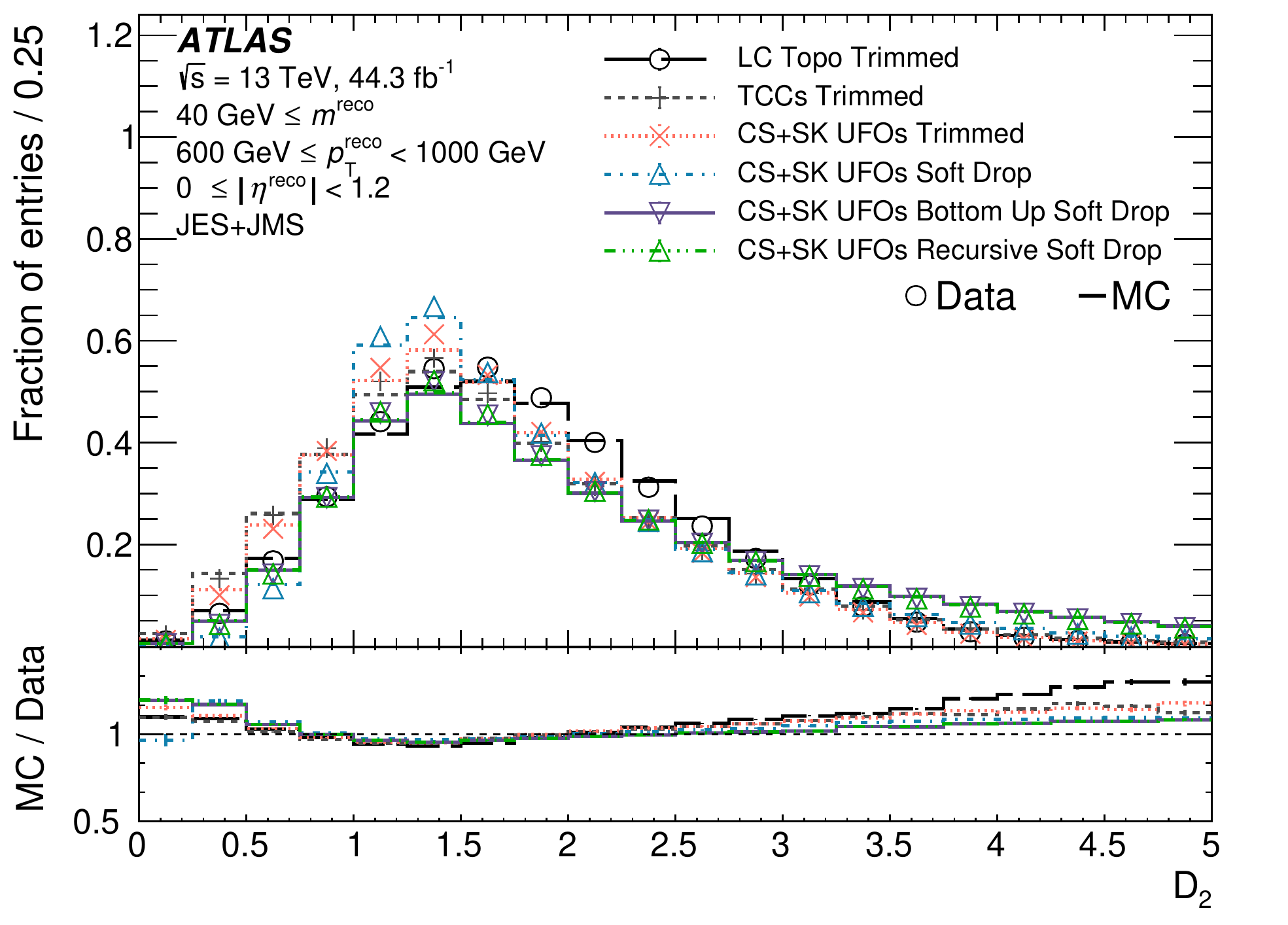}}
\subfigure[]{\includegraphics[width=0.47\textwidth]{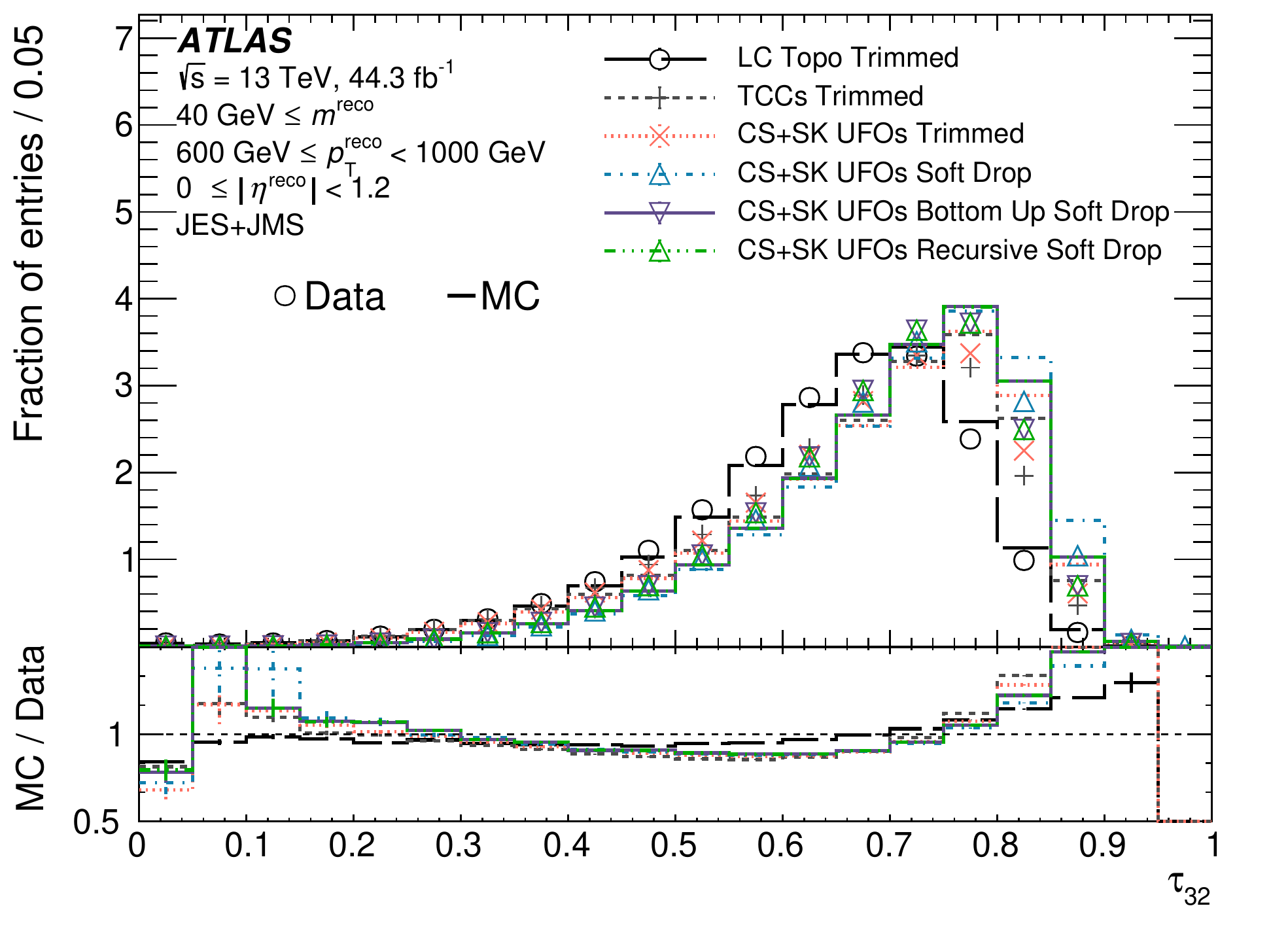}}
\caption{Data-to-simulation comparisons of (a) the groomed jet mass, (b) the number of constituents, (c) the groomed jet $D_{2}$ and (d) the groomed jet $\tau_{32}$. Only statistical uncertainties are displayed, and the statistical uncertainty of the simulation is negligible compared to that of the data. The ratio of simulation to data is provided in the lower panel of each figure.
}\label{fig:dataStudies}
\end{figure}

The background rejection for the mass+JSS taggers described in Section~\ref{sec:results} is shown in Figure~\ref{fig:dataMC:backgroundRejection} as a function of the \largeR jet \pt, where taggers are created for each \pt bin, using the 50\% signal efficiency working point. For the $W$ tagger, agreement between data and simulation is similar for all jet definitions, while for the top taggers, agreement is slightly worse for UFO jets than for the topocluster trimmed definition.
 
\begin{figure}[htbp]
\centering
\subfigure[]{\includegraphics[width=0.47\textwidth]{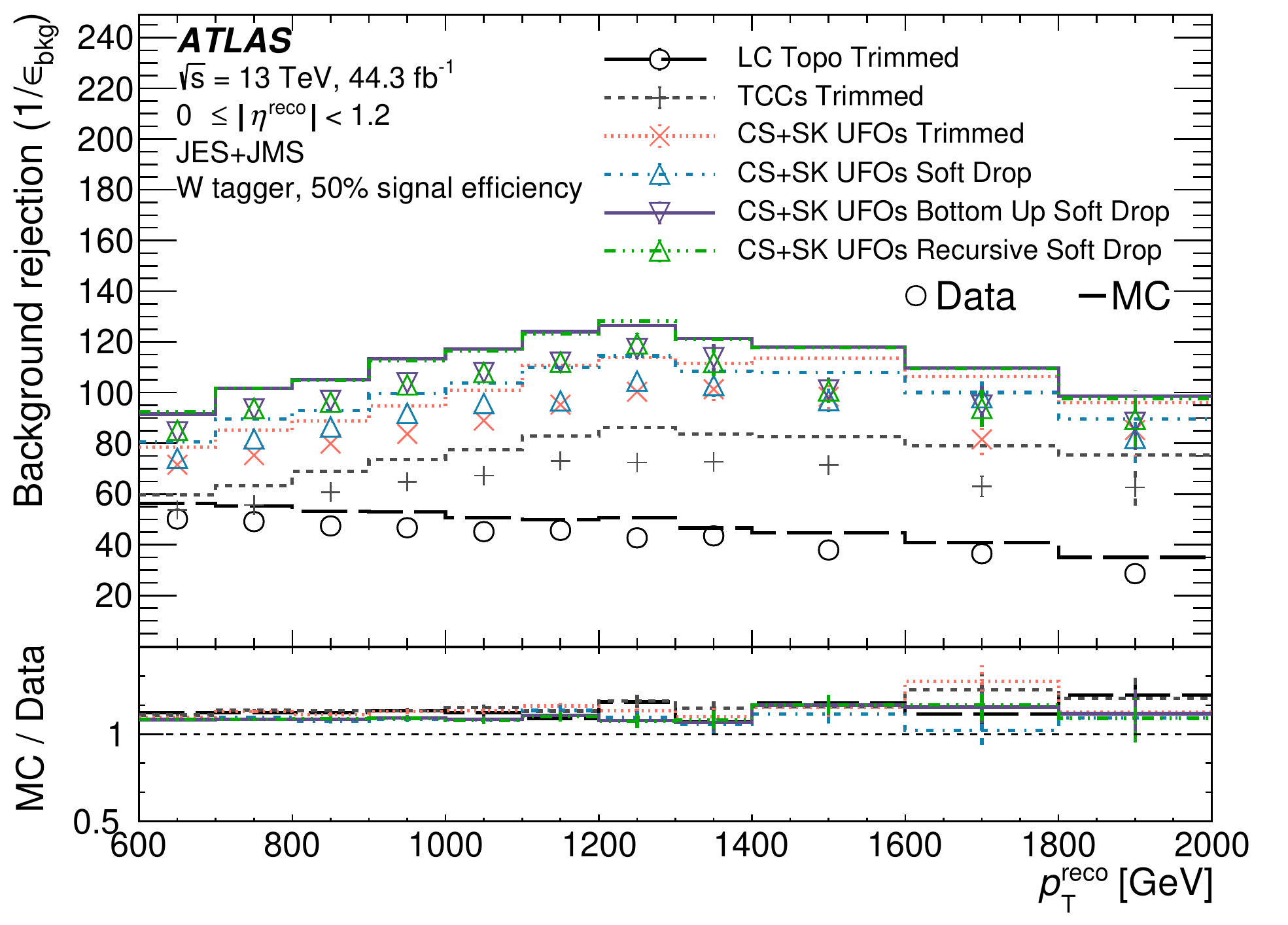}}
\subfigure[]{\includegraphics[width=0.47\textwidth]{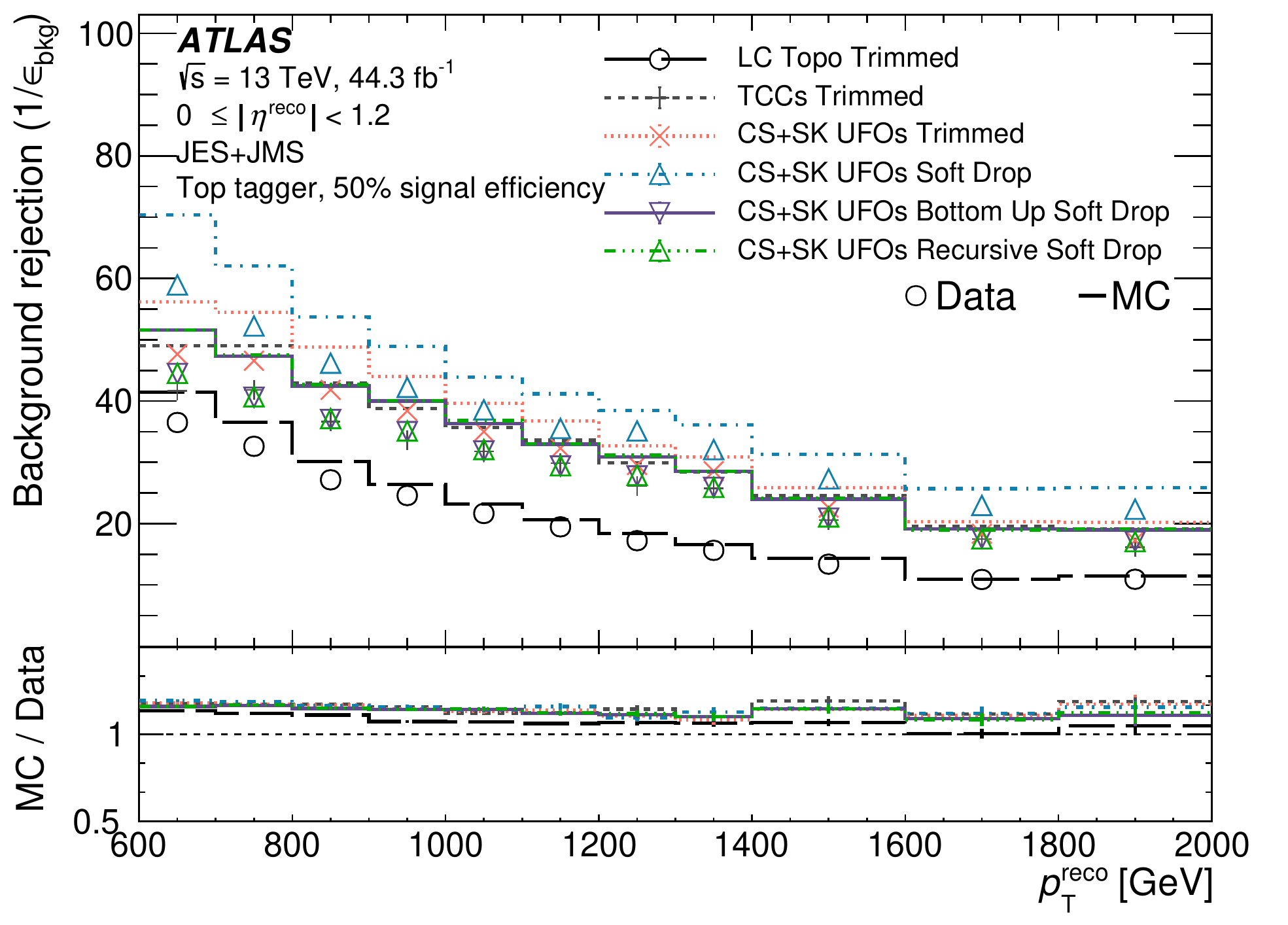}}
\caption{Data-to-simulation comparisons of the background rejection for groomed jets for (a) the mass+\DTwo $W$ tagger, and (b) the mass+\tauthrtwo top tagger.}
\label{fig:dataMC:backgroundRejection}
\end{figure}
 
\FloatBarrier
\FloatBarrier
\section{Concluding remarks}\label{sec:conclusion}
The development of jet substructure techniques has enabled new searches and measurements, boosting the sensitivity of the Large Hadron Collider experiments to the physics of and beyond the Standard Model. This paper has presented a set of performance comparisons in order to determine the most promising \largeR jet definitions for use in future analyses, with a focus on comparing different jet input objects, pile-up mitigation algorithms and jet grooming algorithms.
 
A new type of jet input, called a Unified Flow Object, has been proposed which incorporates tracking information into jet substructure reconstruction by combining particle-flow reconstruction for low-\pt particles and cluster splitting for particles at high \pt and in dense environments. These UFO inputs can increase the background rejection of jet taggers across a wide kinematic range by up to 120\% for a simple $W$ tagger at 50\% signal efficiency, and up to 135\% for a simple top tagger at 50\% signal efficiency when compared with the current baseline trimmed topocluster \largeR jet definition. While the \pt resolution of these jets is degraded relative to the baseline LCW topocluster-based ATLAS \largeR jet definition due to the different topocluster energy scales used as input objects, UFO jets provide an improved jet mass resolution, with up to a 45\% improvement at high \pt for signal jets when compared with existing ATLAS \largeR jet definitions.
 
The application of CS+SK pile-up mitigation has been shown to stabilise and augment performance as a function of the number of pile-up interactions, which will be crucial in the face of the difficult experimental conditions to come during future LHC data-taking periods. Pile-up mitigation increases the number of experimentally viable grooming configurations to include options which do not groom soft radiation aggressively enough to be considered with unmodified jet inputs.
 
Several promising grooming algorithms were compared using \largeR CS+SK UFO jets. Definitions incorporating soft-drop grooming and its extensions, recursive soft-drop and bottom-up soft-drop, all outperform the baseline ATLAS trimming configuration in terms of high-\pt $W$ and top quark tagging using simple taggers. These collections are viable for general-purpose use in the challenging experimental conditions of the LHC only due to the improvements in jet inputs and pile-up mitigation algorithms. The soft-drop definition using $\zcut=0.1$ and angular exponent $\beta=1.0$ outperforms all other candidates when identifying high-\pt top quarks, and is competitive to within 5--10\% of the considered RSD and BUSD options when tagging boosted $W$ bosons. These jets also exhibit good pile-up stability and a tolerable sensitivity to topological effects, according to the metrics studied. This definition provides superior jet mass resolution for low-\pt $W$ jets when compared with RSD and BUSD options. Due to its wide range of applicability, it is concluded that the CS+SK UFO soft-drop ($\beta=1.0$, $\zcut=0.1$) \largeR jet definition provides the best performance for use as a general-purpose jet definition in ATLAS physics analyses.

\clearpage

\FloatBarrier

\section*{Acknowledgements}
 

We thank CERN for the very successful operation of the LHC, as well as the
support staff from our institutions without whom ATLAS could not be
operated efficiently.
 
We acknowledge the support of ANPCyT, Argentina; YerPhI, Armenia; ARC, Australia; BMWFW and FWF, Austria; ANAS, Azerbaijan; SSTC, Belarus; CNPq and FAPESP, Brazil; NSERC, NRC and CFI, Canada; CERN; ANID, Chile; CAS, MOST and NSFC, China; COLCIENCIAS, Colombia; MSMT CR, MPO CR and VSC CR, Czech Republic; DNRF and DNSRC, Denmark; IN2P3-CNRS and CEA-DRF/IRFU, France; SRNSFG, Georgia; BMBF, HGF and MPG, Germany; GSRT, Greece; RGC and Hong Kong SAR, China; ISF and Benoziyo Center, Israel; INFN, Italy; MEXT and JSPS, Japan; CNRST, Morocco; NWO, Netherlands; RCN, Norway; MNiSW and NCN, Poland; FCT, Portugal; MNE/IFA, Romania; JINR; MES of Russia and NRC KI, Russian Federation; MESTD, Serbia; MSSR, Slovakia; ARRS and MIZ\v{S}, Slovenia; DST/NRF, South Africa; MICINN, Spain; SRC and Wallenberg Foundation, Sweden; SERI, SNSF and Cantons of Bern and Geneva, Switzerland; MOST, Taiwan; TAEK, Turkey; STFC, United Kingdom; DOE and NSF, United States of America. In addition, individual groups and members have received support from BCKDF, CANARIE, Compute Canada, CRC and IVADO, Canada; Beijing Municipal Science \& Technology Commission, China; COST, ERC, ERDF, Horizon 2020 and Marie Sk{\l}odowska-Curie Actions, European Union; Investissements d'Avenir Labex, Investissements d'Avenir Idex and ANR, France; DFG and AvH Foundation, Germany; Herakleitos, Thales and Aristeia programmes co-financed by EU-ESF and the Greek NSRF, Greece; BSF-NSF and GIF, Israel; La Caixa Banking Foundation, CERCA Programme Generalitat de Catalunya and PROMETEO and GenT Programmes Generalitat Valenciana, Spain; G\"{o}ran Gustafssons Stiftelse, Sweden; The Royal Society and Leverhulme Trust, United Kingdom.
 
The crucial computing support from all WLCG partners is acknowledged gratefully, in particular from CERN, the ATLAS Tier-1 facilities at TRIUMF (Canada), NDGF (Denmark, Norway, Sweden), CC-IN2P3 (France), KIT/GridKA (Germany), INFN-CNAF (Italy), NL-T1 (Netherlands), PIC (Spain), ASGC (Taiwan), RAL (UK) and BNL (USA), the Tier-2 facilities worldwide and large non-WLCG resource providers. Major contributors of computing resources are listed in Ref.~\cite{ATL-SOFT-PUB-2020-001}.
 

\printbibliography

@article{Cacciari:2010te,
    author = "Cacciari, Matteo and Rojo, Juan and Salam, Gavin P. and Soyez, Gregory",
    title = "{Jet reconstruction in heavy ion collisions}",
    eprint = "1010.1759",
    archivePrefix = "arXiv",
    primaryClass = "hep-ph",
    reportNumber = "CERN-PH-TH-2010-223",
    doi = "10.1140/epjc/s10052-011-1539-z",
    journal = "Eur. Phys. J. C",
    volume = "71",
    pages = "1539",
    year = "2011"
}

@article{Soyez:2012hv,
    author = "Soyez, Gregory and Salam, Gavin P. and Kim, Jihun and Dutta, Souvik and Cacciari, Matteo",
    title = "{Pileup Subtraction for Jet Shapes}",
    eprint = "1211.2811",
    archivePrefix = "arXiv",
    primaryClass = "hep-ph",
    reportNumber = "CERN-PH-TH-2012-300",
    doi = "10.1103/PhysRevLett.110.162001",
    journal = "Phys. Rev. Lett.",
    volume = "110",
    number = "16",
    pages = "162001",
    year = "2013"
}

@article{Berta:2019hnj,
    author = "Berta, P. and Masetti, L. and Miller, D.W. and Spousta, M.",
    title = "{Pileup and underlying event mitigation with iterative constituent subtraction}",
    eprint = "1905.03470",
    archivePrefix = "arXiv",
    primaryClass = "hep-ph",
    doi = "10.1007/JHEP08(2019)175",
    journal = "JHEP",
    volume = "08",
    pages = "175",
    year = "2019"
}

@article{Dokshitzer:1997in,
    author = "Dokshitzer, Yuri L. and Leder, G.D. and Moretti, S. and Webber, B.R.",
    title = "{Better jet clustering algorithms}",
    eprint = "hep-ph/9707323",
    archivePrefix = "arXiv",
    reportNumber = "CAVENDISH-HEP-97-06",
    doi = "10.1088/1126-6708/1997/08/001",
    journal = "JHEP",
    volume = "08",
    pages = "001",
    year = "1997"
}

@inproceedings{Wobisch:1998wt,
    author = "Wobisch, M. and Wengler, T.",
    title = "{Hadronization Corrections to Jet Cross Sections in Deep-Inelastic Scattering}",
    booktitle = "{Workshop on Monte Carlo Generators for HERA Physics (Plenary Starting Meeting)}",
    eprint = "hep-ph/9907280",
    archivePrefix = "arXiv",
    reportNumber = "PITHA-99-16",
    pages = "270--279",
    month = "4",
    year = "1998"
}

@article{Thaler_2011,
   title={Identifying boosted objects with N-subjettiness},
   volume={03},
   ISSN={1029-8479},
   DOI={10.1007/jhep03(2011)015},
   journal={JHEP},
   publisher={Springer Science and Business Media LLC},
   author={Thaler, Jesse and Van Tilburg, Ken},
   year={2011},
   month={Mar}
}

@article{Thaler_2012,
   title={Maximizing boosted top identification by minimizing N-subjettiness},
   volume={02},
   ISSN={1029-8479},
   DOI={10.1007/jhep02(2012)093},
   journal={JHEP},
   publisher={Springer Science and Business Media LLC},
   author={Thaler, Jesse and Van Tilburg, Ken},
   year={2012},
   month={Feb}
}

@article{LUCID2,
  author={G. Avoni and others},
  title={The new LUCID-2 detector for luminosity measurement and monitoring in ATLAS},
  journal={JINST},
  volume={13},
  number={07},
  pages={P07017},
  doi="10.1088/1748-0221/13/07/P07017",
  year={2018},
}

@article{Adam:2020kug,
      author         = "{STAR Collaboration}",
      title          = "{Measurement of Groomed Jet Substructure Observables in
                        \pp Collisions at $\sqrt{s} = 200$ GeV with STAR}",
      collaboration  = "STAR",
      year           = "2020",
      eprint         = "2003.02114",
      archivePrefix  = "arXiv",
      primaryClass   = "hep-ex",
      SLACcitation   = "%%CITATION = ARXIV:2003.02114;%%"
}

@article{Aaij:2017fak,
    author = "{LHCb Collaboration}",
    title = "{Study of J/\ensuremath{\psi} Production in Jets}",
    eprint = "1701.05116",
    archivePrefix = "arXiv",
    primaryClass = "hep-ex",
    reportNumber = "LHCB-PAPER-2016-064, CERN-EP-2017-006",
    doi = "10.1103/PhysRevLett.118.192001",
    journal = "Phys. Rev. Lett.",
    volume = "118",
    number = "19",
    pages = "192001",
    year = "2017"
}

@article{Aaij:2019ctd,
    author = "{LHCb Collaboration}",
    title = "{Measurement of charged hadron production in $Z$-tagged jets in proton-proton collisions at $\sqrt{s}=8$ TeV}",
    eprint = "1904.08878",
    archivePrefix = "arXiv",
    primaryClass = "hep-ex",
    reportNumber = "CERN-EP-2019-065, LHCb-PAPER-2019-012",
    doi = "10.1103/PhysRevLett.123.232001",
    journal = "Phys. Rev. Lett.",
    volume = "123",
    number = "23",
    pages = "232001",
    year = "2019"
}

@article{Zardoshti:2020cwl,
    author = "{ALICE Collaboration}",
    title = "{First Direct Observation of the Dead-Cone Effect}",
    eprint = "2004.05968",
    archivePrefix = "arXiv",
    primaryClass = "hep-ex",
    doi = "10.1016/j.nuclphysa.2020.121905",
    journal = "Nucl. Phys. A",
    volume = "1005",
    pages = "121905",
    year = "2021"
}

@article{Acharya:2019djg,
    author = "{ALICE Collaboration}",
    title = "{Exploration of jet substructure using iterative declustering in pp and Pb\textendash{}Pb collisions at LHC energies}",
    eprint = "1905.02512",
    archivePrefix = "arXiv",
    primaryClass = "nucl-ex",
    reportNumber = "CERN-EP-2019-087",
    doi = "10.1016/j.physletb.2020.135227",
    journal = "Phys. Lett. B",
    volume = "802",
    pages = "135227",
    year = "2020"
}

@article{Frye:2016okc,
      author         = "Frye, Christopher and Larkoski, Andrew J. and Schwartz,
                        Matthew D. and Yan, Kai",
      title          = "{Precision physics with pile-up insensitive observables}",
      year           = "2016",
      eprint         = "1603.06375",
      archivePrefix  = "arXiv",
      primaryClass   = "hep-ph",
      SLACcitation   = "%%CITATION = ARXIV:1603.06375;%%"
}

@article{Frye:2016aiz,
      author         = "Frye, Christopher and Larkoski, Andrew J. and Schwartz,
                        Matthew D. and Yan, Kai",
      title          = "{Factorization for groomed jet substructure beyond the
                        next-to-leading logarithm}",
      journal        = "JHEP",
      volume         = "07",
      year           = "2016",
      pages          = "064",
      doi            = "10.1007/JHEP07(2016)064",
      eprint         = "1603.09338",
      archivePrefix  = "arXiv",
      primaryClass   = "hep-ph",
      SLACcitation   = "%%CITATION = ARXIV:1603.09338;%%"
}

@article{Kang:2018jwa,
      author         = "Kang, Zhong-Bo and Lee, Kyle and Liu, Xiaohui and Ringer,
                        Felix",
      title          = "{The groomed and ungroomed jet mass distribution for
                        inclusive jet production at the LHC}",
      journal        = "JHEP",
      volume         = "10",
      year           = "2018",
      pages          = "137",
      doi            = "10.1007/JHEP10(2018)137",
      eprint         = "1803.03645",
      archivePrefix  = "arXiv",
      primaryClass   = "hep-ph",
      SLACcitation   = "%%CITATION = ARXIV:1803.03645;%%"
}

@article{Kang:2018vgn,
      author         = "Kang, Zhong-Bo and Lee, Kyle and Liu, Xiaohui and Ringer,
                        Felix",
      title          = "{Soft drop groomed jet angularities at the LHC}",
      journal        = "Phys. Lett. B",
      volume         = "793",
      year           = "2019",
      pages          = "41-47",
      doi            = "10.1016/j.physletb.2019.04.018",
      eprint         = "1811.06983",
      archivePrefix  = "arXiv",
      primaryClass   = "hep-ph",
      SLACcitation   = "%%CITATION = ARXIV:1811.06983;%%"
}

@article{Marzani:2017kqd,
      author         = "Marzani, Simone and Schunk, Lais and Soyez, Gregory",
      title          = "{The jet mass distribution after Soft Drop}",
      journal        = "Eur. Phys. J. C",
      volume         = "78",
      year           = "2018",
      number         = "2",
      pages          = "96",
      doi            = "10.1140/epjc/s10052-018-5579-5",
      eprint         = "1712.05105",
      archivePrefix  = "arXiv",
      primaryClass   = "hep-ph",
      SLACcitation   = "%%CITATION = ARXIV:1712.05105;%%"
}

@article{Marzani:2017mva,
      author         = "Marzani, Simone and Schunk, Lais and Soyez, Gregory",
      title          = "{A study of jet mass distributions with grooming}",
      journal        = "JHEP",
      volume         = "07",
      year           = "2017",
      pages          = "132",
      doi            = "10.1007/JHEP07(2017)132",
      eprint         = "1704.02210",
      archivePrefix  = "arXiv",
      primaryClass   = "hep-ph",
      SLACcitation   = "%%CITATION = ARXIV:1704.02210;%%"
}

@article{Cacciari:2008gp,
      author         = "Cacciari, Matteo and Salam, Gavin P. and Soyez, Gregory",
      title          = "{The anti-$k_{t}$ jet clustering algorithm}",
      journal        = "JHEP",
      volume         = "04",
      year           = "2008",
      pages          = "063",
      doi            = "10.1088/1126-6708/2008/04/063",
      eprint         = "0802.1189",
      archivePrefix  = "arXiv",
      primaryClass   = "hep-ph",
      reportNumber   = "LPTHE-07-03",
      SLACcitation   = "%%CITATION = ARXIV:0802.1189;%%"
}

@article{Ball:2012cx,
      author         = "Ball, Richard D. and others",
      title          = "{Parton distributions with LHC data}",
      journal        = "Nucl. Phys. B",
      volume         = "867",
      pages          = "244",
      doi            = "10.1016/j.nuclphysb.2012.10.003",
      year           = "2013",
      reportNumber   = "EDINBURGH-2012-08, IFUM-FT-997, FR-PHENO-2012-014,
                        RWTH-TTK-12-25, CERN-PH-TH-2012-037, SFB-CPP-12-47\",
      SLACcitation   = "%%CITATION = ARXIV:1207.1303;%%",
}

@article{Sjostrand:1984ic,
      author         = "Sj{\"o}strand, Torbjorn",
      title          = "{Jet fragmentation of multiparton configurations in a string framework}",
      journal        = "Nucl. Phys. B",
      volume         = "248",
      year           = "1984",
      pages          = "469-502",
      doi            = "10.1016/0550-3213(84)90607-2",
      reportNumber   = "DESY-T-84-01",
      SLACcitation   = "%%CITATION = NUPHA,B248,469;%%"
}

@article{Andersson:1983ia,
      author         = "Andersson, Bo and Gustafson, G. and Ingelman, G. and
                        Sj{\"o}strand, T.",
      title          = "{Parton fragmentation and string dynamics}",
      journal        = "Phys. Rept.",
      volume         = "97",
      year           = "1983",
      pages          = "31-145",
      doi            = "10.1016/0370-1573(83)90080-7",
      reportNumber   = "LU-TP-83-10",
      SLACcitation   = "%%CITATION = PRPLC,97,31;%%"
}

@article{Sjostrand:2006za,
      author         = "Sj{\"o}strand, T and Mrenna, Stephen and Skands, Peter
                        Z.",
      title          = "{PYTHIA 6.4 physics and manual}",
      journal        = "JHEP",
      volume         = "05",
      year           = "2006",
      pages          = "026",
      doi            = "10.1088/1126-6708/2006/05/026",
      eprint         = "hep-ph/0603175",
      archivePrefix  = "arXiv",
      reportNumber   = "FERMILAB-PUB-06-052-CD-T, LU-TP-06-13",
      SLACcitation   = "%%CITATION = HEP-PH/0603175;%%"
}

@article{Sjostrand:2007gs,
      author         = "Sj{\"o}strand, Torbjorn and Mrenna, Stephen and Skands, Peter
                        Z.",
      title          = "{A brief introduction to PYTHIA 8.1}",
      journal        = "Comput. Phys. Commun.",
      volume         = "178",
      year           = "2008",
      pages          = "852-867",
      doi            = "10.1016/j.cpc.2008.01.036",
      eprint         = "0710.3820",
      archivePrefix  = "arXiv",
      primaryClass   = "hep-ph",
      reportNumber   = "CERN-LCGAPP-2007-04, LU-TP-07-28,
                        FERMILAB-PUB-07-512-CD-T",
      SLACcitation   = "%%CITATION = ARXIV:0710.3820;%%"
}

@article{Dasgupta:2013ihk,
      author         = "Dasgupta, Mrinal and Fregoso, Alessandro and Marzani,
                        Simone and Salam, Gavin P.",
      title          = "{Towards an understanding of jet substructure}",
      journal        = "JHEP",
      volume         = "09",
      year           = "2013",
      pages          = "029",
      doi            = "10.1007/JHEP09(2013)029",
      eprint         = "1307.0007",
      archivePrefix  = "arXiv",
      primaryClass   = "hep-ph",
      reportNumber   = "CERN-PH-TH-2013-145, DCPT-13-86, IPPP-13-43, LPN13-036,
                        MAN-HEP-2013-12",
      SLACcitation   = "%%CITATION = ARXIV:1307.0007;%%"
}

@article{Agostinelli:2002hh,
      author         = "Agostinelli, S. and others",
      title          = "{\textsc{Geant4} -- a simulation toolkit}",
      collaboration  = "GEANT4",
      journal        = "Nucl. Instrum. Meth. A",
      volume         = "506",
      year           = "2003",
      pages          = "250",
      doi            = "10.1016/S0168-9002(03)01368-8",
      reportNumber   = "SLAC-PUB-9350, FERMILAB-PUB-03-339",
      SLACcitation   = "%%CITATION = NUIMA,A506,250;%%"
}

@article{Dasgupta:2007wa,
    author = "Dasgupta, Mrinal and Magnea, Lorenzo and Salam, Gavin P.",
    archivePrefix = "arXiv",
    doi = "10.1088/1126-6708/2008/02/055",
    eprint = "0712.3014",
    journal = "JHEP",
    pages = "055",
    primaryClass = "hep-ph",
    reportNumber = "DFTT-27-2007, MAN-HEP-2007-41",
    title = "{Non-perturbative QCD effects in jets at hadron colliders}",
    volume = "02",
    year = "2008"
}

@phdthesis{Soyez:2018opl,
    author = "Soyez, Gregory",
    archivePrefix = "arXiv",
    doi = "10.1016/j.physrep.2019.01.007",
    eprint = "1801.09721",
    primaryClass = "hep-ph",
    school = "IPhT, Saclay",
    title = "{Pileup mitigation at the LHC: A theorist's view}",
    Type = {habilitation},
    year = "2018"
}

@article{Dreyer:2018tjj,
    author = "Dreyer, Frédéric A. and Necib, Lina and Soyez, Gregory and Thaler, Jesse",
    archivePrefix = "arXiv",
    doi = "10.1007/JHEP06(2018)093",
    eprint = "1804.03657",
    journal = "JHEP",
    pages = "093",
    primaryClass = "hep-ph",
    title = "{Recursive Soft Drop}",
    volume = "06",
    year = "2018"
}

@article{Krohn:2009th,
    author = "Krohn, David and Thaler, Jesse and Wang, Lian-Tao",
    archivePrefix = "arXiv",
    doi = "10.1007/JHEP02(2010)084",
    eprint = "0912.1342",
    journal = "JHEP",
    pages = "084",
    primaryClass = "hep-ph",
    title = "{Jet trimming}",
    volume = "02",
    year = "2010"
}

@article{Ellis:2009me,
    author = "Ellis, Stephen D. and Vermilion, Christopher K. and Walsh, Jonathan R.",
    archivePrefix = "arXiv",
    doi = "10.1103/PhysRevD.81.094023",
    eprint = "0912.0033",
    journal = "Phys. Rev. D",
    pages = "094023",
    primaryClass = "hep-ph",
    title = "{Recombination algorithms and jet substructure: Pruning as a tool for heavy particle searches}",
    volume = "81",
    year = "2010"
}

@article{Butterworth:2008iy,
    author = "Butterworth, Jonathan M. and Davison, Adam R. and Rubin, Mathieu and Salam, Gavin P.",
    archivePrefix = "arXiv",
    doi = "10.1103/PhysRevLett.100.242001",
    eprint = "0802.2470",
    journal = "Phys. Rev. Lett.",
    pages = "242001",
    primaryClass = "hep-ph",
    title = "{Jet Substructure as a New Higgs-Search Channel at the Large Hadron Collider}",
    volume = "100",
    year = "2008"
}

@article{Larkoski:2014wba,
    author = "Larkoski, Andrew J. and Marzani, Simone and Soyez, Gregory and Thaler, Jesse",
    archivePrefix = "arXiv",
    doi = "10.1007/JHEP05(2014)146",
    eprint = "1402.2657",
    journal = "JHEP",
    pages = "146",
    primaryClass = "hep-ph",
    reportNumber = "MIT-CTP-4531, DCPT-14-24, IPPP-14-12",
    title = "{Soft drop}",
    volume = "05",
    year = "2014"
}

@article{Berta:2014eza,
    author = "Berta, Peter and Spousta, Martin and Miller, David W. and Leitner, Rupert",
    archivePrefix = "arXiv",
    doi = "10.1007/JHEP06(2014)092",
    eprint = "1403.3108",
    journal = "JHEP",
    pages = "092",
    primaryClass = "hep-ex",
    title = "{Particle-level pileup subtraction for jets and jet shapes}",
    volume = "06",
    year = "2014"
}

@article{Cacciari:2014gra,
    author = "Cacciari, Matteo and Salam, Gavin P. and Soyez, Gregory",
    archivePrefix = "arXiv",
    doi = "10.1140/epjc/s10052-015-3267-2",
    eprint = "1407.0408",
    journal = "Eur. Phys. J. C",
    number = "2",
    pages = "59",
    primaryClass = "hep-ph",
    reportNumber = "CERN-PH-TH-2014-116",
    title = "{SoftKiller, a particle-level pileup removal method}",
    volume = "75",
    year = "2015"
}

@article{Ellis:1993tq,
    author = "Ellis, Stephen D. and Soper, Davison E.",
    archivePrefix = "arXiv",
    doi = "10.1103/PhysRevD.48.3160",
    eprint = "hep-ph/9305266",
    journal = "Phys. Rev. D",
    pages = "3160--3166",
    reportNumber = "CERN-TH-6860-93",
    title = "{Successive combination jet algorithm for hadron collisions}",
    volume = "48",
    year = "1993"
}

@article{Cacciari:2011ma,
    author = "Cacciari, Matteo and Salam, Gavin P. and Soyez, Gregory",
    archivePrefix = "arXiv",
    doi = "10.1140/epjc/s10052-012-1896-2",
    eprint = "1111.6097",
    journal = "Eur. Phys. J. C",
    pages = "1896",
    primaryClass = "hep-ph",
    reportNumber = "CERN-PH-TH-2011-297",
    title = "{FastJet user manual}",
    volume = "72",
    year = "2012"
}

@article{Bertolini:2014bba,
    author = "Bertolini, Daniele and Harris, Philip and Low, Matthew and Tran, Nhan",
    archivePrefix = "arXiv",
    doi = "10.1007/JHEP10(2014)059",
    eprint = "1407.6013",
    journal = "JHEP",
    pages = "059",
    primaryClass = "hep-ph",
    reportNumber = "FERMILAB-PUB-14-238-PPD, EFI-14-18, MIT-CTP-4558",
    title = "{Pileup per particle identification}",
    volume = "10",
    year = "2014"
}

@article{ApollinariG.:2017ojx,
    editor = "Apollinari, G. and Béjar Alonso, I. and Brüning, O. and Fessia, P. and Lamont, M. and Rossi, L. and Tavian, L.",
    title = "{High-Luminosity Large Hadron Collider (HL-LHC)}: {Technical Design Report V. 0.1}",
    reportNumber = "CERN-2017-007-M",
    doi = "10.23731/CYRM-2017-004",
    volume = "4/2017",
    month = "11",
    year = "2017"
}

@article{Larkoski:2013eya,
    author = "Larkoski, Andrew J. and Salam, Gavin P. and Thaler, Jesse",
    archivePrefix = "arXiv",
    doi = "10.1007/JHEP06(2013)108",
    eprint = "1305.0007",
    journal = "JHEP",
    pages = "108",
    primaryClass = "hep-ph",
    reportNumber = "MIT-CTP-4446, CERN-PH-TH-2013-066, LPN13-026",
    title = "{Energy correlation functions for jet substructure}",
    volume = "06",
    year = "2013"
}

@article{Larkoski:2014gra,
    author = "Larkoski, Andrew J. and Moult, Ian and Neill, Duff",
    archivePrefix = "arXiv",
    doi = "10.1007/JHEP12(2014)009",
    eprint = "1409.6298",
    journal = "JHEP",
    pages = "009",
    primaryClass = "hep-ph",
    reportNumber = "MIT--CTP-4588",
    title = "{Power counting to better jet observables}",
    volume = "12",
    year = "2014"
}

@article{Larkoski:2015kga,
    author = "Larkoski, Andrew J. and Moult, Ian and Neill, Duff",
    archivePrefix = "arXiv",
    doi = "10.1007/JHEP05(2016)117",
    eprint = "1507.03018",
    journal = "JHEP",
    pages = "117",
    primaryClass = "hep-ph",
    reportNumber = "MIT-CTP-4681",
    title = "{Analytic boosted boson discrimination}",
    volume = "05",
    year = "2016"
}

@article{Asquith:2018igt,
    author = "Kogler, Roman and others",
    archivePrefix = "arXiv",
    doi = "10.1103/RevModPhys.91.045003",
    eprint = "1803.06991",
    journal = "Rev. Mod. Phys.",
    number = "4",
    pages = "045003",
    primaryClass = "hep-ex",
    reportNumber = "FERMILAB-PUB-18-123-PPD",
    title = "{Jet substructure at the Large Hadron Collider: Experimental Review}",
    volume = "91",
    year = "2019"
}

@article{Larkoski:2017jix,
    author = "Larkoski, Andrew J. and Moult, Ian and Nachman, Benjamin",
    archivePrefix = "arXiv",
    doi = "10.1016/j.physrep.2019.11.001",
    eprint = "1709.04464",
    journal = "Phys. Rept.",
    pages = "1--63",
    primaryClass = "hep-ph",
    title = "{Jet substructure at the Large Hadron Collider: a review of recent advances in theory and machine learning}",
    volume = "841",
    year = "2020"
}

@article{Hoang:2017kmk,
    author = "Hoang, Andre H. and Mantry, Sonny and Pathak, Aditya and Stewart, Iain W.",
    title = "{Extracting a short distance top mass with light grooming}",
    eprint = "1708.02586",
    archivePrefix = "arXiv",
    primaryClass = "hep-ph",
    reportNumber = "MIT-CTP 4923, UWThPh-2017-25, MIT-CTP-4923, UWTHPH-2017-25",
    doi = "10.1103/PhysRevD.100.074021",
    journal = "Phys. Rev. D",
    volume = "100",
    number = "7",
    pages = "074021",
    year = "2019"
}

@article{Aad:2020zcn,
    author         = "{ATLAS Collaboration}",
    title = "{Measurement of the Lund Jet Plane Using Charged Particles in 13 TeV Proton-Proton Collisions with the ATLAS Detector}",
    eprint = "2004.03540",
    archivePrefix = "arXiv",
    primaryClass = "hep-ex",
    reportNumber = "CERN-EP-2020-030",
    doi = "10.1103/PhysRevLett.124.222002",
    journal = "Phys. Rev. Lett.",
    volume = "124",
    number = "22",
    pages = "222002",
    year = "2020"
}

@article{Aad:2020flx,
    author = "{ATLAS Collaboration}",
    title = "{Jet energy scale and resolution measured in proton-proton collisions at $\sqrt{s}=13$ TeV with the ATLAS detector}",
    eprint = "2007.02645",
    archivePrefix = "arXiv",
    primaryClass = "hep-ex",
    reportNumber = "CERN-EP-2020-083",
    month = "7",
    year = "2020"
}

@inproceedings{Soyez:2009cw,
    author = "Soyez, Gregory",
    title = "{Jet areas as a tool for background subtraction}",
    booktitle = "{25th Winter Workshop on Nuclear Dynamics}",
    eprint = "0905.2851",
    archivePrefix = "arXiv",
    primaryClass = "hep-ph",
    month = "5",
    year = "2009"
}

@article{Cacciari:2008gn,
    author = "Cacciari, Matteo and Salam, Gavin P. and Soyez, Gregory",
    title = "{The catchment area of jets}",
    eprint = "0802.1188",
    archivePrefix = "arXiv",
    primaryClass = "hep-ph",
    reportNumber = "LPTHE-07-02",
    doi = "10.1088/1126-6708/2008/04/005",
    journal = "JHEP",
    volume = "04",
    pages = "005",
    year = "2008"
}

@Article{PERF-2007-01,
    author         = "{ATLAS Collaboration}",
    title          = "{The ATLAS Experiment at the CERN Large Hadron Collider}",
    journal        = "JINST",
    volume         = "3",
    year           = "2008",
    pages          = "S08003",
    doi            = "10.1088/1748-0221/3/08/S08003",
    primaryClass   = "hep-ex",
}

@Article{SOFT-2010-01,
    author         = "{ATLAS Collaboration}",
    title          = "{The ATLAS Simulation Infrastructure}",
    journal        = "Eur. Phys. J. C",
    volume         = "70",
    year           = "2010",
    pages          = "823",
    doi            = "10.1140/epjc/s10052-010-1429-9",
    eprint         = "1005.4568",
    archivePrefix  = "arXiv",
    primaryClass   = "physics.ins-det",
}

@Article{STDM-2011-19,
    author         = "{ATLAS Collaboration}",
    title          = "{Jet mass and substructure of inclusive jets in \(\sqrt{s} = 7\,\text{TeV}\) \(pp\) collisions with the ATLAS experiment}",
    journal        = "JHEP",
    volume         = "05",
    year           = "2012",
    pages          = "128",
    doi            = "10.1007/JHEP05(2012)128",
    reportNumber   = "CERN-PH-EP-2012-031",
    eprint         = "1203.4606",
    archivePrefix  = "arXiv",
    primaryClass   = "hep-ex",
}

@Article{PERF-2012-02,
    author         = "{ATLAS Collaboration}",
    title          = "{Performance of jet substructure techniques for large-\(R\) jets in proton--proton collisions at \(\sqrt{s} = 7\,\text{TeV}\) using the ATLAS detector}",
    journal        = "JHEP",
    volume         = "09",
    year           = "2013",
    pages          = "076",
    doi            = "10.1007/JHEP09(2013)076",
    reportNumber   = "CERN-PH-EP-2013-069",
    eprint         = "1306.4945",
    archivePrefix  = "arXiv",
    primaryClass   = "hep-ex",
}

@Article{PERF-2014-03,
    author         = "{ATLAS Collaboration}",
    title          = "{Performance of pile-up mitigation techniques for jets in \(pp\) collisions at \(\sqrt{s} = 8\,\text{TeV}\) using the ATLAS detector}",
    journal        = "Eur. Phys. J. C",
    volume         = "76",
    year           = "2016",
    pages          = "581",
    doi            = "10.1140/epjc/s10052-016-4395-z",
    reportNumber   = "CERN-PH-EP-2015-206",
    eprint         = "1510.03823",
    archivePrefix  = "arXiv",
    primaryClass   = "hep-ex",
}

@Article{PERF-2014-07,
    author         = "{ATLAS Collaboration}",
    title          = "{Topological cell clustering in the ATLAS calorimeters and its performance in LHC Run~1}",
    journal        = "Eur. Phys. J. C",
    volume         = "77",
    year           = "2017",
    pages          = "490",
    doi            = "10.1140/epjc/s10052-017-5004-5",
    reportNumber   = "CERN-PH-EP-2015-304",
    eprint         = "1603.02934",
    archivePrefix  = "arXiv",
    primaryClass   = "hep-ex",
}

@Article{EXOT-2015-04,
    author         = "{ATLAS Collaboration}",
    title          = "{Search for heavy particles decaying into top-quark pairs using lepton-plus-jets events in proton--proton collisions at \(\sqrt{s} = 13\,\text{TeV}\) with the ATLAS detector}",
    journal        = "Eur. Phys. J. C",
    volume         = "78",
    year           = "2018",
    pages          = "565",
    doi            = "10.1140/epjc/s10052-018-5995-6",
    reportNumber   = "CERN-EP-2018-48",
    eprint         = "1804.10823",
    archivePrefix  = "arXiv",
    primaryClass   = "hep-ex",
}

@Article{PERF-2015-03,
    author         = "{ATLAS Collaboration}",
    title          = "{Identification of boosted, hadronically decaying \(W\) bosons and comparisons with ATLAS data taken at \(\sqrt{s} = 8\,\text{TeV}\)}",
    journal        = "Eur. Phys. J. C",
    volume         = "76",
    year           = "2016",
    pages          = "154",
    doi            = "10.1140/epjc/s10052-016-3978-z",
    reportNumber   = "CERN-PH-EP-2015-204",
    eprint         = "1510.05821",
    archivePrefix  = "arXiv",
    primaryClass   = "hep-ex",
}

@Article{PERF-2015-04,
    author         = "{ATLAS Collaboration}",
    title          = "{Identification of high transverse momentum top quarks in \(pp\) collisions at \(\sqrt{s} = 8\,\text{TeV}\) with the ATLAS detector}",
    journal        = "JHEP",
    volume         = "06",
    year           = "2016",
    pages          = "093",
    doi            = "10.1007/JHEP06(2016)093",
    reportNumber   = "CERN-EP-2016-010",
    eprint         = "1603.03127",
    archivePrefix  = "arXiv",
    primaryClass   = "hep-ex",
}

@Article{PERF-2015-09,
    author         = "{ATLAS Collaboration}",
    title          = "{Jet reconstruction and performance using particle flow with the ATLAS Detector}",
    journal        = "Eur. Phys. J. C",
    volume         = "77",
    year           = "2017",
    pages          = "466",
    doi            = "10.1140/epjc/s10052-017-5031-2",
    reportNumber   = "CERN-EP-2017-024",
    eprint         = "1703.10485",
    archivePrefix  = "arXiv",
    primaryClass   = "hep-ex",
}

@Article{EXOT-2016-25,
    author         = "{ATLAS Collaboration}",
    title          = "{Search for Dark Matter Produced in Association with a Higgs Boson Decaying to \(b\bar{b}\) using \(36\,\text{fb}^{-1}\) of \(pp\) collisions at \(\sqrt s=13\) TeV with the ATLAS Detector}",
    journal        = "Phys. Rev. Lett.",
    volume         = "119",
    year           = "2017",
    pages          = "181804",
    doi            = "10.1103/PhysRevLett.119.181804",
    reportNumber   = "CERN-EP-2017-117",
    eprint         = "1707.01302",
    archivePrefix  = "arXiv",
    primaryClass   = "hep-ex",
}

@Article{SUSY-2016-10,
    author         = "{ATLAS Collaboration}",
    title          = "{Search for supersymmetry in final states with missing transverse momentum and multiple \(b\)-jets in proton--proton collisions at \(\sqrt{s} = 13\,\text{TeV}\) with the ATLAS detector}",
    journal        = "JHEP",
    volume         = "06",
    year           = "2018",
    pages          = "107",
    doi            = "10.1007/JHEP06(2018)107",
    reportNumber   = "CERN-EP-2017-182",
    eprint         = "1711.01901",
    archivePrefix  = "arXiv",
    primaryClass   = "hep-ex",
}

@Article{STDM-2017-04,
    author         = "{ATLAS Collaboration}",
    title          = "{Measurement of the Soft-Drop Jet Mass in \(pp\) Collisions at \(\sqrt{s} = 13\,\text{TeV}\) with the ATLAS detector}",
    journal        = "Phys. Rev. Lett.",
    volume         = "121",
    year           = "2018",
    pages          = "092001",
    doi            = "10.1103/PhysRevLett.121.092001",
    reportNumber   = "CERN-EP-2017-231",
    eprint         = "1711.08341",
    archivePrefix  = "arXiv",
    primaryClass   = "hep-ex",
}

@Article{STDM-2017-14,
    author         = "{ATLAS Collaboration}",
    title          = "{Measurement of the jet mass in high transverse momentum \(Z(\to b\bar{b})\gamma\) production at \(\sqrt{s} = 13\,\text{TeV}\) using the ATLAS detector}",
    journal        = "Phys. Lett. B",
    volume         = "812",
    year           = "2021",
    pages          = "135991",
    doi            = "10.1016/j.physletb.2020.135991",
    reportNumber   = "CERN-EP-2019-115",
    eprint         = "1907.07093",
    archivePrefix  = "arXiv",
    primaryClass   = "hep-ex",
}

@Article{STDM-2017-33,
    author         = "{ATLAS Collaboration}",
    title          = "{Measurement of soft-drop jet observables in \(pp\) collisions with the ATLAS detector at \(\sqrt{s} = 13\,\text{TeV}\)}",
    journal        = "Phys. Rev. D",
    volume         = "101",
    year           = "2020",
    pages          = "052007",
    doi            = "10.1103/PhysRevD.101.052007",
    reportNumber   = "CERN-EP-2019-269",
    eprint         = "1912.09837",
    archivePrefix  = "arXiv",
    primaryClass   = "hep-ex",
}

@Article{SUSY-2017-01,
    author         = "{ATLAS Collaboration}",
    title          = "{Search for chargino and neutralino production in final states with a Higgs boson and missing transverse momentum at \(\sqrt{s} = 13\,\text{TeV}\) with the ATLAS detector}",
    journal        = "Phys. Rev. D",
    volume         = "100",
    year           = "2019",
    pages          = "012006",
    doi            = "10.1103/PhysRevD.100.012006",
    reportNumber   = "CERN-EP-2018-306",
    eprint         = "1812.09432",
    archivePrefix  = "arXiv",
    primaryClass   = "hep-ex",
}

@Article{DAPR-2018-01,
    author         = "{ATLAS Collaboration}",
    title          = "{ATLAS data quality operations and performance for 2015--2018 data-taking}",
    journal        = "JINST",
    volume         = "15",
    year           = "2020",
    pages          = "P04003",
    doi            = "10.1088/1748-0221/15/04/P04003",
    reportNumber   = "CERN-EP-2019-207",
    eprint         = "1911.04632",
    archivePrefix  = "arXiv",
    primaryClass   = "physics.ins-det",
}

@Article{HDBS-2018-31,
    author         = "{ATLAS Collaboration}",
    title          = "{Search for diboson resonances in hadronic final states in \(139\,\text{fb}^{-1}\) of \(pp\) collisions at \(\sqrt{s} = 13\,\text{TeV}\) with the ATLAS detector}",
    journal        = "JHEP",
    volume         = "09",
    year           = "2019",
    pages          = "091",
    doi            = "10.1007/JHEP09(2019)091",
    reportNumber   = "CERN-EP-2019-044",
    eprint         = "1906.08589",
    archivePrefix  = "arXiv",
    primaryClass   = "hep-ex",
}

@Article{JETM-2018-02,
    author         = "{ATLAS Collaboration}",
    title          = "{In situ calibration of large-radius jet energy and mass in \(13\,\text{TeV}\) proton--proton collisions with the ATLAS detector}",
    journal        = "Eur. Phys. J. C",
    volume         = "79",
    year           = "2019",
    pages          = "135",
    doi            = "10.1140/epjc/s10052-019-6632-8",
    reportNumber   = "CERN-EP-2018-191",
    eprint         = "1807.09477",
    archivePrefix  = "arXiv",
    primaryClass   = "hep-ex",
}

@Article{JETM-2018-03,
    author         = "{ATLAS Collaboration}",
    title          = "{Performance of top-quark and \(W\)-boson tagging with ATLAS in Run~2 of the LHC}",
    journal        = "Eur. Phys. J. C",
    volume         = "79",
    year           = "2019",
    pages          = "375",
    doi            = "10.1140/epjc/s10052-019-6847-8",
    reportNumber   = "CERN-EP-2018-192",
    eprint         = "1808.07858",
    archivePrefix  = "arXiv",
    primaryClass   = "hep-ex",
}

@Article{PIX-2018-001,
    author         = "Abbott, B. and others",
    title          = "{Production and integration of the ATLAS Insertable B-Layer}",
    journal        = "JINST",
    volume         = "13",
    year           = "2018",
    pages          = "T05008",
    doi            = "10.1088/1748-0221/13/05/T05008",
    eprint         = "1803.00844",
    archivePrefix  = "arXiv",
    primaryClass   = "physics.ins-det",
}

@Booklet{ATL-SOFT-PUB-2020-001,
    author         = "{ATLAS Collaboration}",
    title          = "{ATLAS Computing Acknowledgements}",
    howpublished   = "{ATL-SOFT-PUB-2020-001}",
    url            = "https://cds.cern.ch/record/2717821",
}

@Report{ATLAS-TDR-2010-19,
    author         = "{ATLAS Collaboration}",
    title          = "{ATLAS Insertable B-Layer Technical Design Report}",
    type           = "ATLAS-TDR-19; CERN-LHCC-2010-013",
    year           = "2010",
    url            = "https://cds.cern.ch/record/1291633",
}

@Article{CMS-JME-13-006,
    author         = "{CMS Collaboration}",
    title          = "{Identification techniques for highly boosted \(W\) bosons that decay into hadrons}",
    journal        = "JHEP",
    volume         = "12",
    year           = "2014",
    pages          = "017",
    doi            = "10.1007/JHEP12(2014)017",
    reportNumber   = "CERN-PH-EP-2014-241",
    eprint         = "1410.4227",
    archivePrefix  = "arXiv",
    primaryClass   = "hep-ex",
}

@Article{CMS-PRF-14-001,
    author         = "{CMS Collaboration}",
    title          = "{Particle-flow reconstruction and global event description with the CMS detector}",
    journal        = "JINST",
    volume         = "12",
    year           = "2017",
    pages          = "P10003",
    doi            = "10.1088/1748-0221/12/10/P10003",
    reportNumber   = "CERN-EP-2017-110",
    eprint         = "1706.04965",
    archivePrefix  = "arXiv",
    primaryClass   = "hep-ex",
}

@Article{CMS-SMP-16-010,
    author         = "{CMS Collaboration}",
    title          = "{Measurements of the differential jet cross section as a function of the jet mass in dijet events from proton--proton collisions at \(\sqrt{s} = 13\,\text{TeV}\)}",
    journal        = "JHEP",
    volume         = "11",
    year           = "2018",
    pages          = "113",
    doi            = "10.1007/JHEP11(2018)113",
    reportNumber   = "CERN-EP-2018-180",
    eprint         = "1807.05974",
    archivePrefix  = "arXiv",
    primaryClass   = "hep-ex",
}

@Article{CMS-B2G-17-018,
    author         = "{CMS Collaboration}",
    title          = "{Search for single production of vector-like quarks decaying to a top quark and a \(W\) boson in proton--proton collisions at \(\sqrt{s} = 13\,\text{TeV}\)}",
    journal        = "Eur. Phys. J. C",
    volume         = "79",
    year           = "2019",
    pages          = "90",
    doi            = "10.1140/epjc/s10052-019-6556-3",
    reportNumber   = "CERN-EP-2018-230",
    eprint         = "1809.08597",
    archivePrefix  = "arXiv",
    primaryClass   = "hep-ex",
}

@Article{CMS-B2G-17-019,
    author         = "{CMS Collaboration}",
    title          = "{Search for production of Higgs boson pairs in the four \(b\) quark final state using large-area jets in proton--proton collisions at \(\sqrt{s} = 13\,\text{TeV}\)}",
    journal        = "JHEP",
    volume         = "01",
    year           = "2019",
    pages          = "040",
    doi            = "10.1007/JHEP01(2019)040",
    reportNumber   = "CERN-EP-2018-195",
    eprint         = "1808.01473",
    archivePrefix  = "arXiv",
    primaryClass   = "hep-ex",
}

@Article{CMS-TOP-17-013,
    author         = "{CMS Collaboration}",
    title          = "{Measurement of jet substructure observables in \(t\bar{t}\) events from proton--proton collisions at \(\sqrt{s} = 13\,\text{TeV}\)}",
    journal        = "Phys. Rev. D",
    volume         = "98",
    year           = "2018",
    pages          = "092014",
    doi            = "10.1103/PhysRevD.98.092014",
    reportNumber   = "CERN-EP-2018-214",
    eprint         = "1808.07340",
    archivePrefix  = "arXiv",
    primaryClass   = "hep-ex",
}

@Article{CMS-B2G-18-002,
    author         = "{CMS Collaboration}",
    title          = "{A multi-dimensional search for new heavy resonances decaying to boosted \(WW\), \(WZ\), or \(ZZ\) boson pairs in the dijet final state at \(13\,\text{TeV}\)}",
    journal        = "Eur. Phys. J. C",
    volume         = "80",
    year           = "2020",
    pages          = "237",
    doi            = "10.1140/epjc/s10052-020-7773-5",
    reportNumber   = "CERN-EP-2019-107",
    eprint         = "1906.05977",
    archivePrefix  = "arXiv",
    primaryClass   = "hep-ex",
}

@Article{CMS-EXO-18-012,
    author         = "{CMS Collaboration}",
    title          = "{Search for low mass vector resonances decaying into quark--antiquark pairs in proton--proton collisions at \(\sqrt{s} = 13\,\text{TeV}\)}",
    journal        = "Phys. Rev. D",
    volume         = "100",
    year           = "2019",
    pages          = "112007",
    doi            = "10.1103/PhysRevD.100.112007",
    reportNumber   = "CERN-EP-2019-176",
    eprint         = "1909.04114",
    archivePrefix  = "arXiv",
    primaryClass   = "hep-ex",
}

@Article{CMS-JME-18-001,
    author         = "{CMS Collaboration}",
    title          = "{Pileup mitigation at CMS in \(13\,\text{TeV}\) data}",
    journal        = "JINST",
    volume         = "15",
    year           = "2020",
    pages          = "P09018",
    doi            = "10.1088/1748-0221/15/09/P09018",
    reportNumber   = "CERN-EP-2020-017",
    eprint         = "2003.00503",
    archivePrefix  = "arXiv",
    primaryClass   = "hep-ex",
}

@Article{CMS-JME-18-002,
    author         = "{CMS Collaboration}",
    title          = "{Identification of heavy, energetic, hadronically decaying particles using machine-learning techniques}",
    journal        = "JINST",
    volume         = "15",
    year           = "2020",
    pages          = "P06005",
    doi            = "10.1088/1748-0221/15/06/P06005",
    reportNumber   = "CERN-EP-2020-037",
    eprint         = "2004.08262",
    archivePrefix  = "arXiv",
    primaryClass   = "hep-ex",
}

@Article{CMS-HIG-19-003,
    author         = "{CMS Collaboration}",
    title          = "{Inclusive search for highly boosted Higgs bosons decaying to bottom quark-antiquark pairs in proton--proton collisions at \(\sqrt{s} = 13\,\text{TeV}\)}",
    journal        = "JHEP",
    volume         = "12",
    year           = "2020",
    pages          = "085",
    doi            = "10.1007/JHEP12(2020)085",
    reportNumber   = "CERN-EP-2020-107",
    eprint         = "2006.13251",
    archivePrefix  = "arXiv",
    primaryClass   = "hep-ex",
}

@Article{CMS-TOP-19-005,
    author         = "{CMS Collaboration}",
    title          = "{Measurement of the jet mass distribution and top quark mass in hadronic decays of boosted top quarks in \(pp\) collisions at \(\sqrt{s} = 13\,\text{TeV}\)}",
    journal        = "Phys. Rev. Lett.",
    volume         = "124",
    year           = "2020",
    pages          = "202001",
    doi            = "10.1103/PhysRevLett.124.202001",
    reportNumber   = "CERN-EP-2019-226",
    eprint         = "1911.03800",
    archivePrefix  = "arXiv",
    primaryClass   = "hep-ex",
}

@Booklet{ATLAS-CONF-2015-037,
    author         = "{ATLAS Collaboration}",
    title          = "{Monte Carlo Calibration and Combination of In-situ Measurements of Jet Energy Scale, Jet Energy Resolution and Jet Mass in ATLAS}",
    howpublished   = "{ATLAS-CONF-2015-037}",
    url            = "https://cds.cern.ch/record/2044941",
    year           = "2015",
}

@Booklet{ATLAS-CONF-2017-065,
    author         = "{ATLAS Collaboration}",
    title          = "{Constituent-level pile-up mitigation techniques in ATLAS}",
    howpublished   = "{ATLAS-CONF-2017-065}",
    url            = "https://cds.cern.ch/record/2281055",
    year           = "2017",
}

@Booklet{ATLAS-CONF-2019-021,
    author         = "{ATLAS Collaboration}",
    title          = "{Luminosity determination in \(pp\) collisions at \(\sqrt{s} = 13\,\text{TeV}\) using the ATLAS detector at the LHC}",
    howpublished   = "{ATLAS-CONF-2019-021}",
    url            = "https://cds.cern.ch/record/2677054",
    year           = "2019",
}

@Booklet{ATL-PHYS-PUB-2014-004,
    author         = "{ATLAS Collaboration}",
    title          = "{Performance of Boosted \(W\) Boson Identification with the ATLAS Detector}",
    howpublished   = "{ATL-PHYS-PUB-2014-004}",
    url            = "https://cds.cern.ch/record/1690048",
    year           = "2014",
}

@Booklet{ATL-PHYS-PUB-2014-021,
    author         = "{ATLAS Collaboration}",
    title          = "{ATLAS Pythia~8 tunes to \(7~\text{TeV}\) data}",
    howpublished   = "{ATL-PHYS-PUB-2014-021}",
    url            = "https://cds.cern.ch/record/1966419",
    year           = "2014",
}

@Booklet{ATL-PHYS-PUB-2015-051,
    author         = "{ATLAS Collaboration}",
    title          = "{Early Inner Detector Tracking Performance in the 2015 Data at \(\sqrt{s} = 13~\text{TeV}\)}",
    howpublished   = "{ATL-PHYS-PUB-2015-051}",
    url            = "https://cds.cern.ch/record/2110140",
    year           = "2015",
}

@Booklet{ATL-PHYS-PUB-2016-017,
    author         = "{ATLAS Collaboration}",
    title          = "{The Pythia~8 A3 tune description of ATLAS minimum bias and inelastic measurements incorporating the Donnachie--Landshoff diffractive model}",
    howpublished   = "{ATL-PHYS-PUB-2016-017}",
    url            = "https://cds.cern.ch/record/2206965",
    year           = "2016",
}

@Booklet{ATL-PHYS-PUB-2017-015,
    author         = "{ATLAS Collaboration}",
    title          = "{Improving jet substructure performance in ATLAS using Track-CaloClusters}",
    howpublished   = "{ATL-PHYS-PUB-2017-015}",
    url            = "https://cds.cern.ch/record/2275636",
    year           = "2017",
}

\clearpage 
 
\begin{flushleft}
\hypersetup{urlcolor=black}
{\Large The ATLAS Collaboration}

\bigskip

\AtlasOrcid[0000-0002-6665-4934]{G.~Aad}$^\textrm{\scriptsize 102}$,    
\AtlasOrcid[0000-0002-5888-2734]{B.~Abbott}$^\textrm{\scriptsize 128}$,    
\AtlasOrcid[0000-0002-7248-3203]{D.C.~Abbott}$^\textrm{\scriptsize 103}$,    
\AtlasOrcid[0000-0002-2788-3822]{A.~Abed~Abud}$^\textrm{\scriptsize 36}$,    
\AtlasOrcid[0000-0002-1002-1652]{K.~Abeling}$^\textrm{\scriptsize 53}$,    
\AtlasOrcid[0000-0002-2987-4006]{D.K.~Abhayasinghe}$^\textrm{\scriptsize 94}$,    
\AtlasOrcid[0000-0002-8496-9294]{S.H.~Abidi}$^\textrm{\scriptsize 167}$,    
\AtlasOrcid[0000-0002-8279-9324]{O.S.~AbouZeid}$^\textrm{\scriptsize 40}$,    
\AtlasOrcid{N.L.~Abraham}$^\textrm{\scriptsize 156}$,    
\AtlasOrcid[0000-0001-5329-6640]{H.~Abramowicz}$^\textrm{\scriptsize 161}$
% ...
\bigskip
\\

% ...
$^{36}$CERN, Geneva; Switzerland.\\
% ...
$^{40}$Niels Bohr Institute, University of Copenhagen, Copenhagen; Denmark.\\
% ...
$^{53}$II. Physikalisches Institut, Georg-August-Universit\"{a}t G\"ottingen, G\"ottingen; Germany.\\
% ...
$^{94}$Department of Physics, Royal Holloway University of London, Egham; United Kingdom.\\
% ...
$^{102}$CPPM, Aix-Marseille Universit\'e, CNRS/IN2P3, Marseille; France.\\
$^{103}$Department of Physics, University of Massachusetts, Amherst MA; United States of America.\\
% ...
$^{128}$Homer L. Dodge Department of Physics and Astronomy, University of Oklahoma, Norman OK; United States of America.\\
% ...
$^{156}$Department of Physics and Astronomy, University of Sussex, Brighton; United Kingdom.\\
% ...
$^{161}$Raymond and Beverly Sackler School of Physics and Astronomy, Tel Aviv University, Tel Aviv; Israel.\\
% ...
$^{167}$Department of Physics, University of Toronto, Toronto ON; Canada.\\
% ...

\end{flushleft}


\end{document}